\documentclass[useAMS,usenatbib,fleqn]{mn2e}

\usepackage{verbatim}

\usepackage{epsfig}
\usepackage{amssymb}
\usepackage{natbib}
\usepackage{aas_macros}
\usepackage{times}

\usepackage{multirow}
\usepackage{xcolor}

\usepackage{soul}

\usepackage{graphics}
\usepackage{amsfonts}
\usepackage{amsmath}
\usepackage{layout}
\usepackage{float}
\usepackage{amssymb}
\usepackage{epsfig}
\usepackage{caption}

\usepackage{hyperref}
\usepackage{orcidlink}

\definecolor{mywhite}{rgb}{0, 0, 0}

\newcommand{\figu}{Figure~}
\newcommand{\figus}{Figures~}
\newcommand{\eq}{Equation~}

\newcommand{\sect}{Section~}

\def\S4G{S$^4$G}

\newcommand{\sis}{$\sigma$}
\newcommand{\sise}{\sigma}

\newcommand{\mbh}{$M_{\rm bh}$}
\newcommand{\mbhe}{M_{\rm bh}}

\newcommand{\mhalo}{$M_{\rm halo}$}

\newcommand{\mstar}{$M_{\rm gal}$}
\newcommand{\mstare}{M_{\rm gal}}
\newcommand{\mgal}{$M_{\rm gal}$}
\newcommand{\mgale}{M_{\rm gal}}
\newcommand{\mbulge}{$M_{\rm sph}$}

\newcommand{\msph}{$M_{\rm sph}$}
\newcommand{\msphe}{M_{\rm sph}}

\newcommand{\msune}{M_{\odot}}

\newcommand{\TNG}{TNG}
\newcommand{\simba}{Simba}

\def\ls{\lower 2pt \hbox{$\;\scriptscriptstyle \buildrel<\over\sim\;$}}
\def\gs{\lower 2pt \hbox{$\;\scriptscriptstyle \buildrel>\over\sim\;$}}

\title[Scalings and Residuals]{Probing the co-evolution of SMBHs and their hosts from scaling relations pairwise residuals: dominance of stellar velocity dispersion and host halo mass}%{The pivotal role of stellar velocity dispersion in setting the scaling relations between Supermassive Black Holes and their galactic hosts: the contrasting view of observational data and state-of-the-art models}

\author[F. Shankar et al.]
{Francesco Shankar$^{\orcidlink{0000-0001-8973-5051},1}$\thanks{E-mail:$\;\;\;\;\;\;$F.Shankar@soton.ac.uk},
Mariangela Bernardi$^{2}$,
Daniel Roberts$^{\orcidlink{0009-0009-7662-0445},1}$,
Miguel Arana-Catania$^{3}$,
\newauthor Tobias Grubenmann$^{4,1}$,
Melanie Habouzit$^{5,6}$,
Amy Smith$^{1}$,
Christopher Marsden$^{1}$,
Karthik \newauthor Mahesh Varadarajan$^{1}$,
Alba Vega Alonso Tetilla$^{\orcidlink{0000-0002-6916-9133}, 1}$,
Daniel Angl\'{e}s-Alc\'{a}zar$^{\orcidlink{0000-0001-5769-4945},7}$,
Lumen Boco$^{8}$,
\newauthor
Duncan Farrah$^{\orcidlink{0000-0003-1748-2010},9,10}$,
Hao Fu$^{\orcidlink{0009-0002-8051-1056},11,1}$,
Henryk Haniewicz$^{1}$,
Andrea Lapi$^{12}$,
Christopher C. \newauthor Lovell$^{13}$,
Nicola Menci$^{14}$,
Meredith Powell$^{15}$,
Federica Ricci$^{16}$
\\
$1$ School of Physics and Astronomy, University of Southampton, Highfield, Southampton, SO17 1BJ, UK\\
$2$ Department of Physics and Astronomy, University of Pennsylvania, 209 South 33rd St, Philadelphia, PA 19104, USA\\
$3$ Digital Scholarship at Oxford, University of Oxford, Oxford, OX1 3BG, UK\\
$4$ School of Computing, Engineering and the Built Environment, Edinburgh Napier University, Edinburgh, EH14 1DJ, UK\\
$5$ Department of Astronomy, University of Geneva, Chemin d’Ecogia, CH-1290 Versoix, Switzerland\\
$6$ Max-Planck-Institut f\"{u}r Astronomie, K\"{o}nigstuhl 17, D-69117 Heidelberg, Germany\\
$7$ Department of Physics, University of Connecticut, 196 Auditorium Road, U-3046, Storrs, CT 06269-3046, USA\\
$8$ Universit\"{a}t Heidelberg, Zentrum f\"{u}r Astronomie, Institut f\"{u}r theoretische Astrophysik, Albert-Ueberle-Str. 2, 69120 Heidelberg, Germany\\
$9$ Department of Physics and Astronomy, University of Hawai`i at M\=anoa, 2505 Correa Rd., Honolulu, HI, 96822, USA\\
$10$ Institute for Astronomy, University of Hawai`i,  2680 Woodlawn Dr., Honolulu, HI, 96822, USA\\
$11$ Center for Astronomy and Astrophysics and Department of Physics, Fudan University, Shanghai 200438, People’s Republic of China\\
$12$ SISSA, Via Bonomea 265, 34136 Trieste, Italy\\
$13$ Institute of Cosmology and Gravitation, University of Portsmouth, Burnaby Road, Portsmouth, PO1 3FX, UK\\
$14$ INAF – Osservatorio Astronomico di Roma, Via Frascati 33, 00078 Monte Porzio, Italy\\
$15$ Leibniz-Institut fur Astrophysik Potsdam (AIP), An der Sternwarte 16, D-14482 Potsdam, Germany\\
$16$ Dipartimento di Matematica e Fisica, Universit\`{a} Roma Tre, via della Vasca Navale 84, I-00146 Roma, Italy
}

\date{Accepted XXX. Received YYY; in original form ZZZ}

\pubyear{2025}

\begin{document}
\label{firstpage}
\pagerange{\pageref{firstpage}--\pageref{lastpage}}
\maketitle

\begin{abstract}
The correlations between Supermassive Black Holes (SMBHs) and their host galaxies still defy our understanding from both the observational and theoretical perspectives. Here we perform pairwise residual analysis on the latest sample of local inactive galaxies with a uniform calibration of their photometric properties and with dynamically measured masses of their central SMBHs. The residuals reveal that stellar velocity dispersion \sis\ and, possibly host dark matter halo mass \mhalo, appear as the galactic properties most correlated with SMBH mass, with a secondary (weaker) correlation with spheroidal (bulge) mass, as also corroborated by additional Machine Learning tests. These findings may favour energetic/kinetic feedback from Active Galactic Nuclei (AGN) as the main driver in shaping SMBH scaling relations. Two state-of-the-art hydrodynamic simulations, inclusive of kinetic AGN feedback, are able to broadly capture the mean trends observed in the residuals, although they tend to either favour \mbulge\ as the most fundamental property, or generate too flat residuals. Increasing AGN feedback kinetic output does not improve the comparison with the data. In the Appendix we also show that the galaxies with dynamically measured SMBHs are biased high in \sis\ at fixed luminosity with respect to the full sample of local galaxies, proving that this bias is not a byproduct of stellar mass discrepancies. Overall, our results suggest that probing the SMBH-galaxy scaling relations in terms of total stellar mass alone may induce biases, and that either current data sets are incomplete, and/or that more insightful modelling is required to fully reproduce observations.
\end{abstract}


\section{Introduction}
\label{sec|intro}

%additional tests to be run *after* submission and *before* resubmission:
%-1- check residuals against Re and n in Sahu's sample.
%-2- check residuals in AGN sample from Nacho
%-3- check residuals in EAGLE (ask Chris)
%-4- check residuals with halo mass at fixed sigma (look at local active and inactive samples)

Supermassive black holes (SMBHs) appear to be ubiquitous in the cores of local galaxies measured with sufficient resolution. Their masses \mbh\ correlate with their host galaxy physical properties such as stellar velocity dispersion \sis\ and bulge \mbulge\ or total galaxy stellar mass \mstar\ \citep[e.g.,][]{FerrareseFord,Beifiori12,Kormendy2013,McConnellMa,Graham2016,Saglia16}. The very existence of these correlations suggests a degree of co-evolution between SMBHs and their hosts. A number of physical processes such as gas accretion, cold flows, fly-bys, mergers, or secular instabilities may have all contributed to this co-evolution by promoting star formation and stellar mass growth in the host galaxies whilst triggering gas accretion and SMBH mergers onto the central SMBHs \citep[e.g.,][]{Granato04,Cattaneo06,Shankar12Mergers,Menci14,Fontanot20}, although SMBHs may have also preceded the formation of their galactic hosts, at least at redshifts $z \gtrsim 5$ \citep[e.g.,][]{Hu22,Ding23,Kokorev23,Maiolino23,Matthee23,Pacucci23,Bogdan24,Greene24,InaIchi24,LiSilverman24}. 

Theoretical models suggest that SMBHs will eventually follow scaling relations between SMBH mass and their host galaxy properties, although the degree of coeval evolution depends on the type of physical processes regulating the mass growth of the two systems \citep[e.g.,][]{SomervilleDave2015,Byrne23}. Accreting SMBHs shining as Active Galactic Nuclei (AGN), could launch powerful winds \citep[e.g.,][]{Cano12,Farrah12,Cicone2014,Carniani16,Fiore17,Musiimenta23} and/or jets that may potentially halt star formation in the host galaxies via removal/heating of the gas, self-regulating black hole growth, predicting a steep and tight correlation with stellar velocity dispersion $\mbhe \propto \sise^{\alpha}$, with $\alpha\sim 3-5$ \citep[e.g.,][]{Silk1998,CavaliereVittorini,King2003,Granato04,DiMatteo2005,Fabian2012,Sijacki2015,Menci23}, or even a correlation with the potential well of the host galaxy of the type $\mbhe \propto \mgale\sigma^2$ \citep[e.g.,][]{Hopkins2007_BHplane}, although strong episodes of AGN feedback are not necessarily a strict prerequisite to yield tight scaling relations between SMBHs and their hosts \citep[e.g.,][]{Granato04,Angles-Alcazar2013}. SMBH mergers could have also played a significant role in shaping SMBH-galaxy scaling relations. Repeated galaxy and SMBH mergers could induce a linear relation preferentially between SMBH mass \mbh\ and host galaxy stellar mass \mstar\ or \mbulge\ \cite[e.g.,][]{Jahnke2011}, modulating their scatter and redshift evolution \citep[e.g.,][]{Robertson06,Hirschmann10}, or generating breaks in particular in the \mbh-\mgal\ relation due to the impact of dry mergers, or in general mergers being more frequent at high masses and/or in specific galaxy subsamples \citep[e.g.,][and references therein]{Graham23HubbleSequenceMbh,Graham23SplittingLentils}. The combined and different impact of supernova and AGN feedback respectively dominating below and above the characteristic mass of $\mstare \sim 3\times 10^{10}\, \msune$, can also generate breaks in the SMBH scaling relations  \citep[e.g.,][]{Cirasuolo05,Shankar06,Fontanot15}, albeit mounting evidence from observational and theoretical works points to a non-negligible role of AGN activity even in low mass galaxies \citep[e.g.,][]{Penny18,Arjona24,Bicha24,Mezcua24}.  

Despite intense theoretical and observational work undertaken in recent decades, the origin, shape, and evolution of the SMBH-galaxy scaling relations remain largely poorly understood. Current calibrations of these relations are usually based on samples of inactive SMBHs with robust dynamical mass measurements limited to around a hundred sources \citep[e.g.,][]{Saglia16,Sahu19}, orders of magnitude smaller than the samples adopted to study the scaling relations controlling the structural and dynamical properties of galaxies \citep[e.g.,][]{Bernardi11curvature,Bernardi11mergers,Marsden22,Figueira24}. The calibrations of the SMBH scaling relations based on AGN sources are extracted from significantly larger samples than those characterizing dynamically measured inactive SMBHs, but still provide different results, sometimes showing significant systematic offsets with respect to the local relations \citep[e.g.,][]{Reines2015,Shankar19,Farrah23}. 
The pairwise residual analysis has been put forward by a series of seminal papers as a powerful tool to dissect the most fundamental relations between SMBH mass and their host galaxy properties, adding critical insight into the physical processes responsible for generating these scaling relations. The correlations between residuals are, in fact, an efficient way of determining whether one variable $Y$ is directly dependent on another variable $X$ or whether the dependence originates from a third variable $Z$. If $Y$ depends exclusively on $X$, then the residuals of the correlations with $X$ should be uncorrelated. The residual analysis is characterised by at least three key features that make this an ideal method to study scaling relations, as demonstrated by extensive Monte Carlo simulations \citep[e.g.,][]{Bernardi07,Hopkins2007_BHplane,Shankar16,Shankar17,Marsden20}: 1) it clearly identifies the dependence of SMBH mass on a single galactic variable, while fixing another one, thus avoiding over-interpreting the dependence of SMBH on a variable in a direct relation; 2) it is as statistically robust as a direct scaling relation, providing equivalent, if not tighter, constraints on the slopes; 3) it is less affected by potential biases that could distort the normalization of the direct scaling relations. Residuals are thus ideal for unveiling the strength of any underlying dependence of SMBH mass on, e.g., stellar velocity dispersion or effective radius at fixed total/bulge stellar mass, therefore setting stringent constraints on the most fundamental galactic property linked to SMBH. Alternatively, it may unveil the existence of any SMBH ``fundamental plane'', where the SMBH mass is linked to two (or more) galactic properties in equal measure \citep[e.g.,][]{Hopkins_2007_FP_observed,Saglia16}.  

Previous work based on residual analysis applied to the local SMBH dynamical samples available at the time, indicated a non-negligible correlation of SMBH mass on both stellar velocity dispersion and galactic stellar or even dynamical mass, possibly supporting the view of an underlying dependence of SMBH on the galactic potential well of the type $\mbhe \propto \mstare \sise^2$ \citep[e.g.,][]{Hopkins2007_BHplane,Iannella21}. Other analyses carried out on more recent and/or more uniform SMBH galaxy samples with detailed Monte Carlo quantification of the statistical uncertainties, still showed a hint for a possible SMBH fundamental plane relation, but with a significantly stronger and steeper dependence on stellar velocity dispersion and a weaker and flatter correlation with total host galaxy stellar mass \citep[e.g.,][]{Bernardi07,Shankar16,Marsden20}. Residual analysis performed by \citet{Shankar19} on uniform local samples of type 1 AGN from \citet{VdB15} and \citet{HoKim14} confirmed these findings. 

Some preliminary tests performed on theoretical models inclusive of AGN feedback struggled to reproduce the strong residual with stellar velocity dispersion at fixed galaxy stellar mass \citep{Barausse17,Menci23}. \citet{Barausse17} showed that their semi-analytic model, once tuned to reproduce the (mean) correlation of black hole mass with velocity dispersion, was not able to simultaneously account for the correlation with stellar mass, in line with what was also found by \citet{Sijacki2015} in the Illustris simulation, a possible signature of biases in the SMBH-galaxy scaling relations. In addition, the residual analyses performed by \citet{Barausse17} on their semi-analytic model and the Horizon-AGN hydrodynamic simulation \citep{Dubois16}, both showed a weak correlation with galaxy stellar mass at fixed stellar velocity dispersion. \citet{Menci23}, more recently, added a new physical treatment of AGN-driven winds in their semi-analytic model of galaxy formation, with outflow expansion and mass outflow rates depending on AGN luminosity, halo circular velocity, and gas fractions. From pairwise residual analysis they still found that the model predicts equally strong correlations of SMBH mass with stellar mass, stellar velocity dispersion, and host halo mass, at variance with some of the empirical results discussed above.    

The purpose of this paper is two-fold. On the one hand, we will be revisiting the pairwise residuals of SMBH scaling relations, making use of the latest SMBH-galaxy sample from \citet{Sahu19} with uniform measurements of the host galaxy and bulge stellar masses, along with other photometric properties. On the other hand, we will also compare with two state-of-the-art hydrodynamic simulations \TNG\ and \simba\ with different implementations of SMBH growth and AGN feedback recipes and efficiencies. We will show that the residuals on the new data clearly point to stellar velocity dispersion, and even possibly to host halo mass, as the dominant host properties linked to SMBH, in line with some of the previous claims. For completeness, we will also compare the residuals outputs with the predictions from a variety of Machine Learning regression algorithms that largely confirm the results of the residuals, although not providing the same level of information. In Appendix~\ref{Appendix:Bias} we will also revisit and confirm the existence of a bias between the local SMBH galactic sample and the larger comparison sample of local galaxies. 

In what follows, wherever relevant, we will adopt a reference cosmology with $h=0.7$, $\Omega_m=0.3$, $\Omega_{\Lambda}=0.7$. The theoretical models may adopt slightly different choices of cosmological parameters, but these differences are small and do not affect any of the results presented in the next sections. Differences in stellar mass estimates between models and data may be present, but they do not affect the residuals analysis, as further discussed below. 

\section{Data}
\label{sec|data}

\subsection{Observational data}
\label{subsec|obsdata}

In this work we make use of the sample of local galaxies with dynamical mass measurements of their central SMBHs by \citet{Sahu19b}, which in total comprises around 150 galaxies. Of these, following \citet{Sahu23}, we retain 73 Early-Type galaxies (ETGs) and 28 Late-Type galaxies (LTGs), which have uniformly calibrated $3.6\mu$m Spitzer photometry and detailed galaxy modelling and decompositions performed by \citet{SavorgnanGraham16}, \citet{Davis19}, and \citet{Sahu19a}. From these two samples we further remove four ETGs, NGC1194, NGC1316, NGC5018, NGC5128, which are classified as mergers by \citep{Kormendy2013}, and two LTGs, NGC4395 and NGC6926, which do not have secure SMBH mass measurements. Our final SMBH sample is reported in Tables\ref{Table 1} and \ref{Table 2}. We convert all galaxy AB total absolute magnitudes to stellar masses adopting as magnitude of the sun $M_{3.6\mu {\rm m},\odot}=6.02$ and a constant mass-to-light ratio at $3.6\mu$m of $(M/\msune)/(L/L_{\odot})=0.6$ as in \citet{Sahu19b}. We note that, as accurately described by \citet{Forbes17}, the mass-to-light ratio at $3.6\mu$m is highly insensitive to metallicity, and only weakly to stellar ages, with a small uncertainty of $\pm0.1$ dex, further supporting our choice of a constant conversion between luminosity and stellar mass. We also stress that the exact value chosen for the (constant) mass-to-light ratio is irrelevant to the residual analysis. To each galaxy we associate a stellar velocity dispersion \sis\ and associated error from the Hyperleda catalogue \citep{Paturel03Hyperleda}, with values corrected to a common aperture of 0.595 kpc. Following \citet{Sahu19a}, we assign a 0.2 mag error to magnitudes (N. Sahu, private communication), which translates into a 0.08 dex error in total galaxy stellar mass when ignoring any error in mass-to-light ratio. In addition, we will also study residuals against stellar spheroidal mass\footnote{In this work we use the word ``spheroid'' as synonym of ``bulge''.}, effective radius $R_e$, and S\'{e}rsic index $n$, all taken from Table A1 in \citet{Sahu2020}. %To each of these quantities we assign a statistical error of, respectively, 18\% for bulges\footnote{The errors on spheroidal masses quoted in \citet{Sahu2020} are sometimes larger than the variance in the distribution of the data and thus cannot be used for the residual analysis.} (as for stellar masses) and 10\% for effective radii and S\'{e}rsic indices. 
As no specific errors have been reported for effective radii and S\'{e}rsic indices, we assign to these galactic properties typical uncertainties of 10\%, noticing that moderate variations to these errors have minimal impact on the pairwise residuals. %Our final samples are reported in additional Tables in electronic format. 
In \sect\ref{sec|discu} we also explore the residuals on the subsample of 41 galaxies from \citet{Sahu23} with dark matter halo mass measurements from \citet{Marasco21}. We assign to all halo masses an error of 0.24 dex, which is the typical error reported by \citet{Marasco21} for most of the galaxies in their sample. Some early-type galaxies have quoted uncertainties significantly larger than 0.24 dex, but if included in the residuals, would generate unrealistic values of the Pearson correlation coefficient (larger than unity). 
In addition, Monte Carlo simulations based on random extractions of the data points, without considering any measurement error but only fitting the raw distribution of points at each iteration, provide very similar results to the full residual analysis inclusive of the assumed statistical uncertainties. We also performed various stability tests by recalibrating the Hyperleda stellar velocity dispersion at different apertures following the prescriptions from both \citet{Bernardi2017} and \citet{Cappellari06} finding that the residuals are stable against increasing the aperture up to twice the spheroidal effective radius quoted in \citet{Sahu2020}. 

\subsection{Hydrodynamical simulations}
\label{subsec|hydrosim}

We compare the data described in \sect~\ref{subsec|obsdata}, in terms of both correlations and residuals, with the predictions from state-of-art hydrodynamical simulations that incorporate AGN feedback in two different ways and thus allow to probe the efficacy of these models in setting the scaling relations between SMBHs and their host galaxies. More specifically, we make use of two simulations: TNG50-1\footnote{\url{https://www.tng-project.org}} \citep[e.g.,][]{Pillepich2018,Springel2018,Nelson19}, and Simba M50N512\footnote{\url{http://simba.roe.ac.uk}} \citep{Dave2019_Simba}. Comparison between observed and predicted SMBH scaling relations, sometimes also considering pairwise residuals, have been carried out previously by adopting specific semi-analytic and hydrodynamical models inclusive of both thermal and kinetic AGN feedback modes \citep[e.g.][]{Hopkins2007_BHplane,Barausse17,Habouzit21,Habouzit22,Habouzit22II,Menci23}, albeit with previous versions of the SMBH data. 

TNG50-1 is the smallest, highest resolution simulation of IllustrisTNG with a volume of $(51.7~{\rm cMpc})^3$ containing $2160^3$ dark matter particles (of mass $m_{\rm DM} = 4.5\times 10^5~{\rm M}_{\odot}$) and $2160^3$ gas cells (with initial baryon mass of $m_{\rm gas} = 8.5 \times 10^4~{\rm M}_{\odot}$). This cosmological volume is evolved using {\sc arepo} \citep{SpringelAREPO} from  $z=127$ to $z=0$ with 100 publically available snapshots for $z\leq20$. The cosmological parameters of TNG50-1 are: $\Omega_{M} = 0.3089$, $\Omega_{b}=0.0486$, $\Omega_{\Lambda} = 0.6911$, $h=0.6774$, $n_{s}=0.9667$, and $\sigma_8 = 0.8159$. We will simply refer to this simulation as \TNG\ from here onwards.

Simba M50N512 is the median box size of the \simba\ simulations with a volume of $(50h^{-1}~{\rm cMpc})^3$ containing $512^3$ dark matter particles (of mass $m_{\rm DM} = 1.2\times 10^7~{\rm M}_{\odot}$) and $512^3$ gas elements (with initial baryon mass of $m_{\rm gas} = 2.88 \times 10^6~{\rm M}_{\odot}$). This cosmological volume is evolved using {\sc gizmo} \citep{Hopkins2015_Gizmo} from $z=100$ to $z=0$ with 151 publically available snapshots for $z\leq20$. The cosmological parameters of \simba\ are: $\Omega_{M} = 0.3$, $\Omega_{b}=0.048$, $\Omega_{\Lambda} = 0.7$, $h=0.68$, $n_{s}=0.97$, and $\sigma_8 = 0.82$. We will simply refer to this simulation as \simba\ from here onwards. We stress that the slight difference in cosmological parameters adopted in the simulations and the reference data does not alter any of our comparison tests in what follows.

It is relevant to briefly note here the main key features characterising the accretion onto the central SMBH and AGN feedback in the two reference hydrodynamical simulations \citep[e.g.,][]{Habouzit2021_Mbh}. In \TNG, the accretion onto the SMBH, which follows the Bondi-Hoyle-Lyttleton formalism (but without any ad-hoc $\alpha$-boost as included in other cosmological simulations), is kernel-weighted over neighbouring cells, ensuring that each accretion episode closely reflects the physical conditions of the gas in the central region of the host. The \TNG\ simulation includes both thermal AGN feedback, where energy is deposited in the surroundings of the accreting SMBH, and kinetic AGN feedback, where momentum is injected in the SMBH surroundings with a direction that is chosen at random at each event, thus generating after a few events a nearly isotropic kinetic feedback. The \simba\ simulation also includes the Bondi–Hoyle–Lyttleton model but only for hot gas above $10^5$ K, while below this temperature the accretion is torque-limited, originating from the gas inflow rate driven by gravitational instabilities to the accretion disc of the SMBH, following the formalism developed by \citet{HopkinsQuataert11} and \citet{Angles-Alcazar2015}. The AGN outflows in \simba\ are bipolar along the angular momentum vector of the stellar disc, and particles are ejected randomly from the black hole accretion kernel with a velocity that varies according to the value of the Eddington ratio, mimicking the effects of radiative- and jet-mode AGN winds (see \citealt{Dave2019_Simba} for full details).
%In addition to the direct output of \TNG, we explored the use of the iMaNGA mock catalogue \citep{Nanni2023}, specifically the value added catalogue\footnote{\url{https://www.tng-project.org/data/docs/specifications/\#sec5\_4}} (VAC). In the VAC, the stellar and gas kinematics are provided by {\sc P}PXF \citep{Cappellari2017}, while the stellar mass is provided by {\sc Firefly} \citep{Wilkinson2017}, and we took $M_{\rm bh}$ from the SubFind catalogue. However, we found there to be insufficient overlap with the SMBH sample.

To pin down the putative impact of AGN feedback, we also make use of the Cosmology and Astrophysics with MachinE Learning Simulations (CAMELS)\footnote{\url{https://camels.readthedocs.io}}, which is a suite of different hydrodynamical simulations \citep[e.g.,][]{CAMELSpresentation,CAMELSrelease}. The simulations in CAMELS are organized into different \emph{suites}. We use the \TNG\ and \simba\ suites. The \TNG\ suite is based on the same subgrid physics as the original IllustrisTNG simulation. Likewise, the \simba\ suite is based on the same subgrid physics as the \simba\ simulation. All simulations within CAMELS have a volume of $(25~\textup{cMpc}/h)^{3}$ with $256^{3}$ dark matter particles and $256^{3}$ gas resolution elements. The simulations span a range from redshift $z=127$ to $z=0$. The cosmological parameters for the simulations are: $\Omega_{b}=0.049, h=0.6711, n_{s}=0.9624, w=-1, M_{\nu}=0.0~\textup{eV}, \Omega_{k}=0.0$. Each suite in the CAMELS repository (\TNG\ and \simba\ in our case) is further split into different \emph{sets}. In this paper, we used the \emph{Extreme} (EX) sets, and in particular we focus on the EX1 set which represents simulation runs with very efficient AGN feedback with respect to the baseline simulations, namely with the AGN wind outflow rate increased by a factor of 100. 

%which is a set of four simulations with the following properties: The first simulation, EX0, can serve as a base-line having fiducial simulation values. The second simulation, EX1, simulates the extreme case having very efficient AGN feedback. The third simulation, EX2, simulates the extreme case having very efficient supernova feedback. Finally, the last simulation, EX3, simulates the extreme case having no feedback at all. In what follows, we will focus on only the renditions EX0 and EX1, which are more relevant for our purposes to pin down the putative effects of AGN feedback on SMBH scaling relations. Prior to moving forward with the CAMELS simulations, we have checked whether the lower resolution and smaller cosmological cubes of the CAMELS are in good agreement with the original simulations Simba M50N512 and TNG50-1. We compared the $M_{\rm bh}-M_{\rm gal}$, $M_{\rm bh}-\sigma_{*}$, $M_{\rm gal}-\sigma_{*}$ relations. We find there is good agreement between the Simba M50N512 and the CAMELS Simba EX0. However, we find that that there is a difference in the normalisation of the $M_{\rm bh}-M_{\rm gal}$ and $M_{\rm bh}-\sigma_{*}$ relations between the TNG50-1 and the CAMELS TNG EX0 simulations due to the differing resolutions, with the CAMELS TNG EX0 scaling relations being in better agreement with those extracted from TNG50-3, which is of comparable resolution. For this reason we limit the discussion of the CAMELS simulations to the impact of increasing the AGN feedback.

For further analysis, we used the CAESAR\footnote{\url{https://caesar.readthedocs.io}} package to extract the necessary information from all the simulations except for \TNG, which uses SubFind \citep[][]{Springel2001} catalogues. In particular, we extracted the following information: Black Hole mass, stellar mass, stellar velocity dispersion, stellar half-mass radius, central or satellite galaxy flags, host halo mass, and bulge-to-total ratio. In what follows, we will only retain the central galaxies in the simulations with a central black hole and a stellar mass $\mgale > 10^{10}\, \msune$, to cover a stellar mass range similar to the data. In the CAESAR catalogue, velocity dispersions are calculated for each particle mass type, gas, stellar, dark matter, black hole, baryonic, and total mass. For each mass type the velocity dispersion in each coordinate axis is computed as the standard deviation of the linear momentum along the chosen axis, divided by the mean mass of the particles
\begin{equation}\label{eq|sigma}
\sigma_x = \sqrt{\frac{\sum\limits_{i=0}^{N}(m_i v_{i,x} - \langle m v_x \rangle)^{2}}{N \langle m \rangle }}.
\end{equation}
From Eq.~\ref{eq|sigma} we calculate the 3D stellar velocity dispersion, $\sigma_{\rm 3D}$, from which, by making the common assumption that the orbits are isotropic, we derive the 1D stellar velocity dispersion $\sigma_{\rm 1D} = \sigma_{\rm 3D}/\sqrt{3}$. By default, CAESAR calculates the stellar velocity dispersion from all stellar particles associated with a galaxy, and therefore, they are not strictly expressed in the same aperture as in the data. We have checked, however, that restricting the calculation of \sis\ in the simulations to a radial distance from the centre equal to the half-mass radius, or even to the aperture of the Hyperleda database, has modest effects on \sis\ with variations contained within $\lesssim 0.1$ dex, in the range of interest to this work. Another possible aperture correction could be applied to \sis\ following the full Jeans modelling as detailed in Appendix C of \citet{Marsden22}. However, this treatment would imply convolving with the full stellar profile of the simulated galaxies and may affect some of the residual analysis when considering velocity dispersions. In what follows, we will thus only show results using the default $\sigma_{\rm 1D}$ from CAESAR, which is in line with the methodology also followed by \citet{Thomas2019}. For \TNG, we extracted the same information from the SubFind catalogues and auxiliary files, as we have done from the CAESAR catalogues for the other simulations. However, due to the velocity dispersion listed in the SubFind catalogue being computed for all particles associated with a galaxy, we recomputed the velocity dispersion for the stellar particles only using Eq.~\ref{eq|sigma}. Finally, we also verified that, in the region of overlap $\sise \lesssim 100$ km/s, the stellar velocity dispersions computed from Eq.~\ref{eq|sigma} are consistent, within 0.1 dex, with those in the iMaNGA mock catalogue \citep{Nanni2023}, inclusive of all relevant observational effects.

\section{Method}
\label{sec|Method}

The main aim of this work is to compute the pairwise residuals between black hole mass and a variety of galactic properties from the latest data and models. The pairwise residuals are a technique specifically designed to isolate the underlying dependence of one variable $Y$ from the variable $Z$ while keeping a third variable $X$ ``fixed''. This step is achieved by computing the correlation between the residuals
\begin{equation}
\Delta(Y|X) \propto \Delta(Z|X)
\label{eq|resid}
\end{equation}
where 
\begin{equation}
\Delta(Y|X)\equiv\log Y-\langle \log Y|\log X \rangle \,
\label{eq|singleresidual}
\end{equation}
is the residual computed in the $Y$ variable (at fixed $X$) from the log-log-linear fit of $Y(X)$ vs $X$, i.e., $\langle \log Y|\log X \rangle$. \eq\ref{eq|resid} calibrates the degree of correlation between $Y$ and $Z$ once the correlations in $X$ of both variables $Y$ and $Z$ are effectively ``factored out''; a perfectly null correlation between the $\Delta(Y|X)$ and $\Delta(Z|X)$ residuals would imply that an apparent correlation between $Y$ and $Z$ is actually a reflection of the underlying dependencies of $Y$ and $Z$ on $X$. In contrast, a strong correlation between $\Delta(Y|X)$ and $\Delta(Z|X)$ would instead imply a minor role of $X$ in setting the correlation between $Y$ and $Z$. Full details on the full pairwise residual analysis formalism which we follow in this work, inclusive of the treatment of errors in the variables and the calculation of the Pearson coefficient, are provided in \citet{ShethBernardi12} and in Appendix B of \citet{Shankar17}. For each pair of variables $Y$ and $Z$, the correlation between their linear residuals (\eq\ref{eq|singleresidual}) is computed a 100 times, and at each iteration 5\% of the objects are removed at random from the original sample. From the complete set of realizations, we then measure the mean slope and its standard deviation. At each iteration, when calculating in the observed data samples the intrinsic slopes and degree of correlations in the residuals, we also take into account the measurement error on each point. Similarly to what we performed with the observational data, for the simulated outputs we follow a stochastic iterative method to compute the residuals, where at each iteration we remove 5\% of the sample, and then calculate the mean slope and Pearson coefficient. However, in the latter case, when computing residuals we do not include observational statistical uncertainties, instead, we simply apply a linear fit to the simulated points to define the residual at each iteration. In addition, in what follows, we also compare the results from the pairwise residual analysis with the outputs extracted from Machine Learning (ML) regression algorithms. We provide details on the latter methods directly in \sect\ref{subsec|ML}.
%describe how we include observational uncertainties in the residuals and also how we do it in the simulations. Also describe Pearson coefficient. Describe the ML and causal models you will be using. 

\begin{figure*}
\begin{center}
\center{{
\epsfig{figure=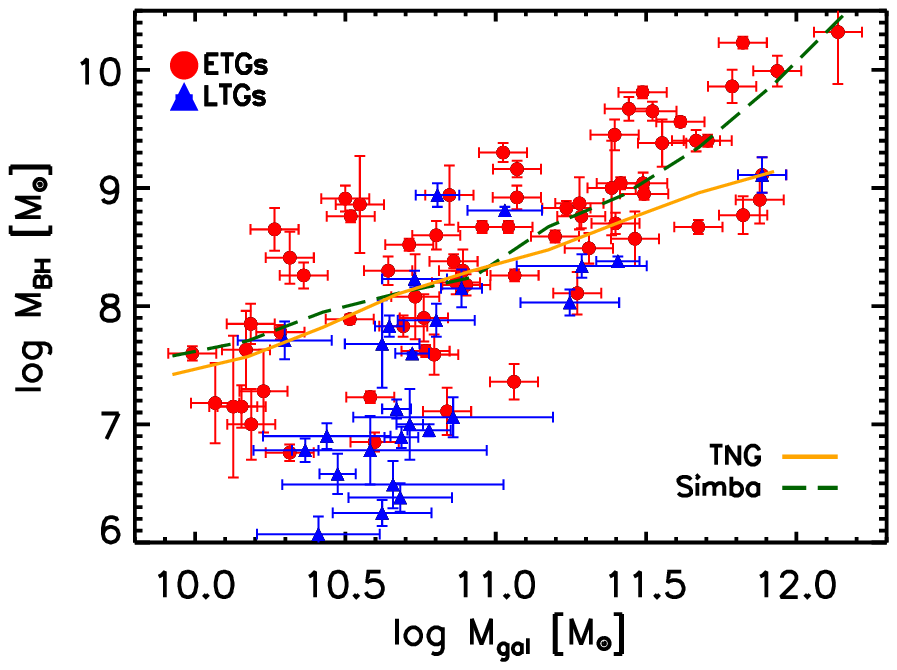,height=6.5cm}\hspace{-0.85cm}
\epsfig{figure=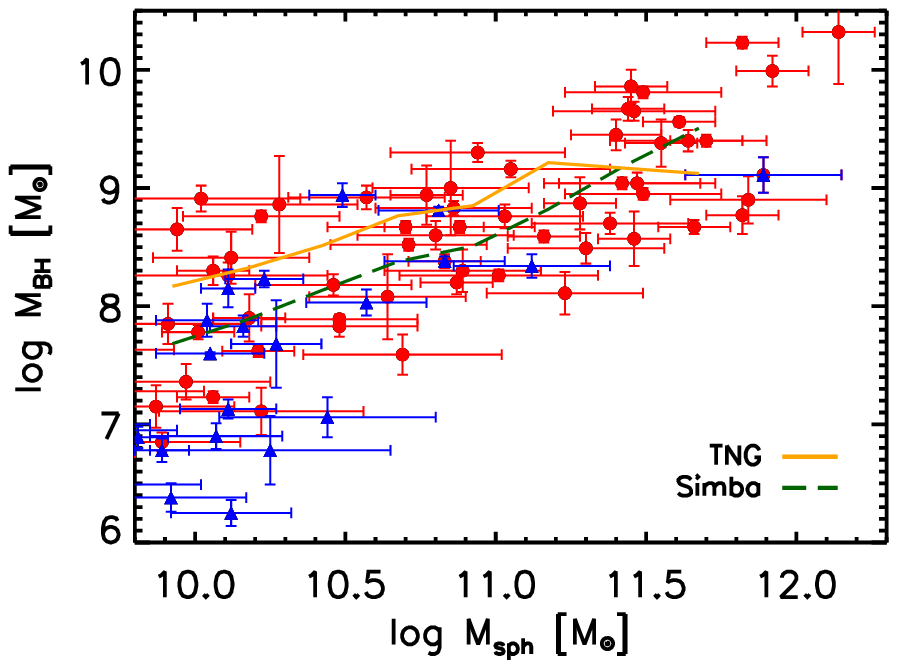,height=6.5cm}
}}
\vspace{-0.52cm}
\center{{
\epsfig{figure=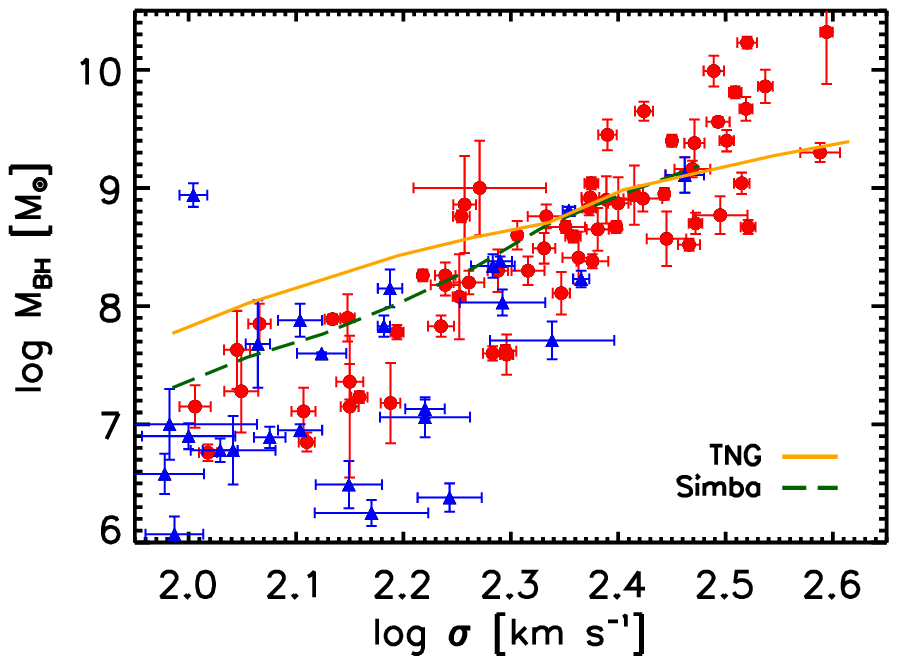,height=6.5cm}\hspace{-0.85cm}
\epsfig{figure=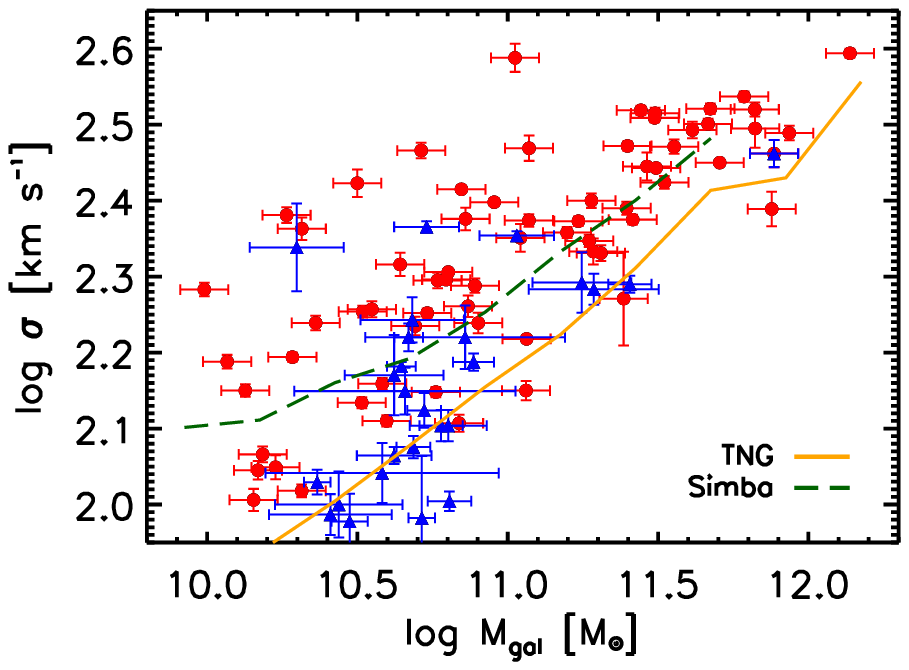,height=6.5cm}
}}
\caption{Top left: Scaling relation between SMBH mass \mbh\ and total galaxy stellar mass \mstar\ for local galaxies with dynamically measured masses of their central SMBHs. Top right: Identical to the top left panel but where only the bulge component of late-type galaxies in the data is included. Bottom left: Corresponding scaling relation between SMBH mass \mbh\ and galaxy stellar velocity dispersion \sis\ (see text for details) for local galaxies. Bottom right: Correlation between \sis\ and galaxy stellar mass \mstar.
Red circles and blue triangles in all panels refer, respectively, to early-type and late-type galaxies. All the data are from \citep{Sahu19}. Solid, orange and green, dashed lines show the predictions from the \TNG\ and \simba\ simulations, respectively. The simulations tend to align with the data in the \mbh-\mstar\ plane but significantly depart from them in the \mbh-\sis\ one.}
        \label{fig|scalings}
\end{center}
\end{figure*}

%\begin{figure*}
%\begin{center}
%\center{{
%\epsfig{figure=ScalingMbhMgal.eps,height=4.5cm}\hspace{-0.85cm}
%\epsfig{figure=ScalingMbhMbulge.eps,height=4.5cm}\hspace{-0.85cm}
%\epsfig{figure=ScalingMbhMbulge.eps,height=4.5cm}
%}}
%\vspace{-0.52cm}
%\center{{
%\epsfig{figure=ScalingMbhSigma.eps,height=4.5cm}\hspace{-0.85cm}
%\epsfig{figure=ScalingSigmaMgal.eps,height=4.5cm}\hspace{-0.85cm}
%\epsfig{figure=ScalingMbhMbulge.eps,height=4.5cm}
%}}
%\caption{Scaling relations for local galaxies with dynamically measured masses of their central SMBHs. Red %circles and blue triangles refer, respectively, to Early-type and Late-type galaxies. All the data are from %\citep{Sahu19}.}
%        \label{scalings}
%\end{center}
%\end{figure*}

\begin{figure*}
\begin{center}
\center{{
\epsfig{figure=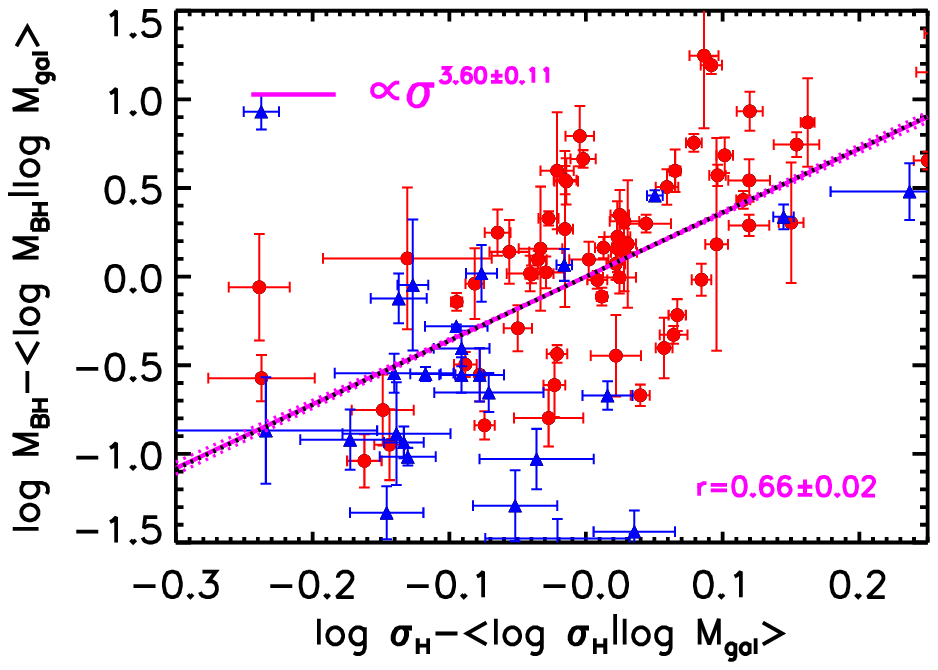,height=6.5cm}\hspace{-0.85cm}
\epsfig{figure=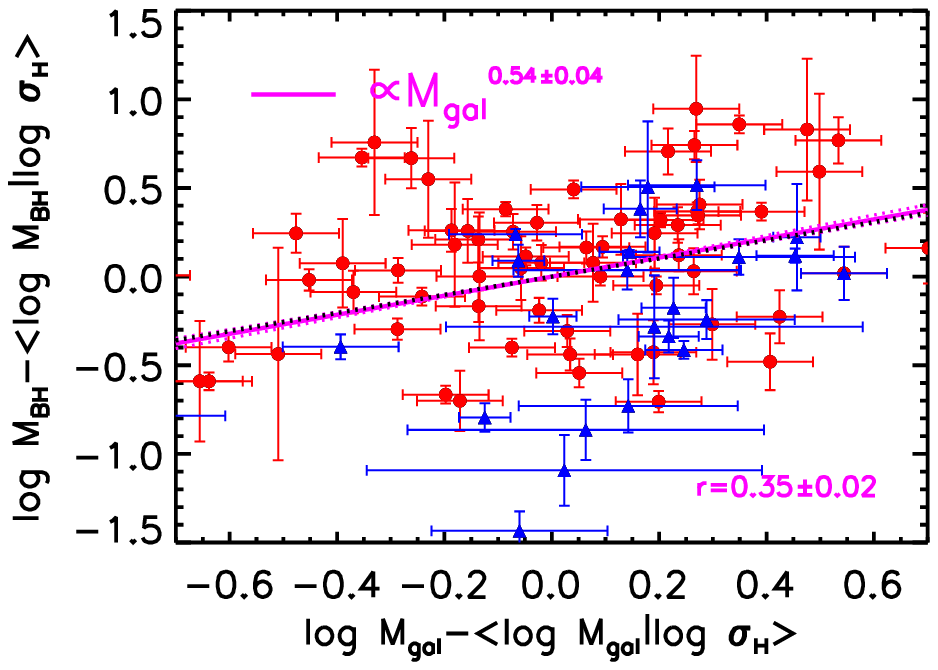,height=6.5cm}
}}
%\vspace{-0.52cm}
%\center{{
%\epsfig{figure=ResidMbhSigmaFixedMsph.eps,height=6.5cm}\hspace{-0.85cm}
%\epsfig{figure=ResidMbhMsphFixedSigma.eps,height=6.5cm}
%}}
\vspace{-0.52cm}
\center{{
\epsfig{figure=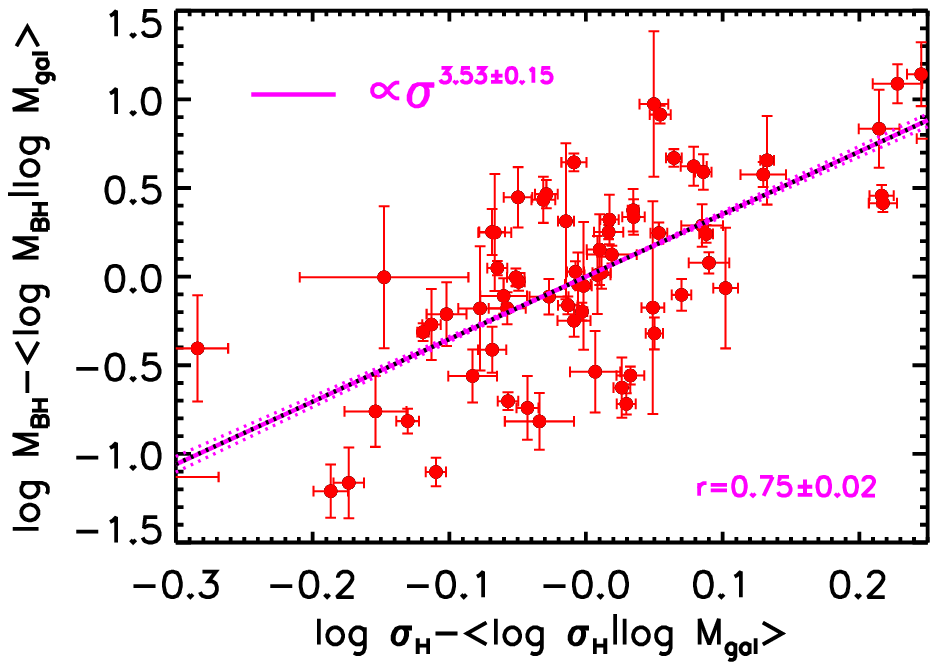,height=6.5cm}\hspace{-0.85cm}
\epsfig{figure=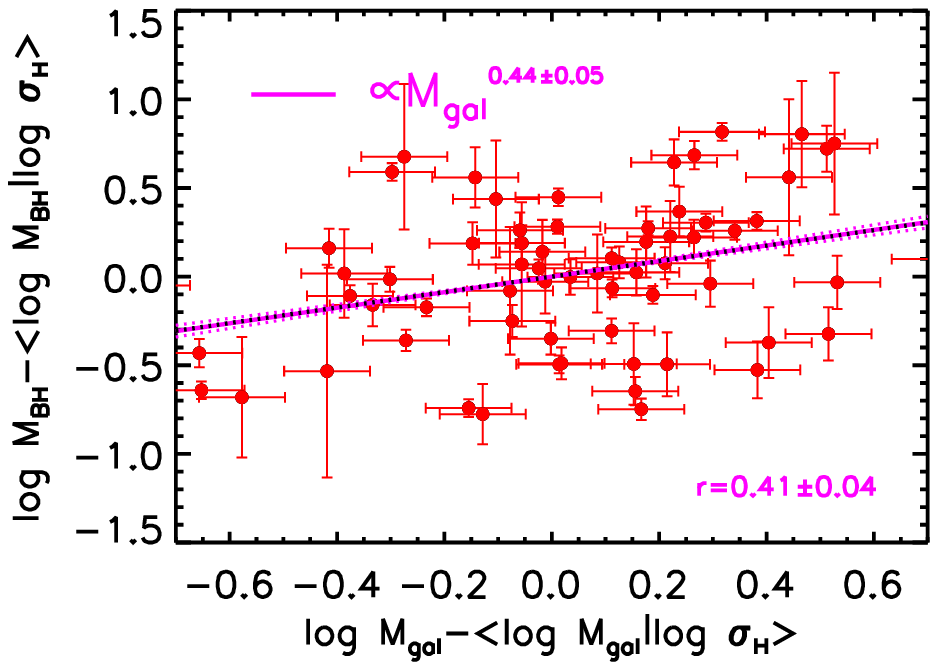,height=6.5cm}
}}
\caption{Residual analysis for the local sample of galaxies with dynamically measured SMBH mass from \figu\ref{fig|scalings}. Left panels: residuals as a function of stellar velocity dispersion \sis\ at fixed galaxy stellar mass \mstar. Right panels: residuals as a function of galaxy stellar mass at fixed stellar velocity dispersion. Top panels use the full galaxy sample, including late-type galaxies, while bottom panels include only early-type galaxies. The residual with stellar velocity dispersion is very significant, while the one with stellar mass is negligible at fixed \sis, fully confirming the primary role of \sis\ compared to total galaxy stellar mass.}
        \label{fig|residuals_sahu}
\end{center}
\end{figure*}

\begin{figure*}
\begin{center}
\center{{
\epsfig{figure=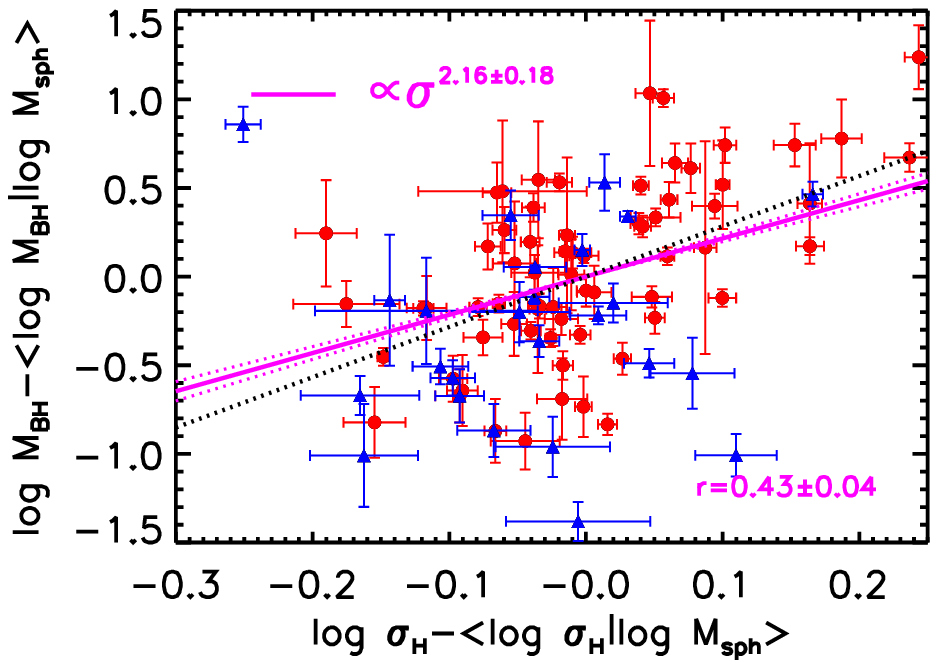,height=6.5cm}\hspace{-0.85cm}
\epsfig{figure=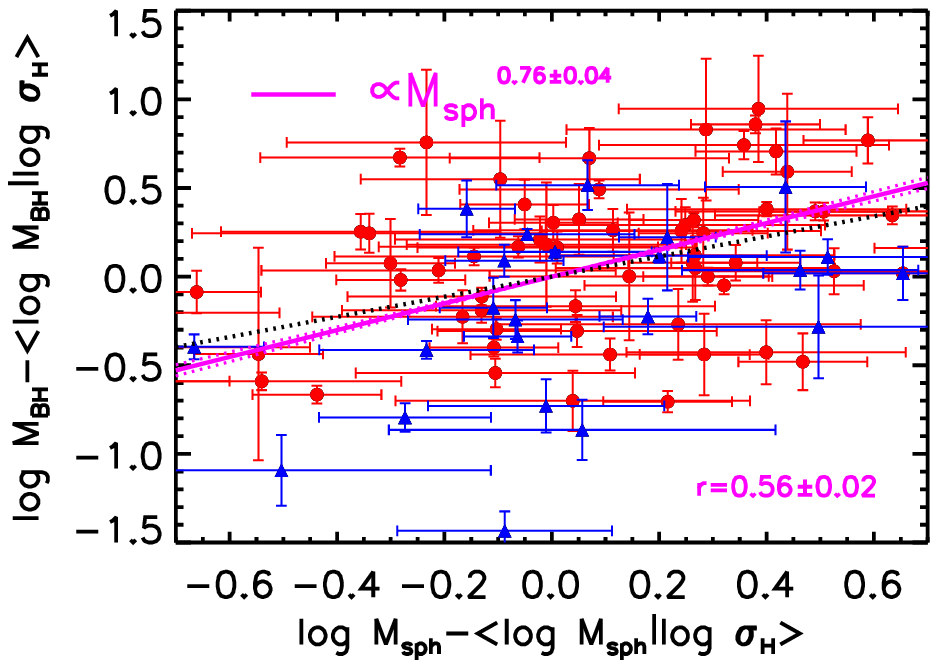,height=6.5cm}
}}
\vspace{-0.52cm}
\center{{
\epsfig{figure=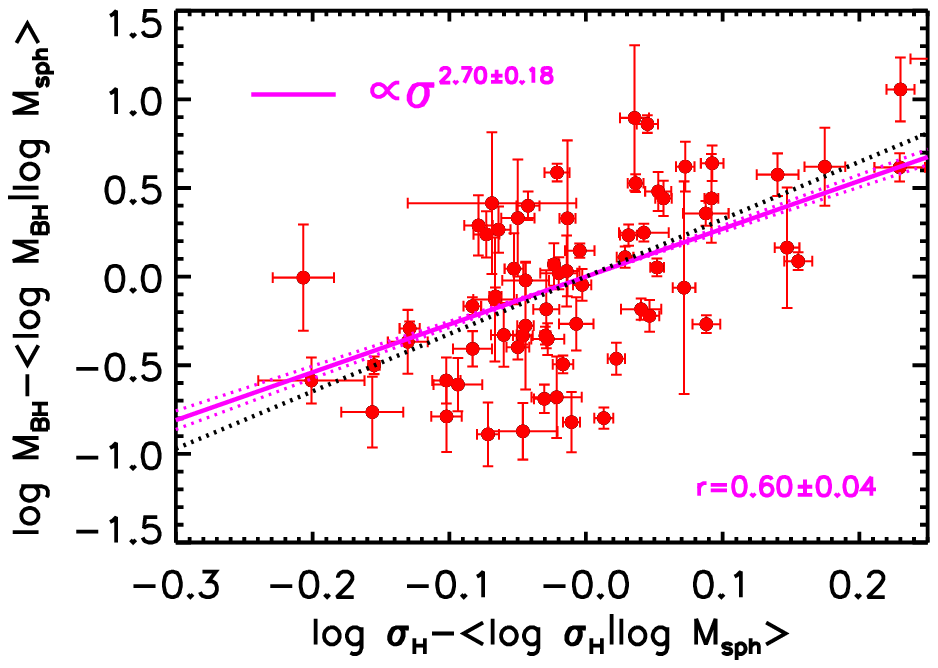,height=6.5cm}\hspace{-0.85cm}
\epsfig{figure=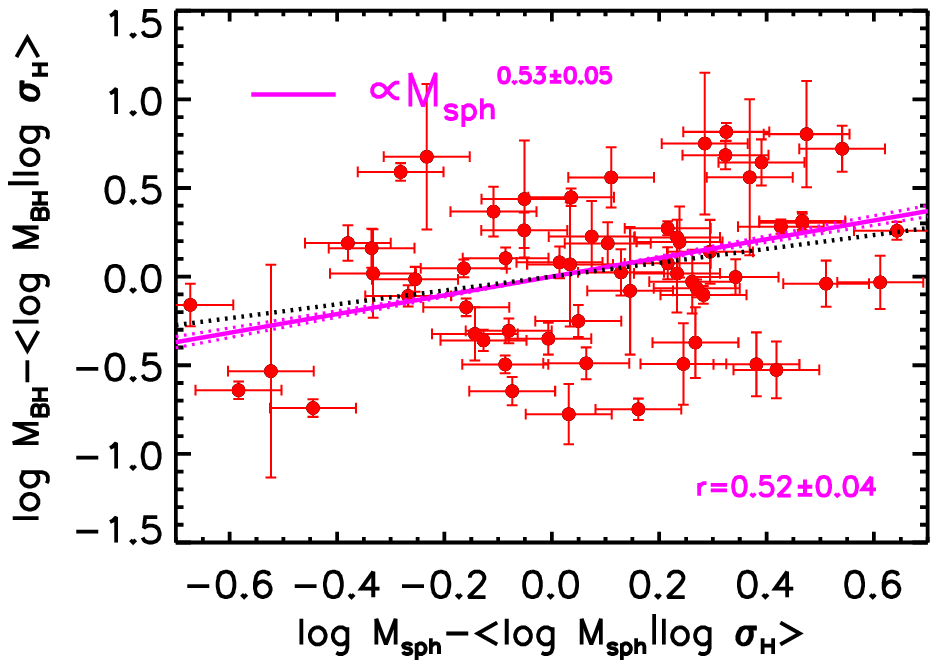,height=6.5cm}
}}
\caption{Same layout as \figu\ref{fig|residuals_sahu} but only considering the spheroidal component of each galaxy. The residual with stellar velocity dispersion appears now equally significant to stellar mass, possibly hinting at the existence of a fundamental plane for the \textit{spheroidal} component only.}
        \label{fig|residuals_sahuMsph}
\end{center}
\end{figure*}

%\begin{figure*}
%\begin{tabular}{@{}ccc}
%\epsfig{figure=ScalingMbhMgal.eps,height=5.6cm} \hspace{-0.2cm}
%\epsfig{figure=ScalingMbhSigma.eps,height=5.6cm} \hspace{-0.2cm}
%\epsfig{figure=ScalingSigmaMgal.eps,height=5.6cm}
%\end{tabular}
%\caption{Top panels: Stellar mass accreted from in-situ star formation as a function of galaxy stellar mass for central galaxies of four different mass bins at $z=0$ ($M_\star \simeq 10^{9.5}, \, 10^{10}, \, 10^{10.5} \, {\rm and} \, 10^{11} \, M_\odot$), as labelled. The blue solid lines and shaded areas show the mean star formation histories and stellar mass variance from the TNG simulation, respectively. The orange dashed lines show the prediction from with the TNG's SMHM relation as input..}
%\label{fig|Fico}
%\end{figure*}

\begin{figure*}
\begin{center}
\center{{
\epsfig{figure=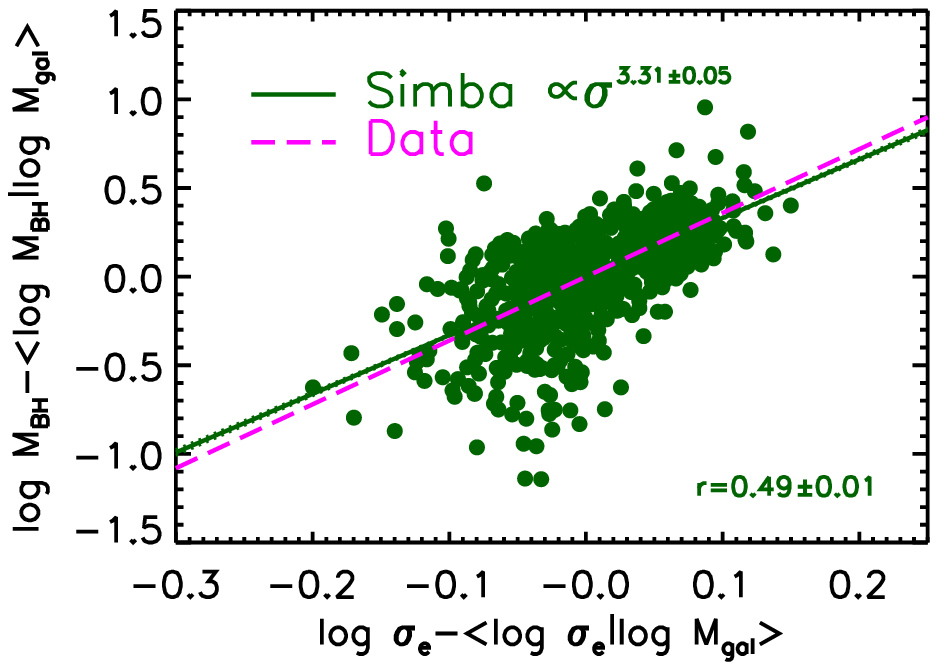,height=6.5cm}\hspace{-0.85cm}
\epsfig{figure=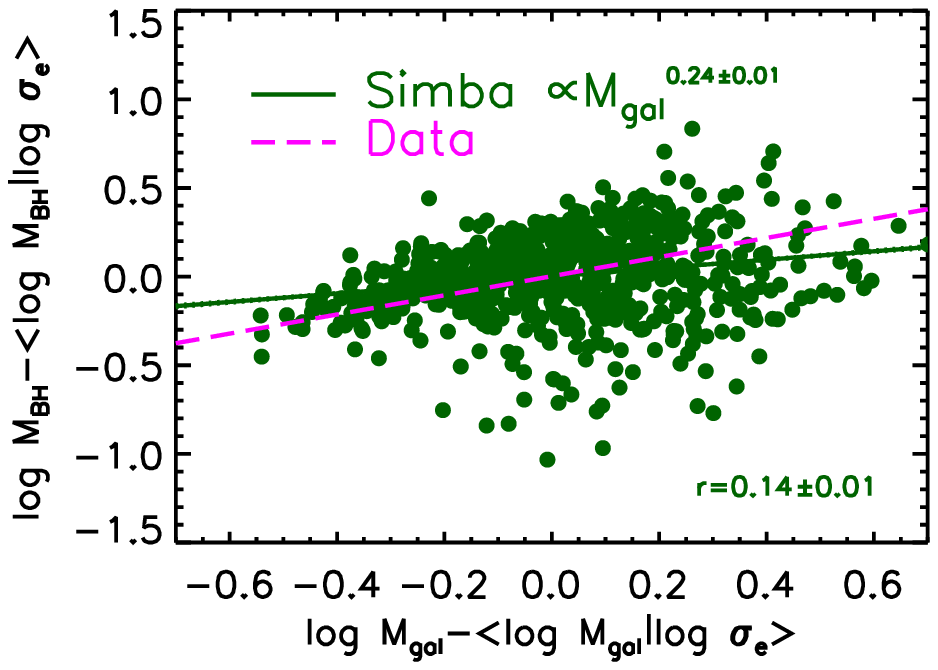,height=6.5cm}
}}
\vspace{-0.52cm}
\center{{
\epsfig{figure=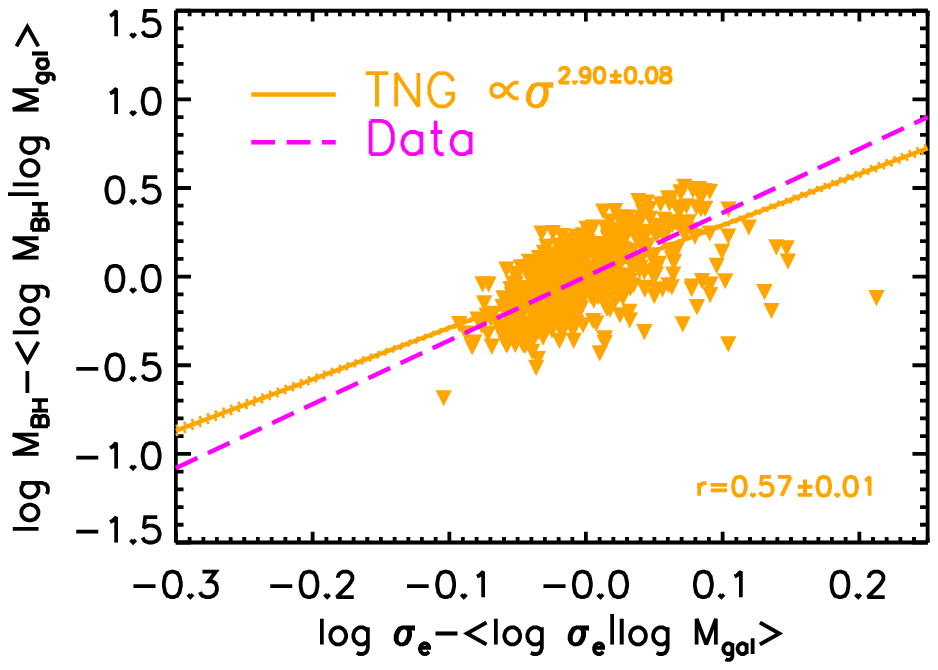,height=6.5cm}\hspace{-0.85cm}
\epsfig{figure=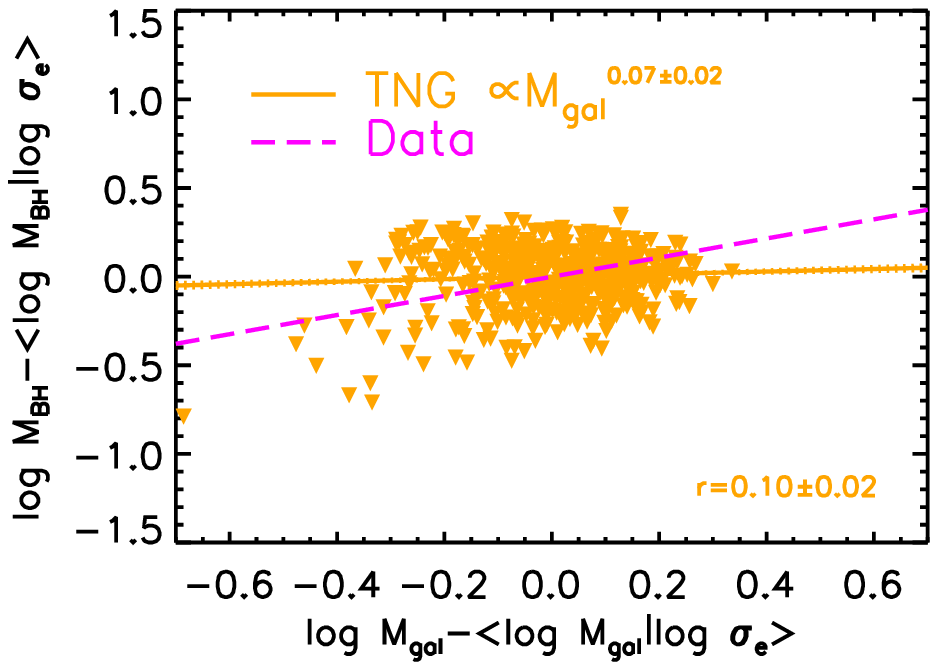,height=6.5cm}
}}
\caption{Residual analysis in the same format as in the top panels of \figu\ref{fig|residuals_sahu}, for the Simba (top panels) and TNG100 (bottom panels) simulations, as labelled, compared to the residuals extracted from the data (magenta dashed lines). Simulations predict weaker correlations with stellar velocity dispersion than in the data.}
        \label{fig|ResidualsSimulations}
\end{center}
\end{figure*}

\begin{figure*}
\begin{center}
\center{{
\epsfig{figure=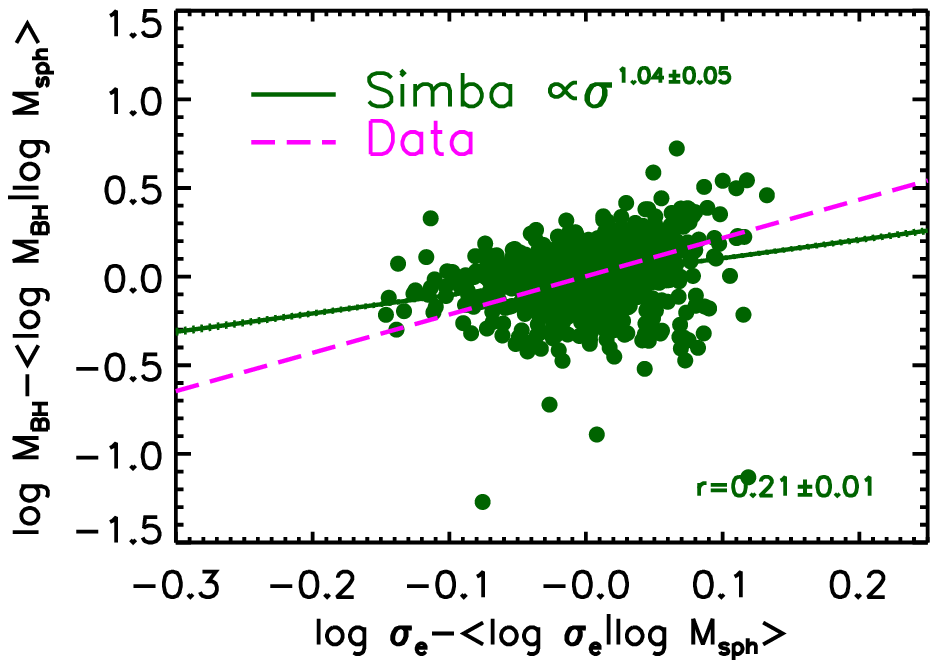,height=6.5cm}\hspace{-0.85cm}
\epsfig{figure=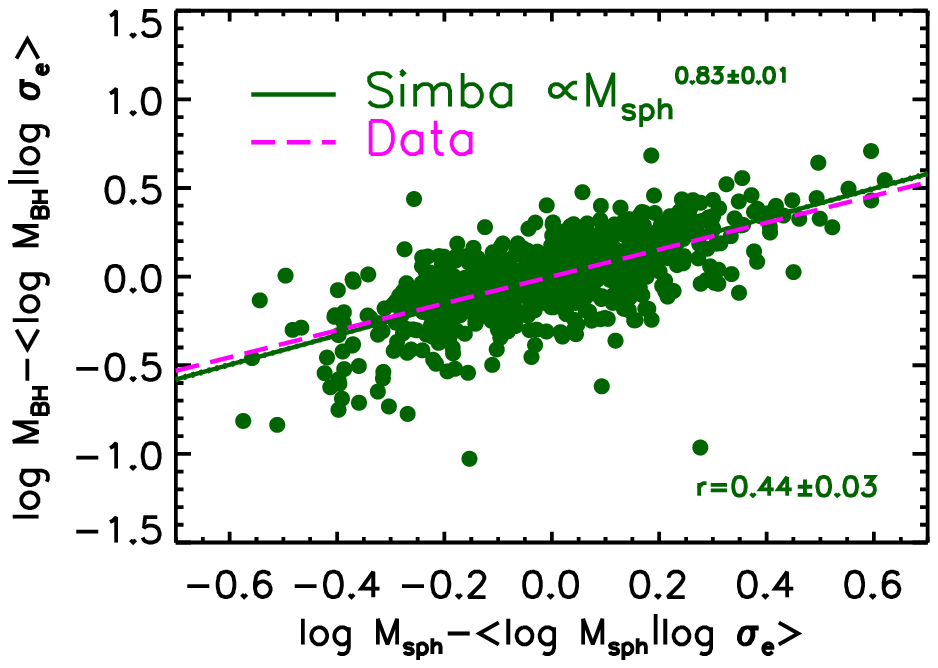,height=6.5cm}
}}
\vspace{-0.52cm}
\center{{
\epsfig{figure=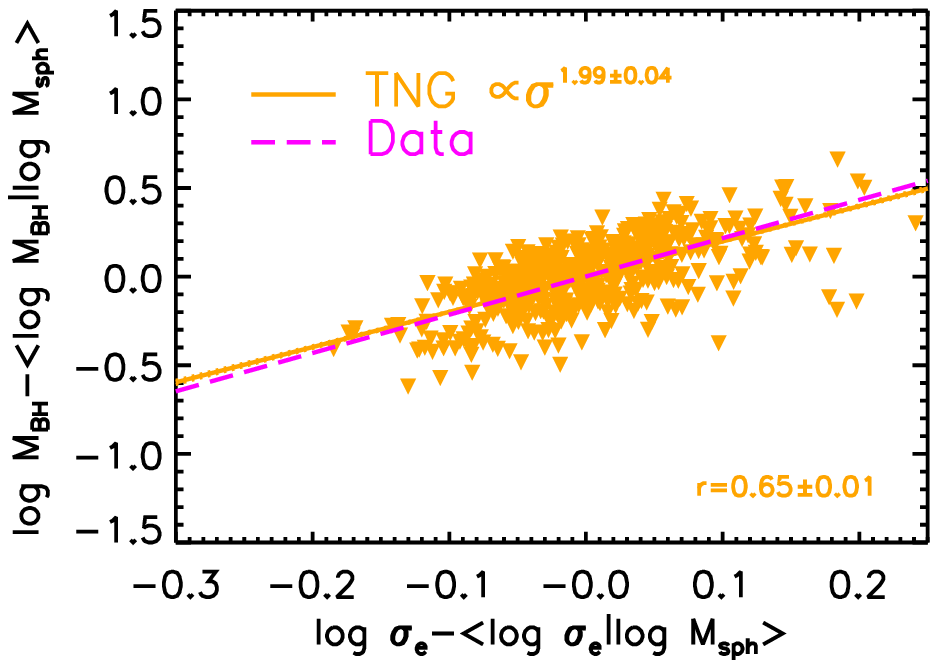,height=6.5cm}\hspace{-0.85cm}
\epsfig{figure=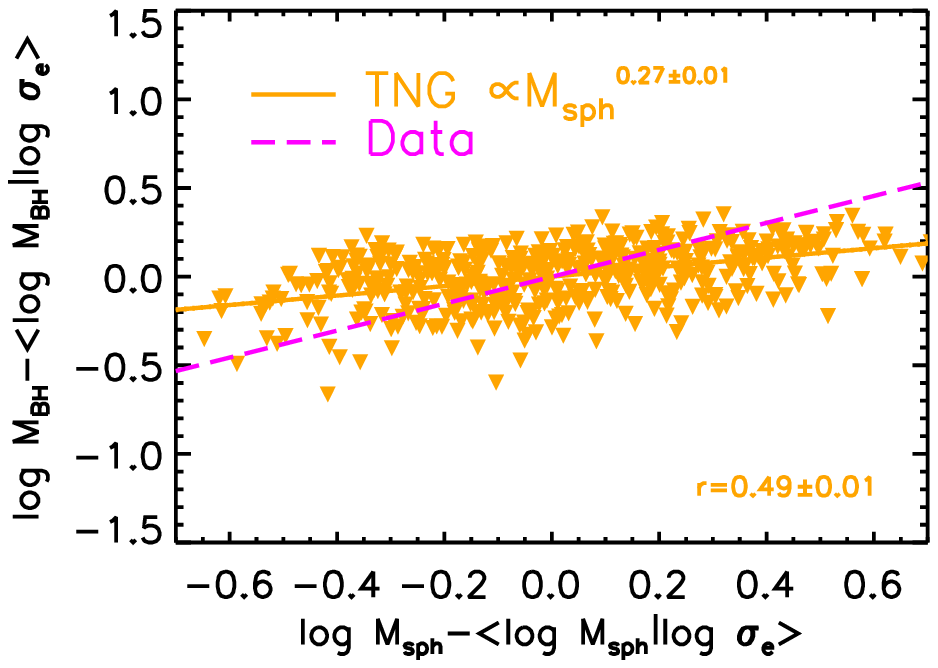,height=6.5cm}
}}
\caption{Same format as \figu\ref{fig|ResidualsSimulations} but replacing total with bulge stellar mass. Simulations predict different results between them and with the data, with \TNG\ being closer to what expected from a correlation between SMBH mass and host binding energy.}
        \label{fig|ResidualsSimulationsMbulge}
\end{center}
\end{figure*}

\begin{figure*}
\begin{center}
\center{{
\epsfig{figure=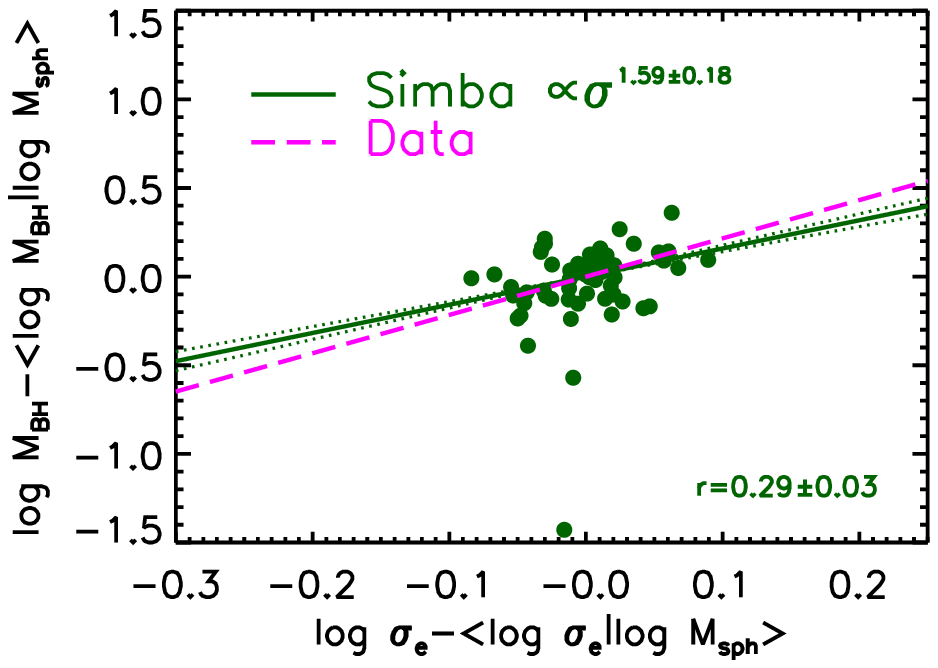,height=6.5cm}\hspace{-0.85cm}
\epsfig{figure=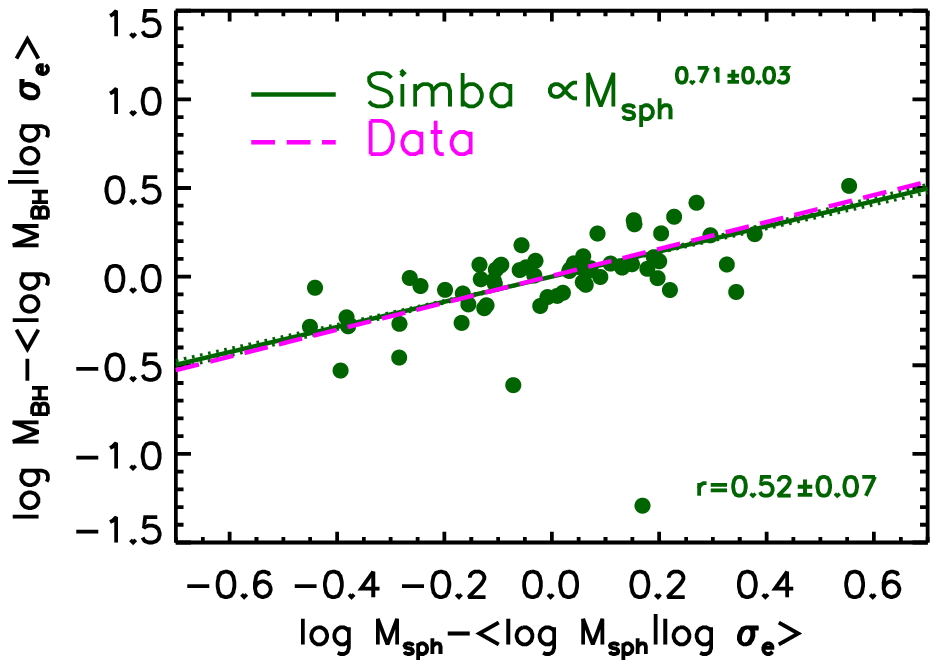,height=6.5cm}
}}
\center{{
\epsfig{figure=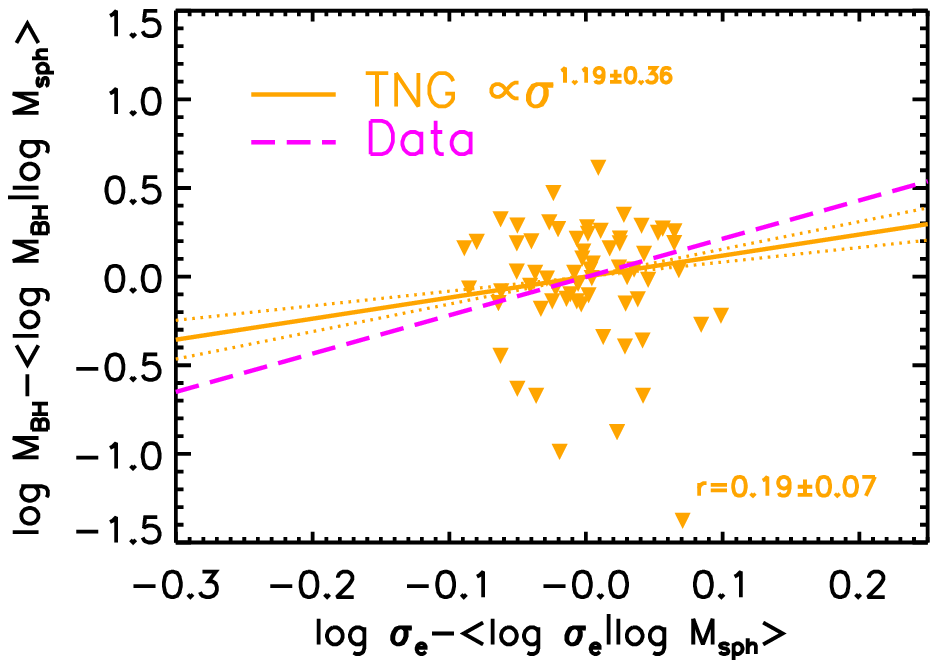,height=6.5cm}\hspace{-0.85cm}
\epsfig{figure=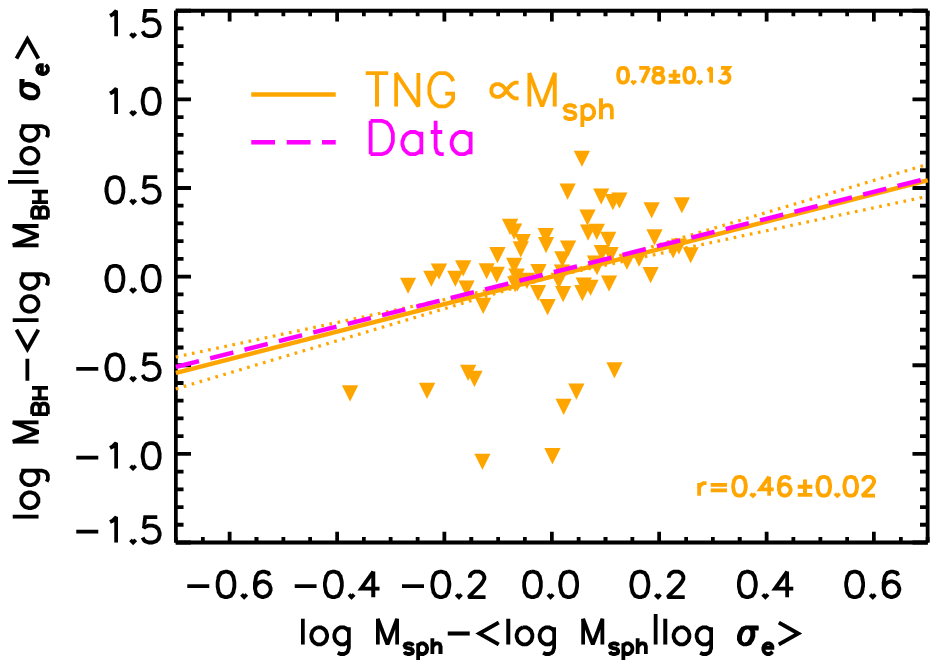,height=6.5cm}
}}
\vspace{-0.52cm}
\center{{
\epsfig{figure=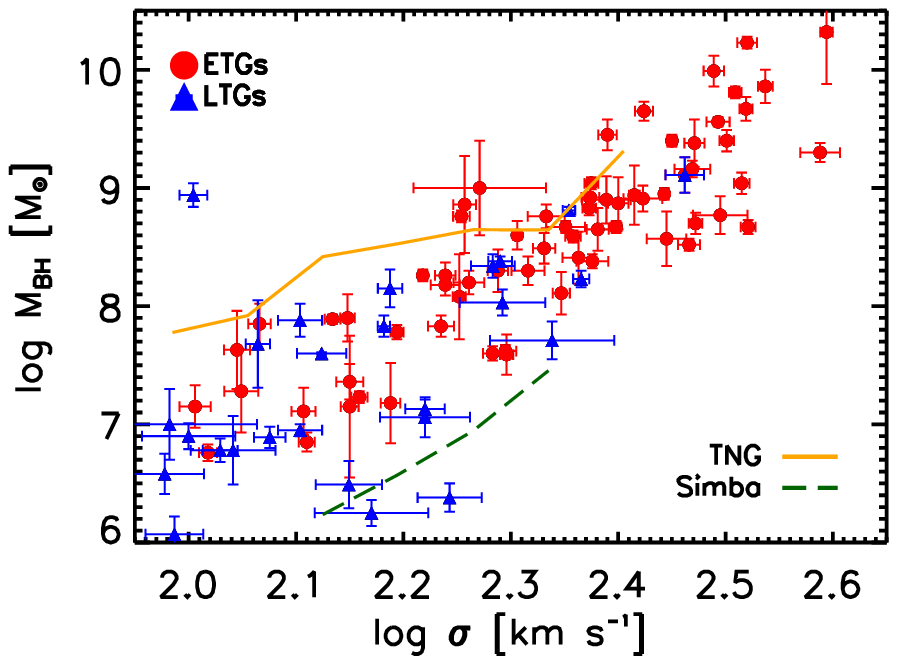,height=6.5cm}\hspace{-0.85cm}
\epsfig{figure=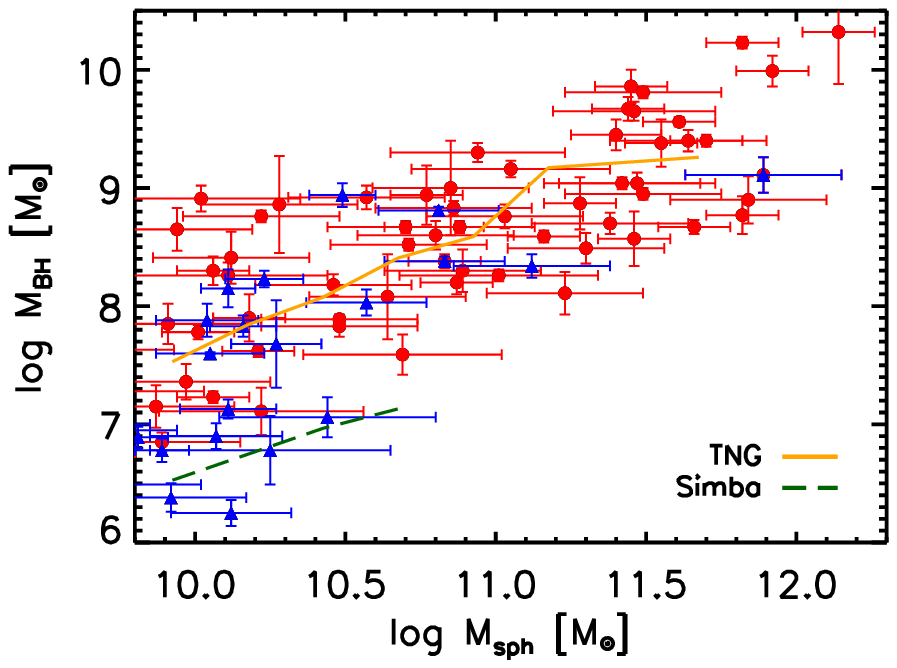,height=6.5cm}
}}
\caption{Same format as \figu\ref{fig|ResidualsSimulations} but with increased AGN feedback efficiency (see text for details). Bottom panels include the predicted SMBH-galaxy scaling relations in the new simulations as in the top panels of \figu\ref{fig|scalings}. An increased AGN feedback efficiency tends to increase the significance of the residuals, at least in the Simba simulation, at the cost of a significantly lower normalization in the scaling relations.}
        \label{figu|ResidualsSimulationsPlus}
\end{center}
\end{figure*}

\begin{figure*}
\begin{center}
%before 6.5, 2.65, 2.12
\center{{
\epsfig{figure=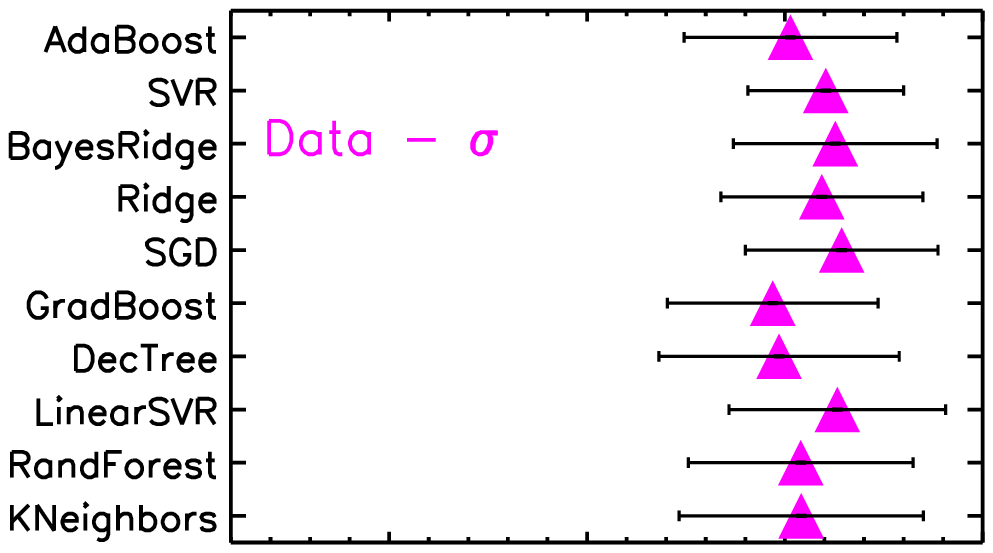,height=5.cm}\hspace{-2.05cm}
\epsfig{figure=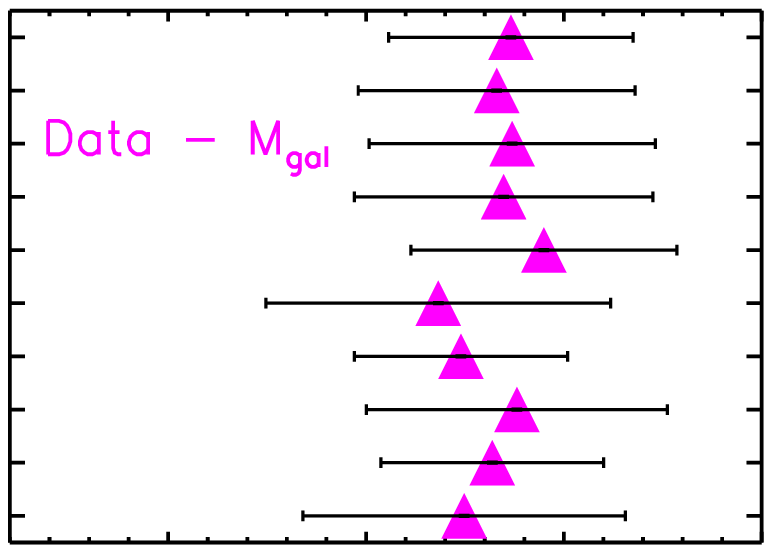,height=5.cm}
\hspace{-2.15cm}
\epsfig{figure=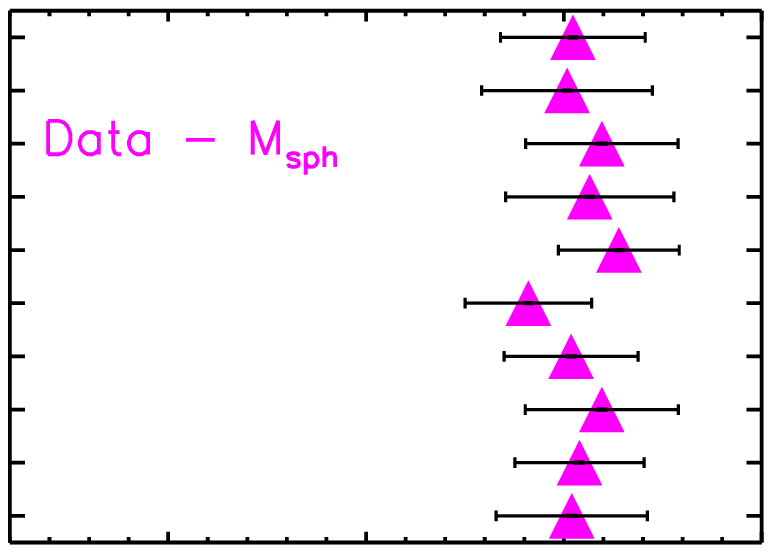,height=5.cm}
}}
\vspace{-1.7cm}
\center{{
\epsfig{figure=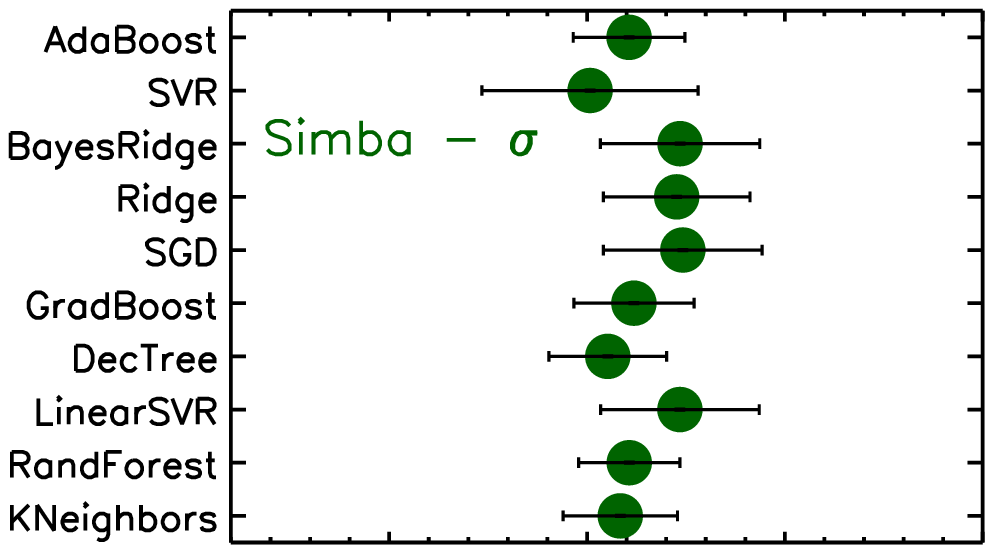,height=5.cm}\hspace{-2.05cm}
\epsfig{figure=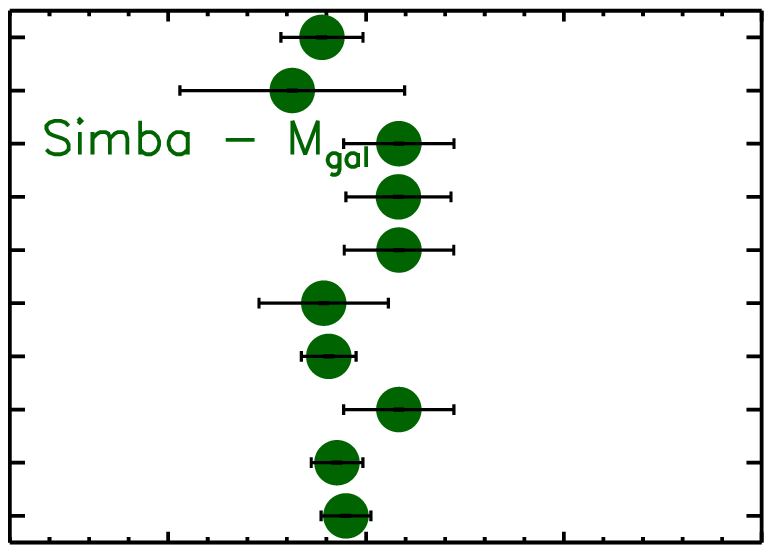,height=5.cm}
\hspace{-2.15cm}
\epsfig{figure=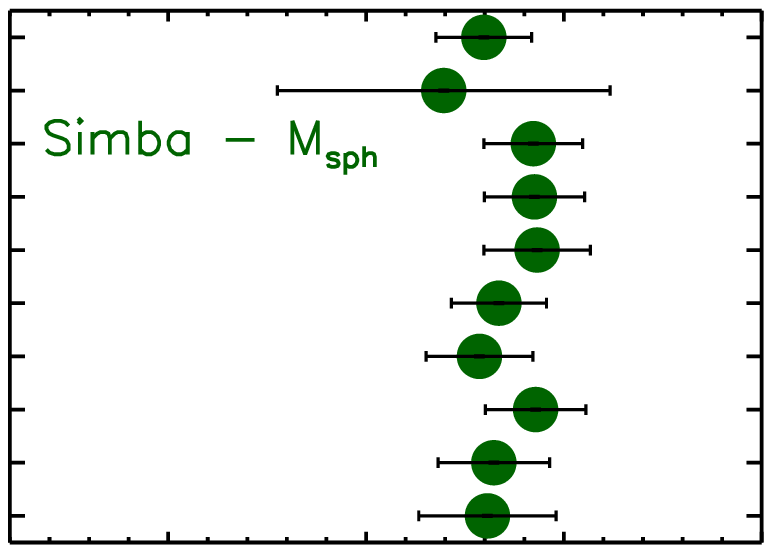,height=5.cm}
}}
\vspace{-1.7cm}
\center{{\hspace{-0.135cm}
\epsfig{figure=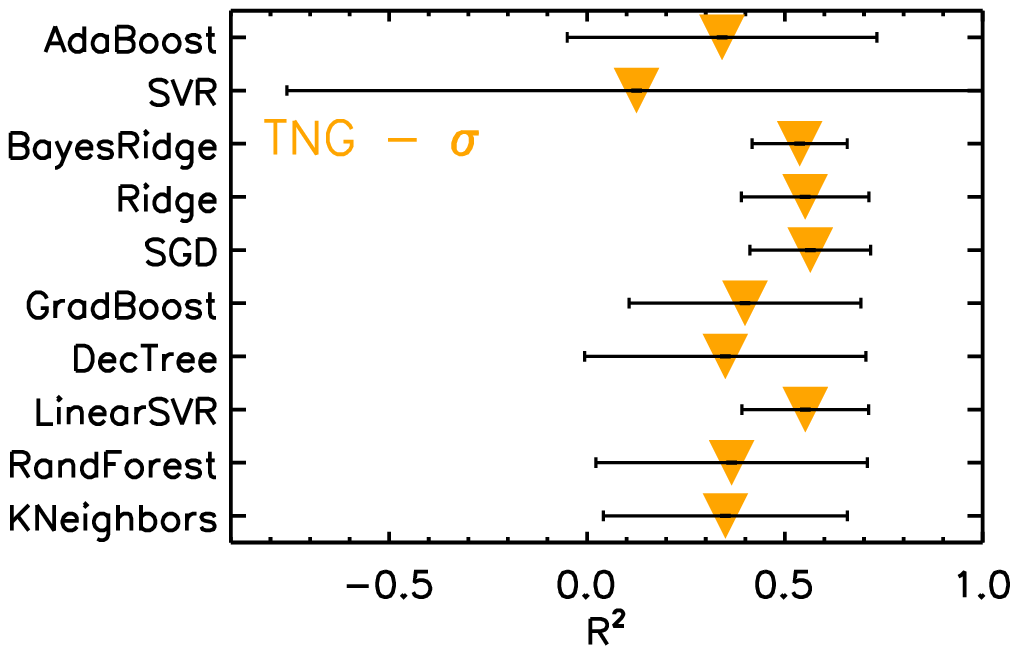,height=5.cm}\hspace{-2.05cm}
\epsfig{figure=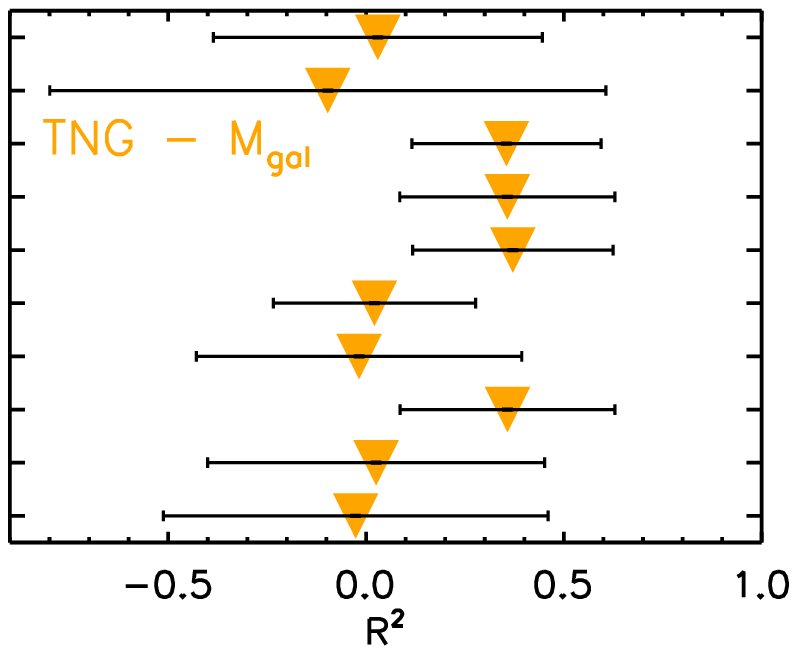,height=5.cm}
\hspace{-2.15cm}
\epsfig{figure=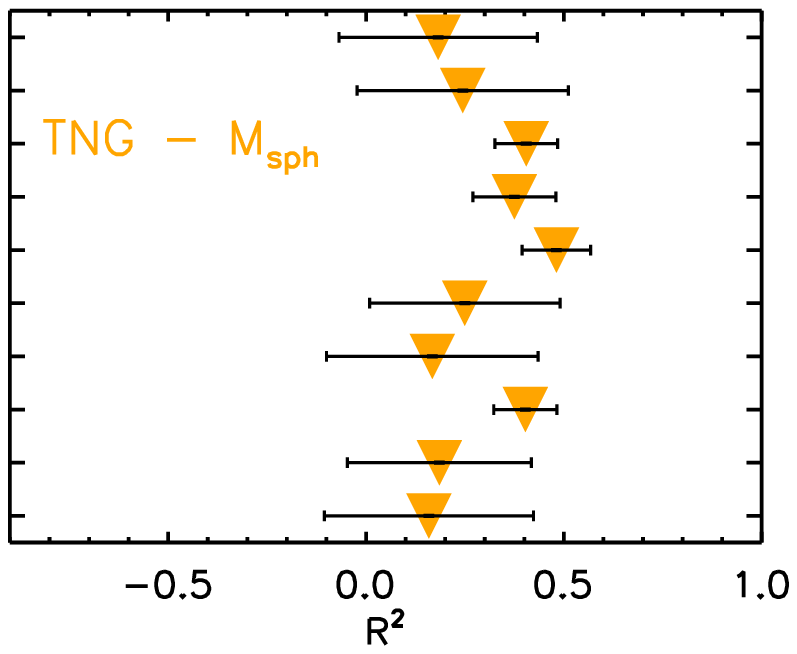,height=5.cm}
}}
\caption{Calculation of the degree of correlation as labelled by the $R^2$ parameter, via a number of machine learning regression algorithms as listed on the left of the Figure. Top, middle, and bottom panels refer to the results of the ML algorithms applied to the full dataset of local galaxies with SMBH mass measurements, the predictions of the (reference) \simba\ and the \TNG\ simulations, respectively. The left, middle and right panels refer to the correlation with velocity dispersion \sis, galaxy total stellar mass \mgal, and galaxy bulge stellar mass \mbulge, respectively. The \simba\ simulation does not predict any strong correlation with any variable, except with \mbulge, while \TNG\ predicts a correlation with \sis\ comparable with what seen in the data and, to a lesser extent, also with \mbulge, in line with also derived from the pairwise residuals (see text for details).}
        \label{figu|CorrelationsMachineLearning}
\end{center}
\end{figure*}

\section{Results}
\label{sec|results}

\subsection{Scaling relations}
\label{subsec|Scalings}

%also include the few galaxies in the simulations above $\mstare\gtrsim 2\times 10^{11}\, \msune$, which contain the most massive SMBHs and are useful to provide broad guidance on the shape of the relations at the largest masses
We start in \figu\ref{fig|scalings} by providing a broad comparison between the scaling relations of local inactive galaxies with dynamically measured SMBHs, from here onwards simply defined as the ``SMBH sample'', and the predictions from hydrodynamic simulations. For the latter, we here use the mean relations predicted by the reference \simba\ and \TNG\ simulations\footnote{When computing the predicted mean relations we only retain bins inclusive of at least five galaxies.} (green, dashed and solid, orange lines, respectively), we will further discuss below the impact of adopting some variations of these simulations. The red circles and blue triangles in all panels refer to Early-Type and Late-Type galaxies, respectively. The top left panel of \figu\ref{fig|scalings} shows the relation between SMBH mass as a function of total galaxy stellar mass \mgal, while in the top right panel we only retain the stellar spheroidal component \mbulge\ of both ETGs and LTGs. It is immediately interesting to note that in the data, in terms of total stellar mass, LTGs define a steeper \mbh-\mgal\ relation than ETGs (top left panel), as already noted by \citet{Sahu19}, while the correlation with SMBH mass becomes more linear when only the stellar spheroidal component is retained, or at least the steepening is shifted to lower stellar masses (top right panel). 

Both simulations align with the ETG data, whilst the LTGs lie mostly below the simulations' predictions when total stellar mass is considered. It is relevant to note that this comparison is only qualitative as the models have been mostly calibrated on local galaxy stellar mass functions from, e.g., \citet{Bernardi13}, which have SDSS photometries and specific choices of mass-to-light ratios that may differ from the ones derived from the 3.6 $\mu$m photometry by \citet{Sahu23}. As discussed in \sect\ref{subsec|hydrosim}, aperture corrections in the stellar velocity dispersions in the simulations are small and have a minor impact on the results in Figure\ref{fig|scalings} as also noted by \citet{Marsden22}. Nevertheless, other factors may affect the comparison with the local scaling relations of SMBHs, in particular in the \mbh-\mstar\ plane, such as the existence of possible biases, which we extensively discuss in Appendix~\ref{Appendix:Bias}. %Following a number of previous works \citep[e.g.,][]{Bernardi07,Batcheldor10,MorabitoDai12}, \citet{Shankar16} discussed that a systematic offset may exist between the SDSS galaxy sample and SMBH galaxy sample at least partly due to selection effects affecting the SMBH galaxy sample, which privileges the inclusion of the SMBHs with the largest gravitational sphere of influence. \citet{Sahu23} suggested instead that there is no selection bias and that the difference in the stellar mass estimates of the two samples is sufficient to explain the offset seen by \citet{Shankar16}. However, their claim requires verification, as it is based on using conversions to 3.6$\mu$m band. As we discuss in Appendix~\ref{Appendix:Bias}, the SMBH sample is biased high also in the $\sigma$-3.6$\mu$m luminosity plane compared to the local galaxy samples in 3.6$\mu$m (S4G, \citealt{Sheth10}). In addition, in the same Appendix we also show that the bias persists in the $r$-band, with the SMBH local sample showing, on average, larger velocity dispersions than the local sample of MaNGA galaxies, further corroborating the initial findings by \citet{Bernardi07}. We conclude that the bias in the local SMBH sample still persists with respect to the larger local galaxy sample of SDSS/MaNGA galaxies (see Appendix~\ref{Appendix:Bias} for full discussion on this topic).

The left bottom panel of \figu\ref{fig|scalings} reports the local SMBH sample in the \mbh-\sis\ plane, compared with predictions from the \simba\ and \TNG\ simulations, similarly to the top panels. %Stellar velocity dispersions in the data are taken from the Hyperleda database \citep{Paturel03Hyperleda}, where all the \sis\ measurements are corrected to a common aperture of 0.595 kpc. The hydrodynamic simulations provide stellar velocity dispersions calibrated on apertures around the stellar half-mass radius of the galaxy, as discussed in \sect\ref{subsec|hydrosim}. %check this is in Section 2.
The \TNG\ simulation tends to produce a slightly flatter relation at higher masses with respect to the distribution of the data, in line with what also found by \citet[][see their Figure 1]{Li20TNG}, while the \simba\ simulation better aligns with the data. In the bottom right panel of \figu\ref{fig|scalings} we report, for completeness, the relation between \sis\ and \mstar\ in the data and in the two simulations. The \simba\ simulation again broadly aligns with the data, whilst the \TNG\ tends to predict $\gtrsim 0.1$ dex systematically lower mean stellar velocity dispersions at fixed stellar mass, as expected given that the simulation is consistent with the \mbh-\mstar\ relation but predicts somewhat lower velocity dispersions at fixed SMBH mass. We stress that alignment in the \sis-\mstar\ plane only ensures internal self-consistency between the model and the SMBH data sample, but not necessarily alignment between the model and the entire local galaxy population if the SMBH sample is affected by selection bias, as suggested in Appendix~\ref{Appendix:Bias}. In other words, current data sets on SMBHs with dynamical mass measurements are still too sparse to be used to securely discern among successful theoretical models. %the same aperture as in the data following \citet{Marsden22}, we verified, would improve the match to the observed galaxy \sis-\mstar\ relation, at least for the \TNG simulation, in line with \citet{Marsden22}, but it would worsen the match to the \mbh-\sis\ relation. Fixing the latter by varying, e.g., the AGN feedback efficiency, would of course inevitably worsen the good match to the \mbh-\mstar\ or \mbh-\mbulge\ relations. 
Additional spurious numerical effects may also affect the comparison between models and data. For example, the simulated central stellar velocity dispersions may be affected by resolution in some galaxies and/or being artificially inflated by the contribution of dark matter particles \citep[e.g.,][]{Schaye2015}. Overall, \figu\ref{fig|scalings} proves that comprehensively and simultaneously reproducing the scaling relations of local SMBHs still represents a non-trivial task even for some of the current state-of-the-art SMBH cosmological models.

\subsection{Residual analysis}
\label{subsec|Residuals}

In the previous \sect\ref{subsec|Scalings} we provided a broad comparison between available local data on galaxies with SMBH mass dynamical mass measurement and the predictions from two state-of-the-art hydrodynamic simulations. We now go deeper into the study of the SMBH scaling relations by analysing the pairwise residuals, following the formalism described in \sect\ref{sec|Method} and in \citet{Shankar17}. \figu\ref{fig|residuals_sahu} shows the residual analysis for the SMBH sample, distinguished in ETGs and LTGs as in the previous Figures. The left panels refer to residuals as a function of stellar velocity dispersion \sis\ at fixed galaxy stellar mass \mstar, while the bottom panels only include early-type galaxies. All panels in \figus\ref{fig|residuals_sahu} and \ref{fig|residuals_sahuMsph} uniformly use stellar velocity dispersion from the Hyperleda database and thus we label it as $\sigma_H$. As discussed in \sect\ref{sec|Method}, each residual plot reports the mean slope and standard deviation (magenta solid and dotted lines), along with its Pearson correlation coefficient $r$ and associated error, which we report in the bottom right corner of each panel. In all panels we also include a direct linear fit to the residuals (black, dotted lines), which are usually very close to the outcome of the iterative method, proving that the outcomes extracted from the residuals are robust against random statistical fluctuations, despite the relatively modest size of the SMBH sample. 

From all panels it is evident that the residual with stellar velocity dispersion is significant, with a correlation coefficient of $r\sim 0.66-0.75$, stronger for the ETG subsample. %When adopting only the bulges of LTGs and the total masses of ETGs the significance of the correlation somewhat drops, which is expected given the uncertainties in computing bulge masses and the fact that ETGs have not been corrected for the bulge component. 
The residual with galaxy stellar mass at fixed \sis\ is instead less significant, in line with what also found by \citet{Bernardi07} and \citet{Shankar16}, quantitatively demonstrating the fundamental importance of stellar velocity dispersion in driving the local SMBH-galaxy scaling relations. We repeat the exercise on the analysis of the residuals substituting total with spheroid stellar mass in \figu\ref{fig|residuals_sahuMsph}. In this case we find, when considering both ETGs and LTGs (top panels), tentative evidence for a ``fundamental plane'' of SMBHs, with SMBH mass being correlated with both stellar velocity dispersion \sis\ and spheroid stellar mass \msph\ in roughly equal strength. The residuals point to a relation of the type $\mbhe \propto \sise^{2.2} \msphe^{0.8}$, which is reminiscent of what predicted by some hydrodynamic models \citep{Hopkins2007_BHplane}, as a consequence of AGN feedback coupling SMBH mass with the binding energy of the host. \citet{Hopkins_2007_FP_observed} also found evidence of a fundamental plane-type relation in the local SMBH samples available at the time, although they were mostly referring to total galaxy stellar mass. 

The bottom panels of \figu\ref{fig|residuals_sahuMsph} show that, when considering the bulge stellar mass of only ETGs, the correlation with stellar velocity dispersion becomes even stronger, whilst the one with bulge mass becomes flatter, suggesting a tilted fundamental plane, as also put forward by \citet{Hopkins_2007_FP_observed}. However, in \figu\ref{figu|AppendixSaglia} we report the results of the residuals with respect to \sis\ and \msph\ extracted from the \citet{Saglia16} SMBH sample\footnote{We checked that the residuals from the \citet{Saglia16} sample still yield similar results even when restricting to the galaxies in common with \citet{Sahu23} and adopting their SMBH masses and stellar velocity dispersions.}, which confirms a strong and steep dependence of SMBH mass on stellar velocity dispersion, close to $\mbhe \propto \sise^4$, but a significantly weaker correlation with spheroid stellar mass, therefore not supporting the existence of a fundamental plane-type relation, as also noted in \citet{Shankar16}. The existence of a fundamental plane for SMBHs is thus not yet unambiguously confirmed, and in any case it is mostly evident when the \textit{spheroid} and not the total stellar mass are considered. 

In \figu\ref{figu|AppendixSersic} we also show the pairwise residuals applied to the \citet{Sahu23} sample between SMBH mass and spheroid S\'{e}rsic index $n$ (left) and spheroid half-light effective radius $R_e$ (right) at fixed stellar velocity dispersion (top) and spheroidal mass (bottom). When fixing stellar velocity dispersion, mild and weak residual correlations are found with $n$ and $R_e$, but these completely disappear when calculating the residuals at fixed spheroid mass \msph\ (bottom panels), further indicating that residual correlations between the SMBH mass and the structural properties of the host galaxy (or at least its spheroidal component) are mostly induced by the underlying correlation between SMBH mass and stellar spheroidal mass, in line with the analysis carried out by \citet{Shankar17}. We note that the anti-correlation with $R_e$ in the lower, right panel is only apparently strong possibly induced by some random error, as the data points are uniformly scattered with negligible correlation, as also indicated by the direct fit to the residuals (black, dotted line).  In summary, the residual analysis applied to the latest local SMBH data sample with dynamical mass measurements continues to support the dominance of stellar velocity dispersion over total or even spheroidal stellar mass as a more fundamental galactic property related to SMBH mass, further supporting the view that \sis\ is a key property in the evolution of galaxies \citep[e.g.,][]{Bernardi11curvature,Bernardi11mergers,Bluck20}.   

\figu\ref{fig|ResidualsSimulations} shows the comparison between the mean residuals derived from the SMBH sample (magenta, long-dashed lines; top panels of \figu\ref{fig|residuals_sahu}) and those predicted from the reference simulations \simba\ and \TNG\ (see \sect\ref{subsec|hydrosim}). For the simulations, stellar velocity dispersions, which we label in the plots as $\sigma_e$ to differentiate them from the ones from Hyperleda, are calculated using Eq~\ref{eq|sigma} over all the stellar particles associated to the host galaxy. The left panels are the residuals with stellar velocity dispersion \sis\ at fixed galaxy stellar mass \mstar, and the right panels plot the residuals with \mstar\ at fixed \sis. We find that the \simba\ simulation provides both residuals as a function of \sis\ and \mstar\ very close to the data, in particular the one with \sis, although the significance of these correlations may not be as strong. The \TNG\ simulation also provides a significant correlation with \sis\ in reasonable agreement with the data. Both simulations interestingly predict a negligible correlation with stellar mass at fixed stellar velocity dispersion, even weaker than in the data. A stronger correlation between SMBH mass and stellar velocity dispersion may be induced by the direct dynamical coupling generated by the kinetic AGN feedback incorporated in both simulations, which we briefly discussed in \sect\ref{subsec|hydrosim}, although this cannot be confirmed at this level of the analysis. 

\figu\ref{fig|ResidualsSimulationsMbulge} shows the same residuals as predicted by the simulations but with total stellar mass replaced by spheroidal stellar mass \mbulge. In \simba\ (top panels) the dependence of residuals with \mbulge\ at fixed \sis\ are more significant than in the case of total stellar mass, whilst the dependence of SMBH mass with \sis\ at fixed \mbulge\ is noticeably reduced, with a low Pearson correlation coefficient of just $r \sim 0.2$. The \TNG\ simulation continues to predict a significant correlation with \sis\ at fixed \mbulge, and also with \mbulge\ at fixed \sis, though flatter than in the \citet{Sahu23} data. %Therefore, despite the simulations predicting residuals in terms of total stellar mass broadly consistent with those observed in the data, they fail in terms of the residuals in stellar bulge mass \mbulge. We also checked that both simulations show a negligible correlation with host halo mass at fixed \sis\ or at fixed \mbulge. 
In other words, within the remit of the tests carried out in this work, the \simba\ simulation tends to point to stellar spheroidal mass as the most fundamental galactic property correlated with SMBH mass in central galaxies, whilst the \TNG\ would still favour \sis\ as the galactic property most correlated to SMBH mass. 
%We have verified that when repeating the selecting from the simulations only bulge stellar mass %For both simulations the errors on the slopes and Pearson coefficients of the residuals derived from the iterative method are very small, once again proving the stability of the correlations, not affected by statistical fluctuations. 
%the residuals with stellar bulge mass do not support a fundamental plane and provide weak dependencies and flatter than what reported by previous AGN feedback models \citep{Hopkins2007_BHplane}, and this persists even in the EX1 simulations with increased AGN kinetic power. 
%comment on residuals with halo mass from simulations and also ML and from data using Marasco21. Also cite Terrazas+21, Bandara09, Kaushala09, Li24. 
The residuals predicted by current state-of-the-art hydrodynamic simulations reported in \figu\ref{fig|ResidualsSimulations} and \ref{fig|ResidualsSimulationsMbulge}, depict SMBH-galaxy scaling relations that, although not yet fully aligned with what suggested by the current (and limited) data, are becoming increasingly closer to what observed, improving on previous comparisons \citep[e.g.,][]{Barausse17}. These hydrodynamic simulations, however, do not suggest the existence of any dynamical fundamental plane of SMBHs, favouring a scenario in which the SMBH is closely correlated to only one single galactic variable, either \mbulge\ or \sis, for \simba\ and \TNG, respectively \citep[see also the discussion in][]{Menci23}. 
%may still not fully capture  as found in \figu\ref{fig|scalings}, depart significantly from these data in terms of the residuals when stellar spheroidal mass is considered, and, at variance with previous theoretical models \citep[e.g.,][]{Hopkins2007_BHplane}, do not point to any clear SMBH fundamental plane, favouring a scenario in which the SMBH is closely correlated to only one single galactic variable, either \mbulge\ or \sis, for \simba\ and \TNG, respectively \citep[see also the discussion in][]{Menci23}. 

On the assumption that the \mbh-\sis\ relation and its residuals are mostly driven by AGN feedback \citep[e.g.,][]{Silk1998,Granato04,DiMatteo2005,Robertson06}, we could test whether an increase in the AGN kinetic feedback could steepen and strengthen the relation with stellar velocity dispersion at fixed bulge mass, in better agreement with the data explored in this work. In the top and middle panels of \figu\ref{figu|ResidualsSimulationsPlus} we report, in the same format as in \figu\ref{fig|ResidualsSimulationsMbulge}, the predicted residuals of, respectively, the \simba\ and \TNG\ models with an AGN kinetic outflow increased by a factor of 100, the so-called ``EX1'' simulation runs in CAMELS \citep{Angles-Alcazar2021}. As discussed in \sect\ref{subsec|hydrosim}, the box of the EX simulations are significantly smaller than the reference ones, and thus the statistical results may be less robust. Nevertheless, it is intriguing to note that the residuals in both \simba\ and \TNG\ are weakly affected by the increase in the AGN feedback kinetic output, generating a similar or even weaker correlations with \sis\ at fixed \msph. More notably, as shown in the bottom panels of \figu\ref{figu|ResidualsSimulationsPlus}, the increase in AGN kinetic output in both simulations inevitably reduces the SMBH mass at fixed host galaxy stellar mass, more markedly for the \simba\ simulation, most probably induced by the self-regulation in SMBH growth, which is reduced proportionally to the availability of gas in the surrounding medium. %Interestingly, the increase in AGN kinetic feedback does not significantly vary the residuals in the \TNG\ simulation (middle panels of \figu\ref{fig|ResidualsSimulations}), although it still reduces its SMBH masses overall (cfr with \figu\ref{fig|scalings}), in better agreement with the \mbh-\sis\ relation.   

\subsection{Machine learning algorithms}
\label{subsec|ML}

In the previous Sections we have used pairwise residuals to define the degree of linear correlation among several SMBH-galaxy scaling relations, finding that in the data \sis\ appears as the most fundamental property linked to SMBH mass. The aim of this \sect\ is to explore putative correlations among these variables by making use of a variety of distinct and complementary approaches, based on Machine Learning (ML) algorithms. We immediately stress that none of the correlation tests performed in this \sect\ provides the wealth of information available from pairwise correlation residuals, which also output the slopes and \textit{relative} strengths of the correlations. Nevertheless, the ones discussed here are useful complementary approaches to probe and confirm any degree of correlation.     

In \figu\ref{figu|CorrelationsMachineLearning} we present the result of applying ML regression techniques to both the observed and simulated data. Each regression has been performed using only one variable in each case as a predictor of \mbh. In this way we can test the predictive power of each variable independently. As it is standard practice in ML regression analysis, we report in the plots the value of the coefficient of determination $R^2$ which indicates the quality of the fit by measuring the total variance of the outcome as indicated by the predictor 
\begin{equation}
    R^2(y,\hat{y})=1-\frac{\sum(y_i-\hat{y}_i)^2}{\sum(y_i-\bar{y})^2}\, ,
    \label{eq|R2}
\end{equation}
where $y_i$ and $\bar{y}$ correspond to the measured and average data values, respectively, while $\hat{y}_i$ are the values predicted by a given model.
%[if you want to include the formula, it is here: https://scikit-learn.org/stable/modules/model_evaluation.html#r2-score ]. 
The closer the coefficient of determination is to 1, the higher the correlation between the two variables, which indicates either a direct causal relationship between the variables or correlations through other intermediate variables. Values of $R^2$ close to zero or negative suggest, instead, no causal relationship between the variables. We note that the $R^2$ coefficient of determination is different from the Pearson correlation coefficient reported in the pairwise residuals. The two statistical indicators are in fact not directly comparable to each other for several reasons: \textit{i}) they are characterised in different ways, with $r$ explicitly defined on both variables $x$ and $y$, while $R^2$ explicitly defined on the $y$ variable and only implicitly on $x$ via the predicted value $\hat{y}$, which depends on $x$; \textit{ii}) the parameter $r$ assumes a linear fit between $x$ and $y$, while the ML regressions adopted here do not follow any particular functional relation between variables, allowing for any type of linear or non-linear relationship; \textit{iii}) the definition of $r$ adopted in this work also includes measurement errors in both variables (see full formalism in \citet{ShethBernardi12} and in Appendix B in \citealt{Shankar17}). Despite these differences, both parameters provide a quantitative estimate of the degree of correlation strength among variables, and our aim here is only to test whether, within a given set of observational or numerical data, the two statistical indicators point to the same variables as being more or less correlated with the mass of the central SMBH.  %The former, defined in \eq\ref{eq|R2}, in practice is a measure of the performance of the model compared to the average value of the data. The latter is formally the ratio of the covariance of both $x$ and $y$ variables and the product of their standard deviations. In addition, in this work the calculation of the Pearson coefficient also includes measurement errors in both variables, following the formalism developed in \citet{ShethBernardi12}, as detailed in Appendix A of \citet{Shankar17}. %When excluding measurements errors and taking into account only the $y$ variable, it can be shown that the Pearson coefficient reduces to the square of $R^2$. 

For each predictor we show the regression results using different techniques. This approach ensures the robustness of the results by confirming that they are independent of the details of any of the specific ML recipe adopted in the analysis. The selected algorithms include a wide and diverse set of methodologies, including linear and non-linear methods, Bayesian approaches, ensemble meta-approaches combining different estimators, gradient optimisation, etc. The techniques applied are the following: AdaBoost \citep{drucker1997improving}, Random Forest \citep{breiman2001random, breiman1998arcing}, Support Vector Machines with linear and non-linear kernels \citep{rong2008liblinear, chang2011libsvm, smola2004tutorial}, Ridge Regression \citep{hoerl1970ridge, rifkin2007notes}, Stochastic Gradient Descent \citep{bottou2012stochastic, tsuruoka2009stochastic, zhang2004solving}, Gradient Tree Boosting \citep{friedman2001greedy, friedman2002stochastic}, Bayesian Ridge Regression \citep{mackay1992bayesian, tipping2001sparse}, Decision Trees \citep{breiman1984classification}, and K-nearest Neighbors \citep{fix1989discriminatory, bentley1975multidimensional, omohundro1989five}. All these techniques have been implemented using the scikit-learn library \citep{pedregosa2011scikit}. The methodology applied is as follows. The data used in each regression are pre-processed by scaling them between zero and one to facilitate the regression. For each regressor, a grid search of its main hyperparameters is conducted in order to select those that obtain the best result evaluated by minimising the mean squared error of the regression. The regression fit and its evaluation are carried out by means of cross-validation with 5 folds. This implies that each regression is carried out 5 times in each case, using different parts of the data to perform the fit and to evaluate the result. This ensures that the choice of which part of the data is used to perform the regression is not biasing the results, in line with the random selection of data we performed when computing the pairwise residuals. In each case 80\% of the data is used for the regression and 20\% for its evaluation. The final result of the coefficient of determination and its error is calculated from the average and standard deviation of these 5 regressions. 

The top, middle, and bottom panels of \figu\ref{figu|CorrelationsMachineLearning} report the results of the adopted ML algorithms, labelled on the $y$-axis, in terms of the $R^2$ parameter on the $x$-axis, applied, respectively, to the full data set of the local SMBH sample (top panels of \figu\ref{fig|scalings}), the (reference) \simba\ and \TNG\ simulations (with reference values for the AGN feedback efficiencies). The first clear result is that all the ML predictors agree in finding a similarly significant correlation in the observational data between \mbh\ and \sis\ and between \mbh\ and \mbulge\ (top left and right panels), supportive of a fundamental plane relation, with values of the coefficient of determination on average $R^2 \sim 0.6$, but not an equally strong relation of \mbh\ with total stellar mass \mgal\ with an average $R^2\sim 0.3$ (top, middle panel). %On the other hand, the correlation between \mbh\ and \mstar\ in the data (top, right) \st{is not significant, with $R^2$ having null or even negative average values} \mac{is much less significant, within the range $R^2 \sim 0.15-0.4$ for the mean values and much larger uncertainties reaching null or even negative average values}. 
These results are in line with what was concluded from the pairwise residuals in the top panels of \figu\ref{fig|residuals_sahuMsph}, where the Pearson coefficient was higher for the correlation with \sis\ and \mbulge, but showing a significantly weaker correlation with \mstar\ at fixed \sis\ (\figu\ref{fig|residuals_sahu}). We also found that when restricting the analysis to only ETGs, the ML algorithms continue to point to a similar correlation with \mbulge, with $R^2 \sim 0.5-0.6$, and an even more marked correlation with \sis, with $R^2 \sim 0.7$, in line with what was retrieved from pairwise residuals in the bottom panels of \figu\ref{fig|residuals_sahuMsph}. We checked that the ML algorithms point to much weaker correlations of SMBH mass with both bulge S\'{e}rsic index and half-light radius, with an average $R^2\sim 0.3$, a trend that, as suggested by \figu\ref{figu|AppendixSersic}, could be entirely ascribed to the underlying dependence of SMBH mass on \mbulge.  Interestingly, the ML predictors do not identify any clear correlation between \mbh\ and \sis\ or \mstar\ in the \simba\ simulation (middle panels), but a moderate one between \mbh\ and \mbulge\ (middle right panel), a result which is consistent with the pairwise residuals analysis performed in the top panels of \figu\ref{fig|ResidualsSimulationsMbulge}, which identified in \simba\ a noticeable correlation of SMBH mass with stellar bulge mass \msph\ at fixed \sis. On the other hand, the \TNG\ simulation, according to the ML predictors, shows a strong, well-defined correlation between \mbh\ and \sis\ with an average $R^2 \sim 0.5$, and, to a lesser extent with \mbulge, again in line with what was concluded from the pairwise residuals in the bottom panels of \figu\ref{fig|ResidualsSimulationsMbulge}.  All in all, the ML tests agree with the results of pairwise residuals, pointing to \mbulge, in the case of \simba, or \sis\ and \mbulge, in the case of \TNG, as the main galaxy properties correlated with the SMBH mass \mbh. The ML algorithms adopted here do not, however, constrain the mathematical relation among the different variables and are thus unable to pin down any specific model of, e.g., AGN feedback. 

\section{Discussion}
\label{sec|discu}

One of the main results of this work is the evidence of a strong correlation between \mbh\ and \sis. This empirical result represents a significant step forward in our understanding of the origin and evolution of SMBH scaling relations. The seminal work by \citet{Bernardi07} had pointed out via residuals the key role played by the stellar velocity dispersion in the SMBH scaling relations, and the residual analysis carried out by \citet{Shankar16}, \citet{Shankar17}, and \citet{Marsden20} confirmed this trend. The exquisite photometric homogeneity achieved by \citet{Sahu19} and \citet{Davis19} on the local SMBH sample, now allows for a more accurate analysis of the residuals, providing further robust evidence for stellar velocity dispersion being more fundamental than total stellar mass or other photometric galactic properties, as shown in Appendix~\ref{Appendix:MoreResiduals}. 

We also found possible evidence for the existence of a fundamental plane between SMBH mass and stellar velocity dispersion and bulge stellar mass, of the type $\mbhe \propto \sise^{2.2} \msphe^{0.8}$, broadly aligned with what claimed by \citet[][see also \citealt{Iannella21}]{Hopkins_2007_FP_observed}. The ``dynamical'' fundamental plane of SMBHs (not to be confused with the fundamental plane of SMBH activity by \citealt{Merloni03}) becomes more tilted when only ETGs are considered (bottom panels of \figu\ref{fig|residuals_sahuMsph}) and evidence for its existence is significantly weakened by the residual analysis performed on the SMBH sample by \citet{Saglia16}, which is also characterised by accurate stellar spheroidal measurements (\figu\ref{Appendix:MoreResiduals}). Uniform and careful analyses on larger SMBH samples are required to unveil the actual existence of a dynamical SMBH fundamental plane, possibly also taking advantage of uniform measurements of host properties and SMBH masses of type 1, low/moderate luminosity AGN, which will soon become a reality with the Euclid \citep{Euclid} and Vera Rubin LSST \citep{Ivezic2019_LSST} surveys. 

%halo mass
%Ferrarese02
%Powell-Shankar20 clustering
The pairwise residuals might favour AGN feedback as a key driver shaping SMBH galaxy scaling relations, either directly via the \mbh-\sis\ relation \citep[e.g.,][]{Silk1998}, or via a tilted correlation involving also spheroid stellar mass. A simultaneous correlation between SMBH mass with both stellar velocity dispersion and stellar spheroid mass, if confirmed, could favour a two-phase SMBH evolution \citep[e.g.,][]{Cook09twophase,Oser10,Boco23}, where SMBH's growth within the proto-spheroid, also possibly triggered by a gas-rich major merger and regulated by AGN feedback, could generate a tilted relation between SMBH mass and host dynamical mass \citep[e.g.,][]{Hopkins2007_BHplane}. The later phase of galaxy evolution, which may include the formation of a surrounding stellar disc \citep[e.g.,][]{Cook09twophase,Hopkins09disks,Cook10}, not necessarily linked with further growth onto the central SMBH, may loosen the correlation between SMBH mass and total galaxy stellar mass, as indeed suggested by current data. 

On the other hand, detailed hydrodynamic simulations such as \simba, inclusive of cutting-edge implementations of AGN outflows, do not necessarily point to an AGN-driven origin of the SMBH scaling relations. \citet[][see also \citealt{Angles-Alcazar2015}]{Angles17} showed that the shape of the \mbh-\mstar\ relation is largely independent of the strength of AGN feedback. Similar normalizations and slopes are recovered in their simulations when moving from a model with no AGN feedback to one with a strong outflow characterized by 20 times the fiducial value (see their Figure 4), with only a modest decrease in normalization by a factor of $\sim 2$ in the latter model. Similar conclusions were also reached by \citet{Menci23} adopting a semi-analytic galaxy evolution model with a new treatment of AGN-driven winds. They showed that all SMBH scaling relations are largely preserved in both normalization and slope when including a significant AGN outflow component, which mostly controls the dispersion around the relations. \citet{Angles17} discussed that in their simulations SMBHs and galaxies grow in lockstep along the \mbh-\mstar\ relation, the normalization of which is largely regulated by the parameters controlling the physics of accretion onto the central SMBHs. \citet[][and references therein]{Datta24} have also discussed how the scaling with galaxy stellar mass arises from the tight relationship between SMBH accretion rate and galaxy star formation rate inducing, in some models like the ASTRID simulation \citep{Bird2022_Astrid}, a nearly linear and redshift independent \mbh-\mstar\ relation. 

The two state-of-the-art hydrodynamic simulations considered in this work, \simba\ and \TNG, which implement different prescriptions for the SMBH gas accretion and AGN feedback (see \sect~\ref{subsec|hydrosim}), align with the previous theoretical studies discussed above, showing that a correlation with galaxy stellar mass, or a steep correlation with stellar velocity dispersion, are not necessarily a direct byproduct of an underlying AGN feedback-regulated process. %We first noticed that both simulations broadly align with the general SMBH-galaxy scaling relations, and also with the residuals at fixed total stellar mass. However, they tend to perform significantly worse in the residuals at fixed bulge stellar mass (\figu\ref{fig|ResidualsSimulationsMbulge}). %There are at least two relevant points that can be derived from \figu\ref{fig|ResidualsSimulationsMbulge}. %Firstly, even though models are often fine-tuned to match similar data sets and produce very similar SMBH scaling relations (cfr. upper panels of \figu\ref{fig|scalings}), they can still predict very different pairwise residuals, a behaviour that in turn reflects the different physics implemented in the models. 
As highlighted by several previous works \citep[e.g.,][]{Barausse17,Shankar17,Menci23}, the apparent match between model predictions and data in the SMBH scaling relations may in fact sometimes simply arise as a result of a combination of different factors (see, e.g., \citealt{CavaliereVittorini} and discussion and appendices in \citealt{Shankar17}). Inspired by the pairwise residuals' results, we could write the global relation between SMBH mass \mbh\ and galaxy properties as $\mbhe\propto\sise^{\beta}\msphe^{\alpha}\propto\sigma^{\beta+\alpha\,\gamma}$, where $\gamma$ comes from the $\msphe\propto\sise^{\gamma}$ relation. The \citet{Sahu23} SMBH sample yields a correlation of the type $\msphe\propto \sise^{3.5}$, which in turn would imply, when combined with the slopes from the top panels of \figu\ref{fig|residuals_sahuMsph}, $\mbhe\propto \sise^{2.2}\msphe^{0.8}$, or $\mbhe\propto \sise^{2.2+0.8\times3.5}\propto \sise^5$. So the SMBH data may still yield a strong and steep correlation with stellar velocity dispersion, but mostly as a byproduct of the additional correlation with the spheroidal component. By repeating the same exercise for the \simba\ simulation we would get $\mbhe\propto \sise^{1+0.8\times3.8}\propto \sise^{4}$ and for the \TNG\ $\mbhe\propto \sise^{2+0.3\times4.2}\propto \sise^{3.3}$, which are close to the correlations seen in the bottom left panel of \figu\ref{fig|scalings}. In other words, both simulations could provide relatively steep \mbh-\sis\ correlations but for somewhat different reasons which are ultimately dictated by the different physics implemented in the models. 

Pairwise residuals, more than the scaling relation themselves \citep[e.g.,][]{Habouzit2021_Mbh}, have the potential to distinguish among the most successful models and ultimately reveal the underlying physics regulating the co-evolution of SMBHs and their host galaxies. It is thus vital to compare theoretical predictions with the data not only in terms of absolute scaling relations but also in terms of their residuals to pin down the true performance of a given model. The statistical tests carried out in \figu\ref{figu|CorrelationsMachineLearning} via a variety of ML regression models, largely support the results from the pairwise residuals, but they are less informative, thus caution needs to be applied when interpreting causality among variables just based upon the outputs of ML-based regression algorithms \citep[e.g.,][]{Bluck20}. 

A tight underlying connection between SMBH mass, stellar velocity dispersion and bulge stellar mass, could still point to a self-regulated SMBH growth, where AGN feedback is controlled by the potential well of its host \citep[e.g.,][and references therein]{Lopez23}. In this context, one would then also expect a correlation between SMBH mass and host dark matter halo mass \mhalo. Several authors have in fact found evidence for a tight correlation between stellar velocity dispersion and the large-scale circular velocity as traced by either dynamical models or HI rotation curves measurements \citep[e.g.,][see also \citealt{Cirasuolo05}]{Ferrarese02,Baes03,Pizzella05}, which, if extrapolated, could indicate an underlying correlation with the host halo mass. 

\begin{figure*}
\begin{center}
\center{{
\epsfig{figure=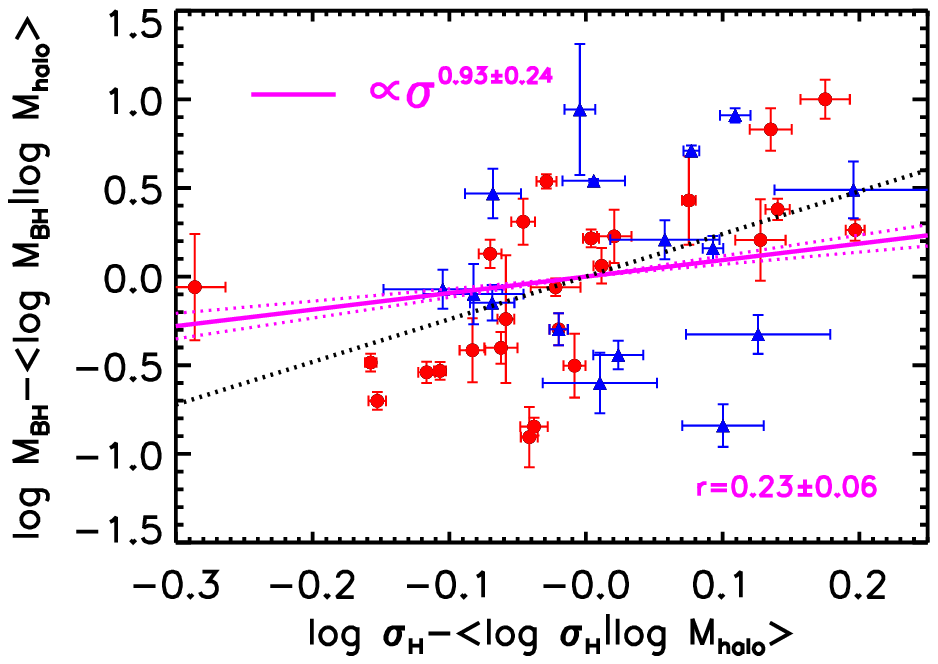,height=6.5cm}\hspace{-0.85cm}
\epsfig{figure=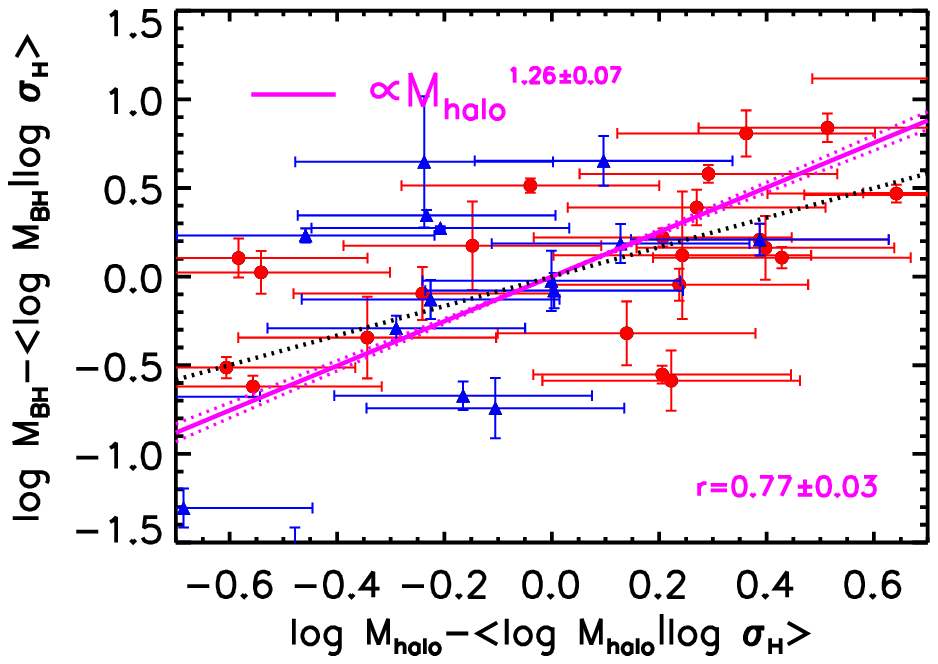,height=6.5cm}
}}
\vspace{-0.52cm}
\center{{
\epsfig{figure=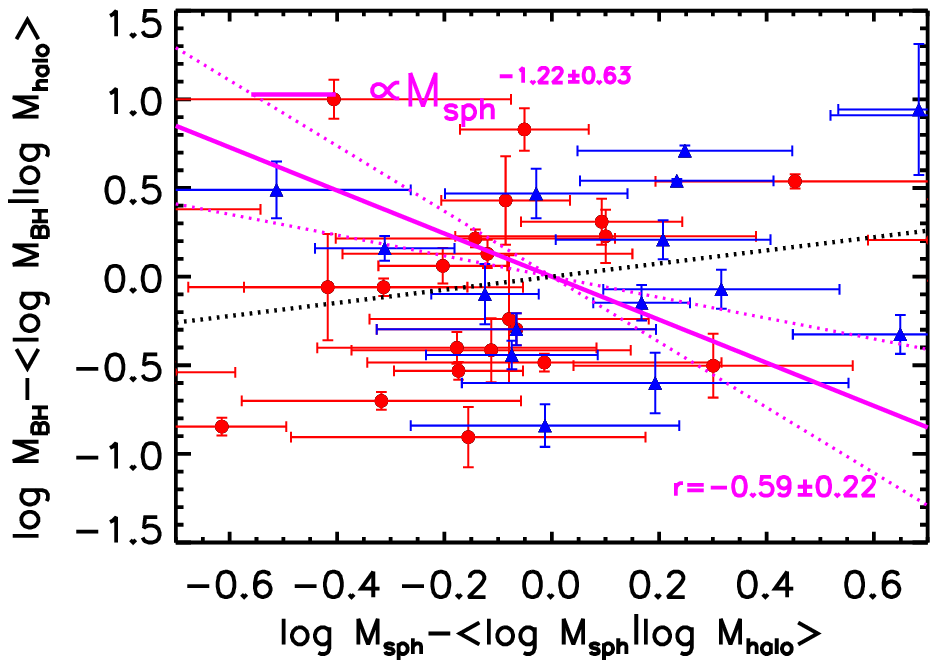,height=6.5cm}\hspace{-0.85cm}
\epsfig{figure=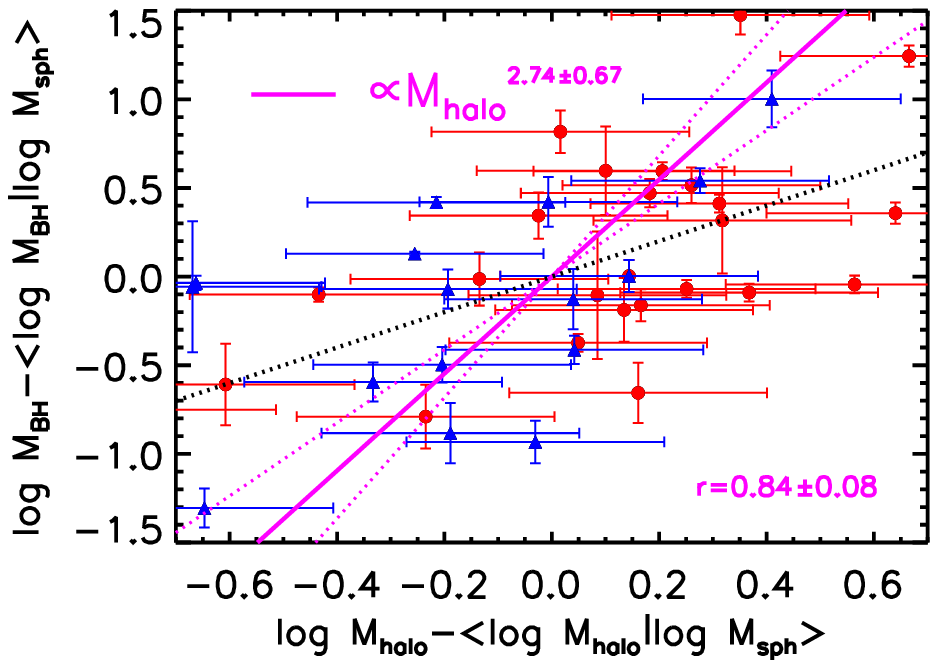,height=6.5cm}
}}
\caption{Residuals as a function of \sis, \msph, and \mhalo\ for the subsample of SMBHs from \citet{Sahu23} with host dark matter halo mass measurement from \citet{Marasco21}. The residuals with host halo mass appear respectively stronger than those with \sis\ and \msph\ at fixed halo mass.}
        \label{fig|residuals_sahuMhalo}
\end{center}
\end{figure*}

To investigate this intriguing possibility using the unique power of pairwise residuals, we have selected the subsample of 41 galaxies from the SMBH parent sample of \citet{Sahu23} with dark matter halo mass measurements from \citet{Marasco21}. The latter have halo masses either derived from globular cluster dynamics coupled with assumptions on the host dark matter halo and stellar profiles from \citet{PostiFall21}, or derived from the circular velocity of the flat part of the rotation curve converted into halo masses using the linear relation \mhalo-$v_{\rm flat}$ from \citet[][see Appendix A in \citealt{Marasco21} for full details]{Posti19}. \figu\ref{fig|residuals_sahuMhalo} shows the residuals of this subsample in terms of \mhalo\ at fixed \sis\ and \mbulge\ (right panels) and in terms of \sis\ and \mbulge\ at fixed \mhalo\ (left panels). It is evident that a significant correlation exists between SMBH mass and host halo mass, whilst the correlation with both \sis\ and with \msph\ are weaker at fixed halo mass which thus appears even more correlated with SMBH mass than either of the galactic variables, possibly because the latter are in turn controlled by the host halo mass (analogously to the case of the residual in $R_e$, the apparent anti-correlation with \mbulge\ at fixed \mhalo\ is a byproduct of some random errors and not significant, as demonstrated by the nearly flat direct fit to the residuals). For completeness, we also checked the residual correlation between SMBH mass and potential well of the spheroidal component $W \propto \sise^2 \msphe$ at fixed halo mass, finding a very weak, slightly negative correlation with $r\sim -0.33$, while still a strong correlation with halo mass at fixed $W$, with $r \sim 0.66$. Interestingly, we do not detect any significant residual correlation with halo mass in \simba\ while a moderate correlation with halo mass is found in \TNG. Caution has to be taken in over-interpreting the results in \figu\ref{fig|residuals_sahuMhalo} as the halo mass measurements are still based on some ad-hoc assumptions and the sample is nearly half of the original one from \citet{Sahu23}, which is already quite small. 

Nevertheless, the conclusions drawn from \figu\ref{fig|residuals_sahuMhalo} would also align with the independent results from some AGN clustering measurements. \citet{Powell22} have shown from forward modelling of the Swift/BAT AGN Spectroscopic Survey that AGN clustering measurements prefer models with a direct correlation between SMBH mass and host halo mass at fixed stellar mass, and their best-fit \mbh-\mhalo\ relation is in line with what retrieved from other groups using independent clustering measurements and different AGN samples \citep[e.g.,][]{Shankar20Nat,Allevato21}. On the other hand, from pairwise residuals analysis we checked that the \simba\ and \TNG\ simulations do not show any noticeable dependence between \mbh\ and \mhalo\ at fixed \sis\ ($r \lesssim 0.15$), further corroborated by the ML regression algorithms, which also point to a negligible correlation with \mhalo\ ($R^2 \lesssim 0.25$). 

Beyond host halo mass, other galactic properties have been put forward in the literature along the years as potentially providing a stronger correlation to SMBH mass than stellar velocity dispersion. For example, several groups have remarked that in cored ellipticals one of the strongest correlations is between the size of the central stellar core and SMBH mass, possibly as a result of the scouring of stars by mergers
of binary SMBHs \citep[e.g.,][and references therein]{Rusli13}. We verified from pairwise residuals that, indeed, in the sample of ellipticals collected and studied by \citet[][]{Rusli13}, also inclusive of NGC1272 from the latest work by \citet{Saglia24}, the SMBH mass appears significantly more strongly correlated with core radius than with stellar velocity dispersion (taken from the Hyperleda database, for consistency with the analysis carried above on the \citet{Sahu23} SMBH sample). However, this result mostly stems from the fact that the limited subsample of 24 cored galaxies with dynamical mass measurements of their central SMBHs (most of which are already included in the \citet{Sahu23} SMBH sample) is characterized by an overall poor correlation between SMBH mass and host galaxy stellar velocity dispersion.%, it is thus expected that the residuals will not detect any strong dependence on \sis. Whilst these cored galaxies, taken in isolation, may indeed departure from the \mbh-\sis\ relation and contribute to the scatter seen in the bottom left panel of \figu\ref{fig|scalings}, when considered with the other galaxies with SMBH dynamical mass measurements, they still follow similar residuals.   

The possible overall dominance of stellar velocity dispersion (or possibly host potential well or host halo mass or even core radius) in driving the SMBH-galaxy scaling relations has profound consequences in the way observers and modellers approach the study of the co-evolution of these two systems. The slope and normalization of the \mbh-\mstar\ relation must reflect the shape and evolution of the \mbh-\sis, \mbh-\mbulge\ or \mbh-\mhalo\ relations \citep[e.g.,][]{Menci23}, thus the interpretation of SMBH evolution only through the lens of the \mbh-\mgal\ plane may be incomplete, affected by large scatter and/or biases (see Appendix \ref{Appendix:Bias}), and possibly even misleading. \citet{Terrazas20} for example studied the effects of AGN feedback on the \mbh-\mgal\ relation coupled to specific star formation rate, but stellar velocity dispersion and, possibly, spheroid stellar mass could represent more robust variables to test recipes for AGN feedback both observationally and theoretically.  

On a similar vein, observational evidence is accumulating for a significant evolution in the \mbh-\mstar\ relation, at least when comparing similar types of galaxies at different epochs \citep[e.g.,][]{Farrah23}, while new high-redshift measurements from local and high-redshift AGN from, e.g., JWST, are showing larger scatters and/or non-trivial evolution \citep[e.g.,][]{Reines2015,Pacucci23}. As described by \citet{Maiolino24Xray}, collectively the new JADES high-z SMBHs tend to better align with the local \mbh-\sis\ relation \citep{Maiolino23,Juod24}, suggesting that the latter is more fundamental, in line with the findings in this work, and has a relatively weak evolution \citep[e.g.,][]{Shankar09MbhSigma,Shen15}. Indeed, many studies carried out on serendipitous AGN samples tend to favour a much weaker evolution in both scaling relations \citep[e.g.,][]{Suh20,Shen15,ZhuangHo,Marsden22,Lopez23,Tanaka24}, as also suggested by arguments based on the time integrated emissivity of AGN converted to SMBH mass densities via a mean radiative efficiency \citep[e.g.,][]{Shankar09MbhSigma}. It is also interesting to remark that SMBH accretion models based on the time integration of empirical (mostly X-ray/IR based) Eddington ratio distributions $P(\lambda \propto L/\mstare,z)$, continue to point to mean \mbh-\mstar\ scaling relations significantly below those traced by local dynamically measured inactive SMBHs when adopting standard radiative efficiencies of $\gtrsim 10\%$, further suggesting some tensions in the demography of SMBHs in the \mbh-\mstar\ relation \citep[e.g.,][]{ReinesVolonteri,Yang17,Shankar19,Carraro20,Shankar20,Shankar20Nat, Suh20,Carraro22,Farrah23,Zou24,Terrazas24}, which largely disappear when considering SMBH accretion models performed on the \mbh-\sis\ plane \citep[e.g.,][]{RicarteNatarajan18a,RicarteNatarajan18b,Marsden22}.

SMBH mergers could also be playing a non-negligible role in shaping the SMBH-galaxy scaling relations \citep[e.g.,][]{JahnkeMaccio,Hirschmann10,Shankar12Mergers,Zou24}, but they must act in ways to preserve the \mbh-\sis\ relation first \citep[e.g.,][]{Robertson06}, which is also the primary relation that should be considered when preparing galaxy-AGN mocks \citep[e.g.,][]{Allevato21}. Our results also have implications for the interpretation of the stochastic gravitational wave background extracted from the 15 yr pulsar timing array (PTA) data set collected by the North American Nanohertz Observatory for Gravitational Waves (NANOGrav) collaboration \citep{Afzal23}. As discussed by \citet[][]{Afzal23}, several SMBH binary populations are able to reproduce both the amplitude and shape of the observed low-frequency GW spectrum \citep[][and references therein]{Sesana16,Agazie23}. However, as pointed out by \citet{LacyFarrahSMBHsGWs}, the peaks of the posterior distributions in \citet{Agazie23} point to a SMBH mass density up to an order of magnitude larger than many of the previous determinations of the local SMBH mass function \citep[e.g.,][]{Marconi04,Shankar04,Tundo07,Shankar09,Shankar13demography,Shankar20,Shankar20Nat,Sicilia22}. 

\citet{LiepoldMa24} have recently attempted a new measurement of the local SMBH function confirming the long-standing issue of a systematic discrepancy between the SMBH mass densities inferred from the \mbh-\mstar\ and \mbh-\sis\ relations, with the latter usually providing a significantly lower value \citep{YuTremaine2002,Shankar04,Tundo07,Shankar09review}. \citet{Shankar16} showed that this discrepancy could be reconciled in light of a possible bias (see Appendix~\ref{Appendix:Bias}) that preferentially affects the local \mbh-\mstar relation. %, which could preferentially push the \mbh-\mstar\ relation to higher values more than the \mbh-\sis\ one, if the SMBH mass has an underlying stronger dependence on \sis\ than on \mgal. 
\citet{Burke24} also recently estimated a local SMBH mass function consistent, at least at the high-mass end, with \citet{LiepoldMa24} when adopting the \mbh-\mstar\ relation, although significantly below it at lower SMBH masses. \citet{LiepoldMa24} suggested that the (higher) value of the SMBH mass function implied by the \mbh-\mstar\ relation should be preferred to the one arising from the \mbh-\sis\ relation, as the former is more consistent with the large masses of the SMBH binaries required by the PTA independent measurements. However, the pairwise residual results from this work would indicate that the \mbh-\sis\ relation is a more solid tracer of SMBH mass, and thus a more secure route towards a more robust census of SMBHs. The residuals would thus suggest that more accurate models of the SMBH mass function and SMBH merger rates should be anchored to the \mbh-\sis\ relation to provide a more insightful comparison with current and future GW measurements. Interestingly, at present, the large masses of the SMBH binaries derived from current PTA measurements are in tension with those extracted from the \mbh-\sis\ relation, as recently emphasized by \citet{Sara25}, highlighting once more the well-known systematic discrepancy between the SMBH mass densities inferred from the two scaling relations with stellar velocity dispersion and stellar mass \citep[e.g.,][]{Shankar04,Tundo07,LiepoldMa24,Burke25}. 

%evolution of scaling relations Mbh-Mgal-Mbulge see
%https://ui.adsabs.harvard.edu/abs/2023MNRAS.tmp.1825S/abstract 
%https://ui.adsabs.harvard.edu/abs/2023NatAs.tmp..175Z/abstract 
%Mbh-Mhalo

\section{Conclusions}
\label{sec|conclu}

The apparently strong correlations between SMBH masses and their host galaxy properties carry the imprint of a possible co-evolution of these two systems. It is of vital importance to unveil the underlying nature and significance of these correlations if we want to truly advance in our knowledge of the growth of SMBHs and their putative impact onto the mass, structural, and dynamical evolution of their hosts. In this work, we have collected the latest sample of local galaxies from \citet{Sahu23} with a uniform calibration of their photometric properties and with dynamically measured masses of their central SMBHs, analysed the scaling relations with pairwise residuals and ML regression algorithms, and compared the results with the outputs from two state-of-the-art hydrodynamic simulations, \simba\ and \TNG. Our main results can be summarised as follows:
\begin{itemize}
    \item The latest local SMBH data with uniform photometric calibrations and accurate bulge-to-total decompositions, point to some degree of correlation between SMBH mass \mbh\ and their host galaxy stellar velocity dispersion \sis\ and stellar mass \mstar. LTGs follow a steeper \mbh\--\mstar\ correlation than ETGs, while both ETGs and LTGs align better with spheroidal mass in the \mbh\--\msph\ relation. The models broadly follow these trends (\figu\ref{fig|scalings}).
    \item The pairwise residuals confirm that \mbh\ correlates more strongly with \sis\ than with total stellar mass \mstar, spheroidal effective radius, or S\'{e}rsic index (\figus\ref{fig|residuals_sahu} and \ref{figu|AppendixSersic}).
    \item When the spheroid stellar mass is considered, the local SMBH sample hints at a possible fundamental plane-type correlation of the type $\mbhe \propto \sise^{2.2} \msphe^{0.8}$ (\figu\ref{fig|residuals_sahuMsph}), although this evidence is significantly weakened when switching to the \citet{Saglia16} SMBH sample (\figu\ref{figu|AppendixSaglia}).
    \item The hydrodynamic simulations considered in this work, \TNG\ and \simba, tend to align with the data in the residuals of \mbh\ versus \sis\ at fixed \mstar\ and predict a negligible correlation with \mstar\ at fixed \sis\ (\figu\ref{fig|ResidualsSimulations}). 
    \item In addition, \simba\ favours \mbulge\ as the galactic variable most closely linked to SMBH mass, while \TNG\ points to \sis\ (\figu\ref{fig|ResidualsSimulationsMbulge}). Increasing the AGN kinetic output in these simulations does not change these trends (\figu\ref{figu|ResidualsSimulationsPlus}).
    %\item A variety of other algorithms bases on Machine Learning regression methods and causal discovery techniques all support the results of the pairwise residuals, supporting a strong dependence of \mbh\ on \sis\ but not on \mstar\ in the data; the regression methods considered here do not provide any information on the type of the correlation, in terms of, e.g., slope (\figus\ref{figu|CorrelationsMachineLearning} and \ref{fig|CorrelationsCausality}).
    \item Another strong underlying correlation is found between \mbh\ and host halo mass \mhalo\ (\figu\ref{fig|residuals_sahuMhalo}), albeit on a significantly smaller sample and with some assumptions on the derivation of the halo masses. %The simulations do not predict any significant underlying correlation with halo mass at fixed \sis. 
    \item We show that the sample of local galaxies with dynamically measured SMBHs is biased high in \sis\ with respect to the (significantly) larger sample of local galaxies as measured in the 3.6$\mu$m or in the SDSS bands, especially towards lower luminosities (\figu\ref{figu|AppendixBias}). We discuss this important point in detail in Appendix~\ref{Appendix:Bias}.  
\end{itemize}

Our results on pairwise residuals favour stellar velocity dispersion and host halo mass as more important global variables regulating the connection between SMBHs and their hosts. The comprehensive models explored here are able to broadly capture these trends, although they still require extreme fine-tuning in their physical prescriptions to simultaneously reproduce the observed SMBH scaling relations \textit{and} their pairwise residuals, in particular with bulge stellar mass \mbulge\ and possibly host halo mass \mhalo. Future models and data must continue to probe the strong link visible in the local Universe between \mbh, \sis\ and potentially even host halo mass \mhalo, both as a reference for SMBH mass measurements, as well as a more robust probe of the evolutionary link between SMBHs and their hosts and the implied stochastic gravitational wave background. 

\section*{Acknowledgments}
FS acknowledges partial support from the European Union's Horizon 2020 research and innovation programme under Marie
Sk{\l}odowska-Curie grant agreement No 860744 ``Big Data Applications for Black
Hole Evolution Studies'' (BiD4BESt; Coordinator F. Shankar).
We acknowledge Nandini Sahu, Alister Graham, and Benjamin Davis for sharing some of their data and for useful clarifications on their data. FS also thanks Romeel Dav\'{e} and Roberto Saglia for interesting discussions. 
HF acknowledges support from the Shanghai Super Post-doctoral Excellence Program grant No. 2024008.

\section*{DATA AVAILABILITY}
All the data and simulations adopted in this paper are publicly available. The software underlying the statistical analysis can be made available upon reasonable request to the corresponding author.

\bibliographystyle{mn2e_Daly}
\bibliography{Refs}

\appendix

\section{Investigating the presence of a bias in the local scaling relations of SMBHs and their host galaxies}
\label{Appendix:Bias}

\begin{figure*}
\begin{center}
\center{{
\epsfig{figure=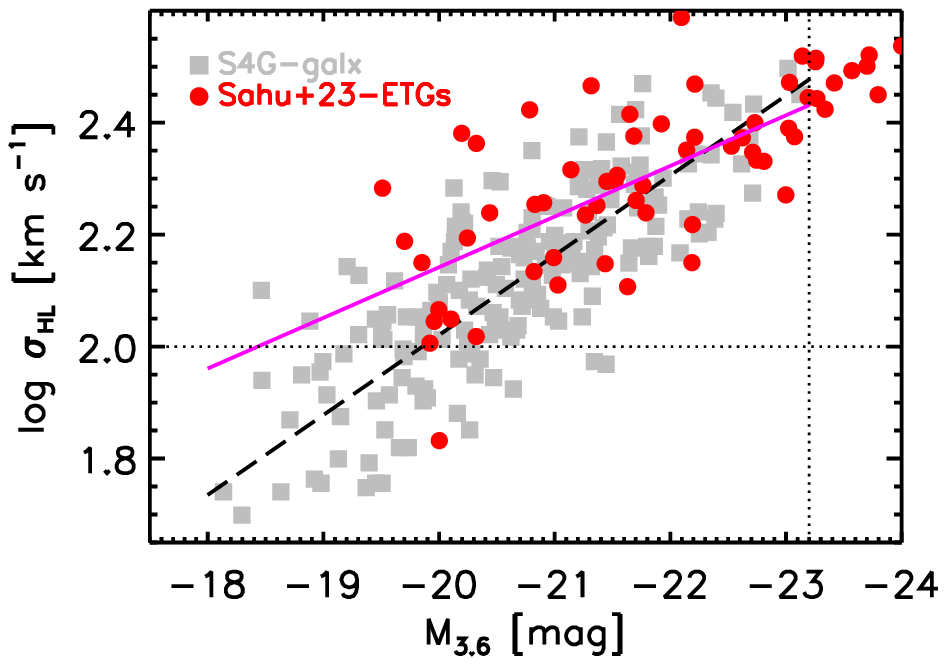,height=6.5cm}\hspace{-0.85cm}
\epsfig{figure=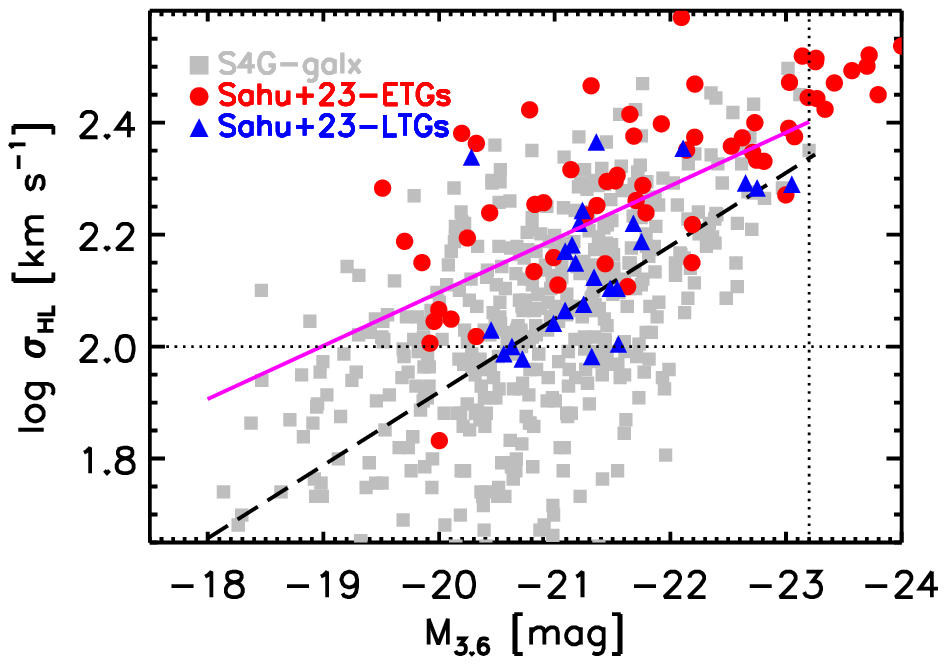,height=6.5cm}
}}
\caption{Scaling relation between the galaxy stellar velocity dispersion and galaxy magnitude at 3.6 $\mu$m for the local sample of early-type galaxies from the S$^4$G sample with \sis\ measurements from the Hyperleda database (filled black circles and long-dashed lines), compared with the SMBH sample collected by \citep{Sahu23} at 3.6 $\mu$m sample with Spitzer photometry. The left panel only includes ETG galaxies from \citep{Sahu23} and ETGs from the S$^4$G sample with T-type $< 1$, while the right panel includes also the LTGs from \citep{Sahu23} and T-type $< 6.5$ from the S$^4$G sample. The SMBH sample has a tendency to be characterized, on average, by larger stellar velocity dispersions at fixed luminosity.}
\label{figu|AppendixBias}
\end{center}
\end{figure*}

In this Appendix we revisit the long-standing issue of possible selections biases affecting the SMBH-galaxy scaling relations in the local Universe \citep[e.g.,][]{Batcheldor10}. Uncovering such biases is vital to pin down the true shape and time evolution of SMBH-galaxy scaling relations \citep[e.g.,][]{Shankar16,Farrah23}. \citet{Shankar16} also showed via extended Monte Carlo tests that selection effects are not expected to significantly impact the residual analysis, which therefore represents a powerful tool to extract the intrinsic correlations (in terms of slopes) between SMBH mass and galaxy properties. %In this Appendix, we revisit the issue of the existence of a possible selection effect between inactive galaxies with SMBH dynamical mass measurements and the larger galaxy population. 

Following the seminal papers by \citet{Bernardi07}, \citet{Gultekin11}, and \citet{MorabitoDai12}, \citet{Shankar16} showed that several local samples of galaxies with dynamically measured masses of their central SMBHs tend to show, on average, higher stellar velocity dispersions at fixed galaxy stellar mass than what predicted by the mean distribution of galaxies in the SDSS galaxy sample. This result was derived by comparing the local sample of SDSS galaxies with four independent SMBH galaxy samples with distinct galactic photometries from \citet{Savorgnan16}, \citet{Beifiori12}, \citet{Laesker14}, and \citet{McConnellMa}. SDSS galaxy stellar masses were derived from S\'{e}rsic \citep{Sersic} plus Exponential fits by \citet{Bernardi14} and colour-dependent mass-to-light ratios from \citet{Bell03}. To infer host galaxy luminosities, \citet{Savorgnan16} adopted 3.6$\mu$m \textit{Spitzer} images with S\'{e}rsic profiles plus, wherever relevant, additional components such as bars and rings. \citet{Shankar16} also included in their analysis galaxies from the original \citet{McConnellMa} sample with 3.6 $\mu$m luminosities derived from the S\'{e}rsic plus Exponential fits by \citet{Sani11}. \citet{Beifiori12} instead extracted homogeneous host galaxy luminosities from bulge-to-disc decompositions of SDSS $i$-band images, from which \citet{Shankar16} derived stellar masses using the same colour-dependent mass-to-light ratios from \citet{Bell03} self-consistently adopted for the comparison SDSS galaxy sample. \citet{Laesker14} extracted galaxy $K$-band photometries from deep, high spatial resolution images obtained from the wide-field WIRCam imager at the Canada–France–Hawaii–Telescope, and luminosities were then converted to stellar masses by \citet{Shankar16} using an average standard mass-to-light ratio. 

\begin{figure}
\begin{center}
\center{{
\epsfig{figure=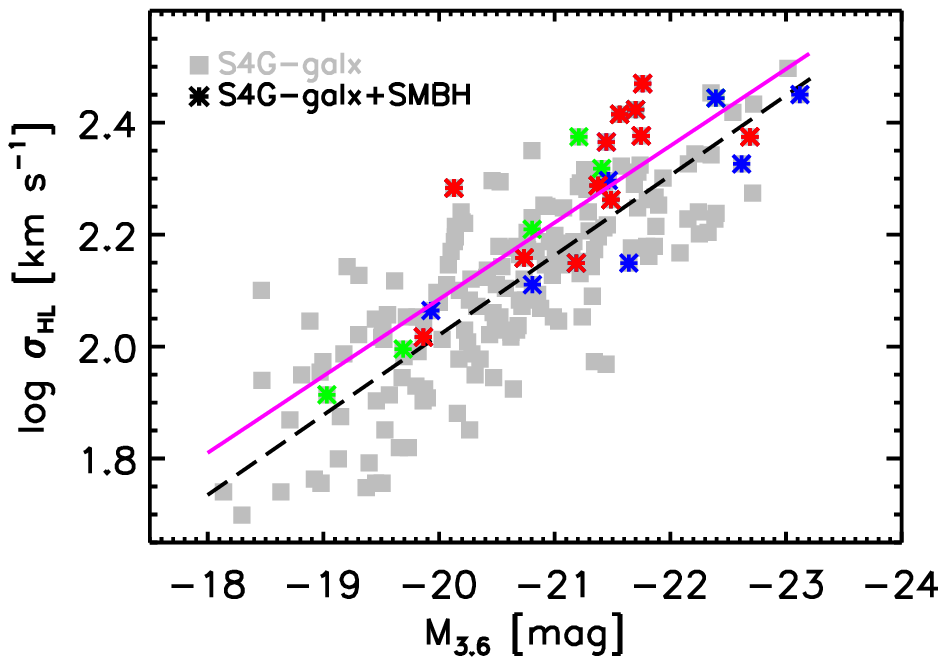,height=6.5cm}
}}
\caption{Scaling relation between the galaxy stellar velocity dispersion and galaxy magnitude at 3.6 $\mu$m for the local sample of early-type galaxies from the S$^4$G sample with \sis\ measurements from the Hyperleda database (filled black circles and long-dashed lines). The coloured symbols mark the galaxies from the original samples adopted by \citet{Shankar16} which are in common to both Hyperleda and the ETG S$^4$G samples, namely from \citet[][green stars]{Beifiori12} and \citet[][red stars]{Savorgnan16}, and some additional galaxies from \citet[][blue stars]{Sahu23}. It is evident that the original sample adopted by \citet{Shankar16} shows a sharp offset in stellar velocity dispersion at fixed magnitude, which is slightly reduced when including the new galaxies from \citet[][blue stars]{Sahu23}.}
        \label{figu|AppendixBiasFewGalaxies}
\end{center}
\end{figure}

\begin{figure*}
\begin{center}
\center{{
\epsfig{figure=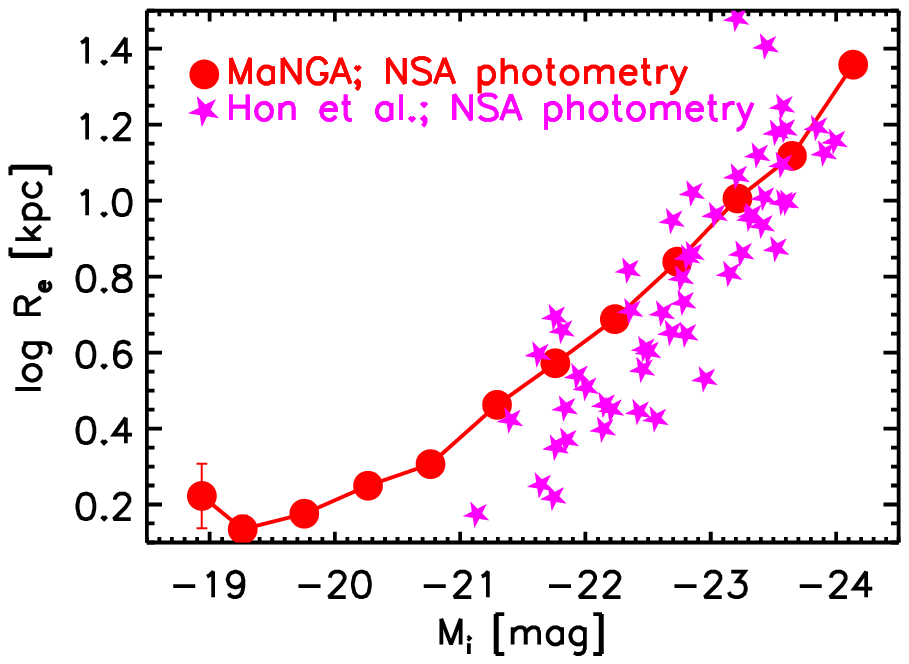,height=6.5cm}\hspace{-0.85cm}
\epsfig{figure=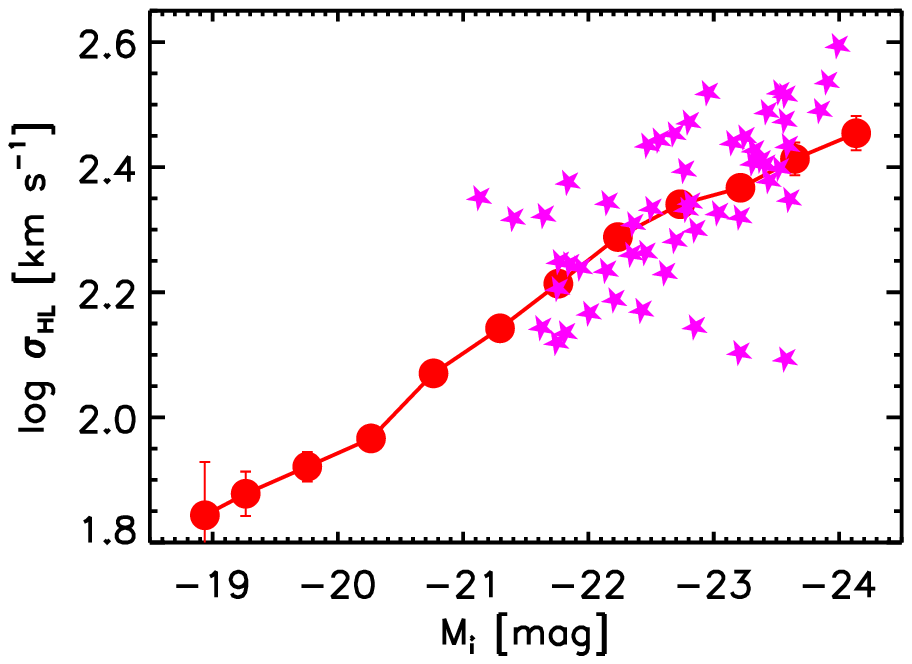,height=6.5cm}
}}
\caption{\textit{Left panel}: Mean scaling between effective radius and $i$-band magnitude extracted from the local MaNGA sample with NSA photometry (filled, red circles) against the \citet{Hon22} data sets of local compact galaxies with the same NSA photometry (magenta stars). \textit{Right panel}: Same format as the left panel but in the stellar velocity dispersion (rescaled to the Hyperleda aperture) vs $i$-band magnitude plane. The \citet{Hon22} galaxies tend to be more compact than the average MaNGA galaxies and with slightly larger stellar velocity dispersion, in particular at bright luminosities.}
        \label{figu|AppendixBiasHon}
\end{center}
\end{figure*}

\citet{Sahu23} criticised the \citet{Shankar16} result of a systematic bias between SMBH and galaxy samples reducing it to a simple byproduct of systematic discrepancies in the stellar mass scales between SDSS galaxies and the SMBH galaxy samples, rather than induced by offsets in stellar velocity dispersion at fixed host galaxy stellar mass. To support their claim, \citet{Sahu23} focused specifically on the local sample of SMBHs by \citet{Savorgnan16} at 3.6 $\mu$m, further enriched by \citet{Sahu19a}, i.e., the one also adopted in this work, and ignored the other local samples of SMBHs with different photometries included in \citet{Shankar16} and summarised above. To estimate the mean stellar mass correction between the stellar masses derived from 3.6 $\mu$m photometry and SDSS imaging, \citet{Sahu23} followed two steps. First, they took a sample of 43 galaxies in common between SDSS and the S$^4$G local sample of galaxies with \textit{Spitzer} photometry, and estimated a mean discrepancy of $\sim 0.2$ dex between the two stellar mass systems. The SDSS galaxy subsample they chose for this step had total galaxy luminosities derived by \citet{Hon22} from the SDSS $i$ band. \citet{Sahu23} then checked that the 37 galaxies in common between the S$^4$G sample and the \citet{Sahu19a} SMBH sample have very similar photometry, and thus the same stellar mass correction between SDSS and S$^4$G could be applied between SDSS and the \citet{Sahu19a} SMBH sample, an offset which is sufficient to remove any apparent systematic offset (``bias'') between the two \sis-\mgal\ relations followed by the SMBH galaxy and SDSS samples. 

%while the right panel also includes spirals up to T-type $\le 3$. In \figu\ref{figu|AppendixBias} we compare the \S4G\ sample (gray squares) with the SMBH sample collected by \citet{Sahu23}, only ETGs in the left panel and both ETGs and LTGs in the right panel (red circles and blue triangles, respectively). 
The methodology followed by \citet{Sahu23} makes the fundamental and untested assumption that the S$^4$G and local SMBH samples follow the same identical stellar velocity dispersion-3.6 $\mu$m luminosity relation, and thus the same stellar mass correction can be applied to both the S$^4$G and SMBH samples when comparing with SDSS galaxies. However, we have verified, that this is not the case. We have cross-correlated the S$^4$G sample with the Hyperleda database and the resulting sample is reported in \figu\ref{figu|AppendixBias} with gray, filled squares. The left panel only includes S$^4$G ETGs with T-type $\le 1$ to include S0/a and Ellipticals, as in the reference ETG sample from \citet{Sahu23}. It is clear that the ETG SMBH sample does \textit{not} strictly follow the distribution of the S$^4$G sample. The vast majority of ETGs below $M_i \lesssim -22$ (approximately $\mstare\sim 10^{11}\, \msune$, using the \citet{Sahu23} mass-to-light ratio at $3.6\mu$m of $M_{\odot}/L_{\odot}$=0.6) lie above the mean relation traced by the \S4G\ sample. To remark this point, we show two linear fits to the \citet{Sahu23} and S$^4$G samples (solid magenta and long-dashed black line, respectively), where we have assumed the same statistical errors on magnitude and $\log \sise$ of 0.2 mag and 0.1 dex, respectively. The ETG SMBH sample tends to be skewed towards larger stellar velocity dispersions, in particular there are virtually no galaxies with SMBH masses below 100 km/s (horizontal dotted lines). %These lines are simply meant to guide the eye, as a proper fit would require detailed measurement errors of the stellar velocity dispersions in both samples, but still show a tendency for the SMBH sample to be skewed towards larger stellar velocity dispersions, in particular there are virtually no galaxies with SMBH masses below 100 km/s (horizontal dotted lines). 

The discrepancy between the S$^4$G and the SMBH samples becomes even more pronounced in the right panel of \figu\ref{figu|AppendixBias} where we include all the late-type galaxies of \citet{Sahu23} and all the corresponding \S4G\ galaxies matched in morphology with T-type $< 6.5$, e.g., down to Sc galaxies. The two samples reach an average offset of up to $\sim 0.2$ dex in $\log \sise$ at fixed luminosity. In addition, the S$^4$G sample with velocity dispersion measurements in the Hyperleda database is around 1/4 of the original sample, and thus it could suffer from incompleteness. We stress that the photometric systems and definitions of total host galaxy luminosity adopted in the \S4G\ and the \citet{Sahu23} samples are not strictly identical and could further bias the comparison. A dedicated study tailored at applying the same measurement techniques and assumptions in a (complete) reference galaxy sample and the local SMBH sample is required to truly advance in our understanding of systematic biases in the SMBH scaling relations traced by the local sample of dynamically measured SMBHs. 

We also note that the subsample of SDSS galaxies chosen by \citet{Sahu23} to calibrate the galaxy stellar mass offset with the S$^4$G local sample is not ideal, being itself biased with respect to the full galaxy population as described by the MaNGA survey, being more compact at fixed host galaxy luminosity, and with a tendency to have larger stellar velocity dispersion, especially at $M_i \lesssim -22.5$, as seen in \figu\ref{figu|AppendixBiasHon}.  

It is true that with an increased size in the SMBH sample and more accurate photometry the offset in stellar velocity dispersion between the SMBH sample compared to the galaxy sample is somewhat reduced, as also noted by \citet{Sahu23}. Nonetheless, the offset is still noticeable at lower luminosities, approaching $\Delta \log \sise \sim 0.1$ at $M_i \lesssim -22$ when comparing the two full samples, and even larger at fainter luminosities. Note that an offset of $\gtrsim 0.2$ dex in $\log \sise$ at fixed luminosity would correspond to a noticeable average offset in SMBH mass of a factor of $\sim 5$, if $\mbhe \propto \sise^{3.6}$, as inferred from the residual analysis presented in the top left panel of \figu\ref{fig|residuals_sahu}. For completeness, \figu\ref{figu|AppendixBiasFewGalaxies} compares the S$^4$G ETG-Hyperleda sample with the fraction of inactive ETGs with SMBHs in common to both S$^4$G and Hyperleda from the original SMBH sample collected by \citet[][red stars]{Shankar16}, plus some galaxies from the SMBH sample of \citet[][green stars]{Beifiori12} and some additional galaxies with SMBH from \citet[][blue stars]{Sahu23}. It is clear that, overall, the original sample adopted by \citet{Shankar16}, from \citet{Savorgnan16}, is clearly biased high in stellar velocity dispersion at fixed luminosity. The addition of new galaxies from \citet{Sahu23} tends to lower the mean offset compared to the original sample adopted by \citet{Shankar16}.

%Some of the offset between the S$^4$G-Hyperleda sample and the \citet{Sahu23} SMBH sample may be induced by underlying differences in the definition of total host galaxy luminosity. A more robust test for dissecting the possible presence of a bias in any of the SMBH-galaxy scaling relation, should be carried out by measuring uniformly the full local SMBH sample (with over 100 galaxies) with the same exact methodology applied to the larger comparison galaxy sample (ideally SDSS or MaNGA), which is not yet available at the moment. 

We conclude that a bias between the local SMBH sample and the larger sample of galaxies without dynamical SMBH mass measurements, persists in the form of an offset in stellar velocity dispersion at fixed host galaxy luminosity, which is thus not a byproduct of different stellar mass-to-light ratios between the two samples. Such an offset in stellar velocity dispersion at fixed luminosity was also identified by \citet{Bernardi07}, \citet{VdB15}, and, more recently, by \citet{Kormendy20}, who reported in their Figure 2 a clear offset in \sis\ at fixed $L_V$ for both core and coreless galaxies. In addition, many local AGN appear to sit significantly below the scaling relations of dormant SMBHs, as discussed by, e.g., \citet{Reines2015} and \citet{Shankar19}, when adopting reasonable assumptions for the virial factors, and it is evident even among AGN samples calibrated with the same photometry as in the comparison galaxy sample. This result appears common to all local AGN independently of their host galaxy morphology, although some dependence of the offset on the Eddington ratio may be present \citep[e.g.,][]{Farrah23}.  

More recently, \citet{Byrne23} adopted an alternative approach to probe the existence of a selection bias in the SMBH mass-host galaxy stellar mass relation by computing the SMBH masses in a mass-complete sample of 18 ETGs from the Virgo cluster. They were able to extract SMBH masses in 11 out of the 18 galaxies, and thus, on the assumption that the remaining galaxies with undetected SMBHs do not contain a SMBH, they were able to place a conservative lower limit to the mean SMBH mass in their sample by dividing the sum of all their detected SMBH masses by 18 (instead of 11). They claimed a mean lower lower limit of $\mbhe =3.7\times 10^7\, \msune$ for host galaxies with mean total stellar mass $\mstare=(1.8\pm 1.1)\times 10^{10}\, \msune$. At face value, this value of the mean SMBH mass would be significantly higher than what predicted by, e.g., Model I in \citet[][their Eq. 6]{Shankar16}. However there are several caveats to be considered here before drawing any definitive conclusion. First off, the results by \citet{Shankar16} were based on Monte Carlo simulations applied to thousands of galaxies from the SDSS survey, and thus one would need to check the consistency between the (small) galaxy sample from \citet{Byrne23} with the much larger SDSS galaxy sample, in particular on the \sis-\mstar\ plane. We verified that, when cross-correlating the 18 galaxies in the \citet{Byrne23} sample with the Hyperleda database, the latter yields an average $\log \sigma_{\rm HL}\sim 2.02$, which, at face value, would be tentatively $\sim 0.1$ dex higher than the mean $\log \sigma_{\rm HL}$ calculated by \citet{Shankar16} at an average stellar mass of $\mstare=1.8 \times 10^{10}\, \msune$. However, possible systematic differences between the ATLAS dynamical mass measurements and the SDSS-based stellar masses from \citet{Shankar16} prevent a robust comparison between two samples. Secondly, several other indicators and/or independent measurements of SMBH masses tend to point to lower masses, as also highlighted by \citet{Byrne23} in their Figure 5, and as also stressed by \citet{Shankar19}. Last but not least, the Monte Carlo method put forward by \citet{Shankar16} did not include the larger SMBH sample by \citet{Sahu23} which, as we discussed above, tends to show a lower degree of bias, comparable to $\Delta \log \sise \sim 0.1$ dex at the mass scale probed by \citet{Byrne23}, as suggested by the left panel of \figu\ref{figu|AppendixBias}.%, corresponding to an average offset of $\sim 3$ in mean SMBH mass on the assumption of a relation of the type $\mbhe \propto \sise^5$.

%\onecolumn
\section{Additional relevant pairwise residual correlations}
\label{Appendix:MoreResiduals}

\begin{figure*}
\begin{center}
\center{{
\epsfig{figure=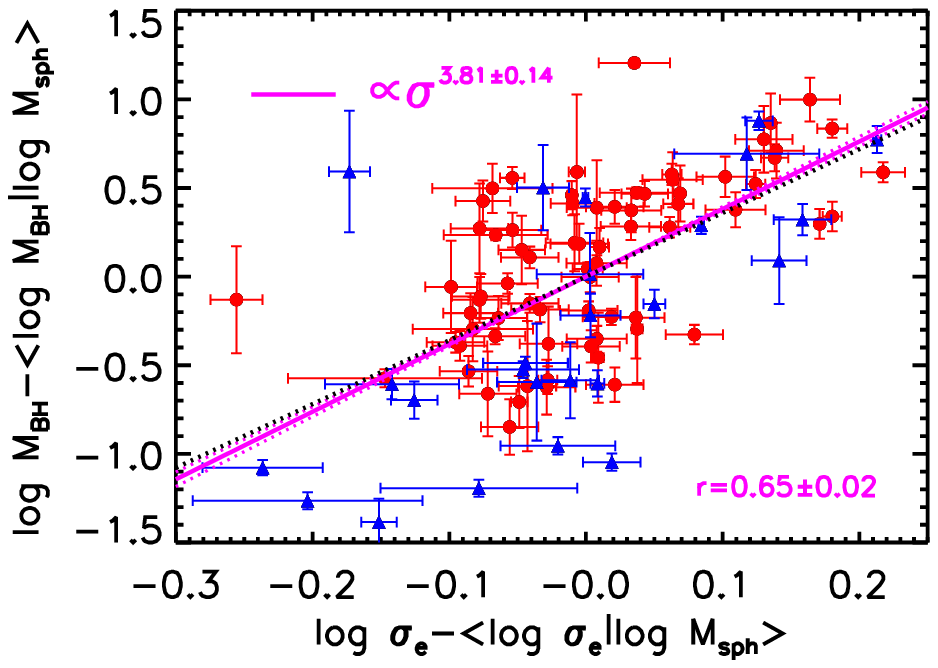,height=6.5cm}\hspace{-0.85cm}
\epsfig{figure=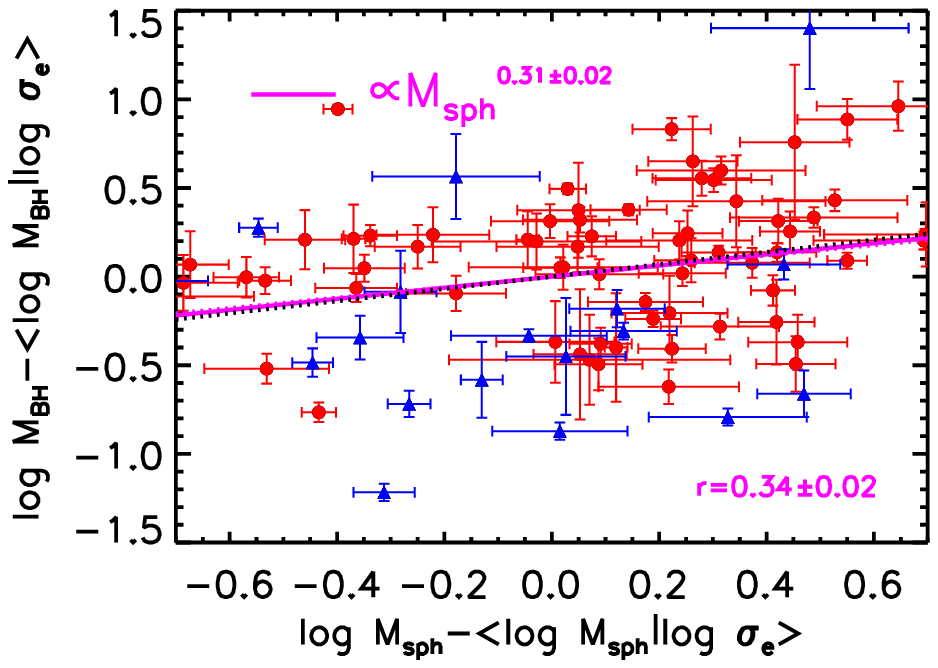,height=6.5cm}
}}
\caption{Pairwise residuals as a function of stellar velocity dispersion at fixed spheroidal mass (left) and vice versa (right) for the \citet{Saglia16} SMBH sample which only considers bulge stellar mass and central velocity dispersions measured at the half-light radius. Even in this sample we still find a stronger dependence on stellar velocity dispersion confirming the results retrieved from the \citet{Sahu23} sample when considering only the bulge component (\figu\ref{fig|residuals_sahuMsph}).}
        \label{figu|AppendixSaglia}
\end{center}
\end{figure*}

Here we present additional residuals derived for the \citet{Saglia16} SMBH sample, with accurate measurements of the stellar velocity dispersion within the effective radius and the stellar bulge component (\figu\ref{figu|AppendixSaglia}), along with additional residuals extracted from the \citet{Sahu23} SMBH sample as a function of half-light effective radius and S\'{e}rsic index. 

\begin{figure*}
\begin{center}
\center{{
\epsfig{figure=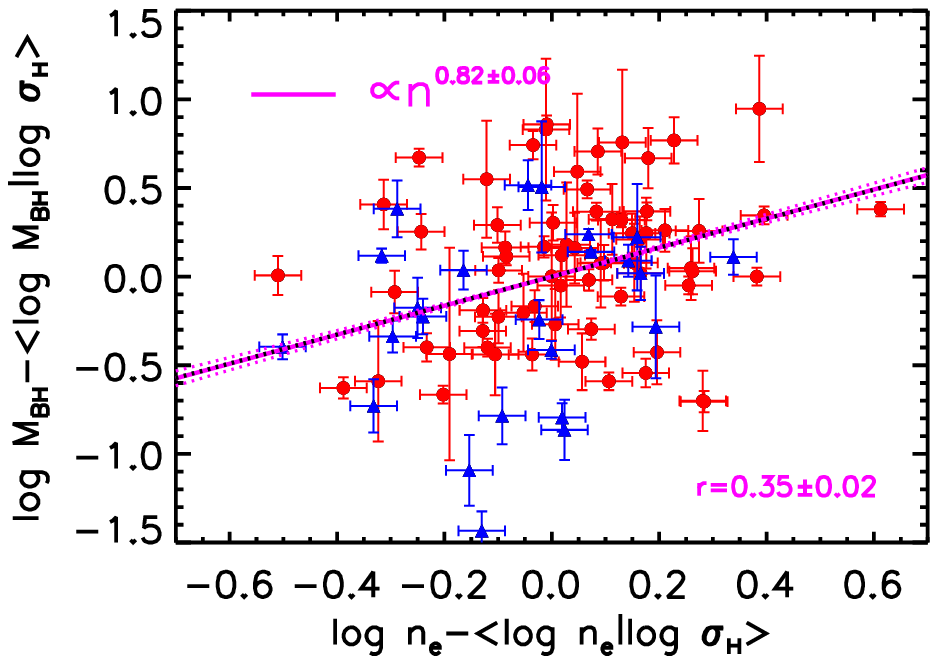,height=6.5cm}\hspace{-0.85cm}
\epsfig{figure=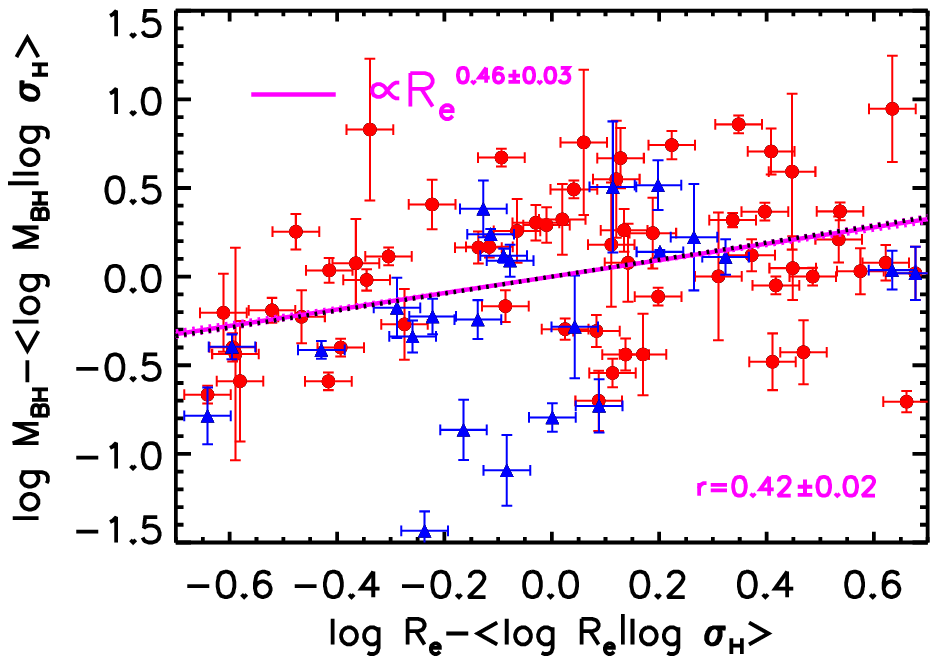,height=6.5cm}
}}
\vspace{-0.52cm}
\center{{
\epsfig{figure=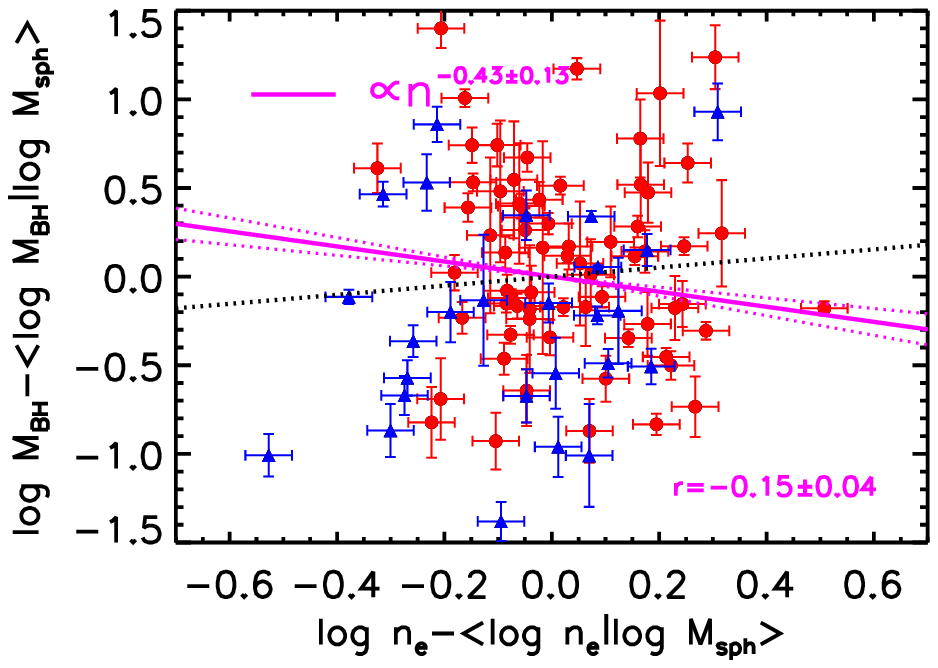,height=6.5cm}\hspace{-0.85cm}
\epsfig{figure=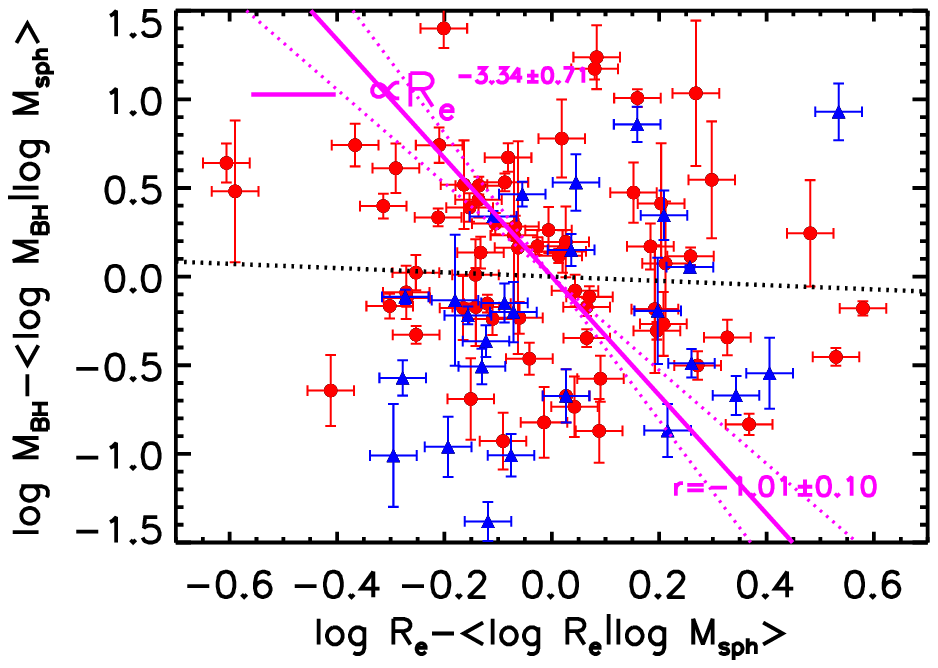,height=6.5cm}
}}
\caption{Pairwise residuals as a function of S\'{e}rsic index $n$ (left) and bulge effective radius $R_e$ (right) at fixed stellar velocity dispersion (top) and spheroidal mass (bottom) for the \citet{Sahu23} SMBH sample. We do not find any significant residual dependence of SMBH mass on any of these quantities when the underlying dependence on stellar velocity dispersion and/or spheroidal (bulge) mass is subtracted.}
        \label{figu|AppendixSersic}
\end{center}
\end{figure*}

\clearpage 

%\label{lastpage}
%\end{document}

%\newpage
%\restylefloat{table}

\section{Data for the SMBH sample adopted in this work}

\vspace{-5cm}

\begin{table*}%[H]
    \centering
    \caption{Data for the 69 Early Type Galaxies (ETGs). $\sigma$ and $\delta\sigma$ have units of km\,s$^{-1}$. All magnitudes are AB at 3.6 $\mu$m. Both effective radius and S\'{e}rsic index are the ``equivalent'' measures from \citet{Sahu2020} (continued over the next page)}
    \begin{tabular}{|c|c|c|c|c|c|c|c|c|c|}
        \hline
        Name & $\log\left(\frac{\rm M_{\text{BH}}}{{\rm M_{\odot}}}\right)$ & $\delta\log\left(\frac{\rm M_{\text{BH}}}{{\rm M_{\odot}}}\right)$ & $\log\left(\frac{\rm M_{\text{sph}}}{{\rm M_{\odot}}}\right)$ & $\sigma$ & $\delta\sigma$ & $\text{R}_e(")$ & $n$ & Distance (Mpc) & Mag \\
        \hline
IC4296 & 9.04 & 0.09 & 11.47 & 327.3 & 5.4 & 41.1 & 3.82 & 40.7 & -23.3 \\
IC1459 & 9.38 & 0.20 & 11.55 & 295.8 & 6.4 & 57.3 & 7.00 & 28.4 & -23.4 \\
NGC0404 & 4.85 & 0.13 & 7.96 & 34.6 & 3.1 & 3.89 & 0.90 & 3.1 & -17.3 \\
NGC0524 & 8.92 & 0.10 & 10.57 & 236.6 & 4.5 & 8.35 & 2.16 & 23.3 & -22.2 \\
NGC0821 & 7.59 & 0.17 & 10.69 & 197.7 & 2.8 & 18.9 & 6.10 & 23.4 & -21.5 \\
NGC1023 & 7.62 & 0.05 & 10.21 & 197.2 & 4.6 & 7.4 & 2.00 & 11.1 & -21.5 \\
NGC1275 & 8.90 & 0.20 & 11.84 & 244.9 & 12.8 & 53.6 & 4.31 & 72.9 & -24.2 \\
NGC1332 & 9.16 & 0.07 & 11.05 & 294.4 & 11.3 & 18.0 & 3.70 & 22.3 & -22.2 \\
NGC1374 & 8.76 & 0.05 & 10.22 & 179.5 & 3.2 & 11.74 & 1.65 & 19.2 & -20.8 \\
NGC1399 & 8.67 & 0.06 & 11.66 & 331.9 & 5.3 & 338.1 & 10.00 & 19.4 & -23.7 \\
NGC1407 & 9.65 & 0.08 & 11.46 & 265.5 & 5.1 & 47.29 & 3.89 & 28 & -23.3 \\
NGC1600 & 10.23 & 0.05 & 11.82 & 331.1 & 7.0 & 49.58 & 5.08 & 64 & -24.1 \\
NGC2549 & 7.15 & 0.60 & 9.59 & 141.3 & 2.7 & 3.1 & 1.50 & 12.3 & -19.9 \\
NGC2778 & 7.18 & 0.34 & 9.41 & 154.2 & 3.2 & 2.2 & 1.20 & 22.3 & -19.7 \\
NGC2787 & 7.60 & 0.06 & 9.13 & 191.9 & 3.9 & 2.88 & 1.27 & 7.3 & -19.5 \\
NGC3091 & 9.56 & 0.04 & 11.61 & 311.2 & 7.7 & 51.2 & 6.60 & 51.2 & -23.6 \\
NGC3115 & 8.94 & 0.25 & 10.77 & 260.0 & 3.0 & 34.4 & 5.10 & 9.4 & -21.6 \\
NGC3245 & 8.30 & 0.12 & 10.06 & 207.0 & 7.3 & 2.4 & 1.70 & 20.3 & -21.1 \\
NGC3377 & 7.89 & 0.04 & 10.48 & 136.1 & 2.3 & 91.7 & 9.20 & 10.9 & -20.8 \\
NGC3379 & 8.60 & 0.12 & 10.8 & 202.3 & 1.8 & 50.9 & 5.30 & 10.3 & -21.5 \\
NGC3384 & 7.23 & 0.05 & 10.06 & 144.2 & 2.5 & 5.6 & 1.80 & 11.3 & -21 \\
NGC3414 & 8.38 & 0.06 & 10.83 & 237.7 & 8.1 & 25.5 & 4.50 & 24.5 & -21.7 \\
NGC3489 & 6.76 & 0.07 & 9.54 & 104.2 & 2.0 & 1.7 & 1.30 & 11.7 & -20.3 \\
NGC3585 & 8.49 & 0.13 & 11.3 & 214.3 & 5.1 & 86.3 & 6.30 & 19.5 & -22.8 \\
NGC3607 & 8.11 & 0.18 & 11.23 & 222.3 & 4.1 & 65.5 & 5.60 & 22.2 & -22.7 \\
NGC3608 & 8.30 & 0.18 & 10.89 & 194.1 & 4.2 & 43.4 & 5.70 & 22.3 & -21.8 \\
NGC3665 & 8.76 & 0.10 & 11.03 & 215.3 & 8.5 & 12.78 & 2.74 & 34.7 & -22.7 \\
NGC3842 & 9.99 & 0.13 & 11.92 & 308.3 & 6.7 & 73.6 & 8.20 & 98.4 & -24.4 \\
NGC3923 & 9.45 & 0.13 & 11.4 & 245.5 & 4.9 & 78.78 & 4.77 & 20.9 & -23 \\
NGC3998 & 8.91 & 0.11 & 10.02 & 264.9 & 11.0 & 4.8 & 1.30 & 13.7 & -20.8 \\
NGC4026 & 8.26 & 0.11 & 10.11 & 173.4 & 3.8 & 2.35 & 3.98 & 13.2 & -20.4 \\
NGC4261 & 8.70 & 0.09 & 11.38 & 296.5 & 4.3 & 47.3 & 4.30 & 30.8 & -23 \\
NGC4291 & 8.52 & 0.05 & 10.71 & 292.4 & 6.9 & 15.4 & 5.90 & 25.5 & -21.3 \\
NGC4339 & 7.63 & 0.33 & 9.67 & 110.9 & 3.1 & 6.42 & 1.40 & 16.0 & -20 \\
NGC4342 & 8.65 & 0.18 & 9.94 & 240.4 & 5.7 & 4.69 & 3.99 & 23.0 & -20.2 \\
NGC4350 & 8.86 & 0.41 & 10.28 & 180.7 & 4.4 & 19.45 & 3.97 & 16.8 & -20.9 \\
NGC4371 & 6.85 & 0.08 & 9.89 & 128.8 & 2.2 & 8.9 & 3.19 & 16.9 & -21 \\
NGC4374 & 8.95 & 0.05 & 11.49 & 277.3 & 2.4 & 129.8 & 7.90 & 17.9 & -23.3 \\
NGC4429 & 8.18 & 0.09 & 10.46 & 173.4 & 5.4 & 11.29 & 2.31 & 16.5 & -21.8 \\
NGC4434 & 7.85 & 0.17 & 9.91 & 116.4 & 2.8 & 5.31 & 2.93 & 22.4 & -20 \\
NGC4459 & 7.83 & 0.09 & 10.48 & 171.8 & 4.8 & 13.0 & 2.60 & 15.7 & -21.3 \\
NGC4472 & 9.40 & 0.05 & 11.7 & 281.8 & 2.9 & 135.3 & 5.40 & 17.1 & -23.8 \\
NGC4473 & 8.08 & 0.36 & 10.64 & 178.6 & 2.5 & 36.9 & 2.90 & 15.3 & -21.4 \\
NGC4486 & 9.81 & 0.05 & 11.49 & 322.8 & 4.3 & 87.1 & 5.90 & 16.8 & -23.3 \\
NGC4526 & 8.67 & 0.05 & 10.7 & 224.4 & 9.4 & 14.88 & 2.96 & 16.9 & -22.1 \\
NGC4552 & 8.67 & 0.05 & 10.88 & 250.0 & 2.9 & 71.5 & 5.36 & 14.9 & -21.9 \\
NGC4564 & 7.78 & 0.06 & 10.01 & 156.3 & 2.2 & 6.0 & 3.00 & 14.6 & -20.2 \\
NGC4578 & 7.28 & 0.35 & 9.77 & 111.9 & 4.1 & 6.32 & 1.99 & 16.3 & -20.1 \\
NGC4596 & 7.90 & 0.20 & 10.18 & 140.6 & 2.2 & 9.0 & 3.00 & 17.0 & -21.4 \\
NGC4621 & 8.59 & 0.05 & 11.16 & 228.0 & 3.8 & 90.9 & 8.80 & 17.8 & -22.5 \\
NGC4649 & 9.67 & 0.10 & 11.44 & 330.4 & 4.6 & 80.59 & 5.21 & 16.4 & -23.1 \\
        \hline
    \end{tabular}
\label{Table 1}
\end{table*}
%\label{lastpage}
%\end{document}

%\newpage

\begin{table*}
\ContinuedFloat
    \centering
    \caption{Data for the 69 Early Type Galaxies (ETGs) - continued. $\sigma$ and $\delta\sigma$ have units of km\,s$^{-1}$. All magnitudes are AB at 3.6 $\mu$m. Both effective radius and S\'{e}rsic index are the ``equivalent'' measures from \citet{Sahu2020}.}
    \begin{tabular}{|c|c|c|c|c|c|c|c|c|c|}
        \hline
        Name & $\log\left(\frac{\rm M_{\text{BH}}}{{\rm M_{\odot}}}\right)$ & $\delta\log\left(\frac{\rm M_{\text{BH}}}{{\rm M_{\odot}}}\right)$ & $\log\left(\frac{\rm M_{\text{sph}}}{{\rm M_{\odot}}}\right)$ & $\sigma$ & $\delta\sigma$ & $\text{R}_e(")$ & $n$ & Distance (Mpc) & Mag \\
        \hline
NGC4697 & 8.26 & 0.05 & 11.01 & 165.2 & 1.6 & 226.4 & 6.70 & 11.4 & -22.2 \\
NGC4742 & 7.15 & 0.18 & 9.87 & 101.4 & 3.4 & 3.41 & 3.20 & 15.5 & -19.9 \\
NGC4762 & 7.36 & 0.15 & 9.97 & 141.3 & 4.1 & 2.24 & 1.85 & 22.6 & -22.2 \\
NGC4889 & 10.32 & 0.44 & 12.14 & 392.6 & 5.3 & 60.8 & 6.80 & 103.2 & -24.9 \\
NGC5077 & 8.87 & 0.22 & 11.28 & 251.2 & 5.5 & 23.0 & 5.70 & 41.2 & -22.7 \\
NGC5252 & 9.00 & 0.40 & 10.85 & 186.6 & 26.5 & 1.47 & 2.95 & 96.8 & -23 \\
NGC5419 & 9.86 & 0.14 & 11.45 & 344.3 & 5.4 & 16.83 & 2.62 & 56.2 & -24 \\
NGC5576 & 8.20 & 0.10 & 10.87 & 182.4 & 6.0 & 49.3 & 3.70 & 24.8 & -21.7 \\
NGC5813 & 8.83 & 0.06 & 10.86 & 236.0 & 3.4 & 14.16 & 3.65 & 31.3 & -22.6 \\
NGC5845 & 8.41 & 0.22 & 10.12 & 230.7 & 7.9 & 5.29 & 3.27 & 25.2 & -20.3 \\
NGC5846 & 9.04 & 0.05 & 11.42 & 237.1 & 3.5 & 83.4 & 5.70 & 24.2 & -23.1 \\
NGC6251 & 8.77 & 0.16 & 11.82 & 312.6 & 18.2 & 30.1 & 5.60 & 104.6 & -24.1 \\
NGC6861 & 9.30 & 0.08 & 10.94 & 387.3 & 16.5 & 20.13 & 3.52 & 27.3 & -22.1 \\
NGC7052 & 8.57 & 0.23 & 11.46 & 278.6 & 11.8 & 20.04 & 3.46 & 66.4 & -23.2 \\
NGC7332 & 7.11 & 0.20 & 10.22 & 127.9 & 3.3 & 2.43 & 2.15 & 24.9 & -21.6 \\
NGC7457 & 7.00 & 0.30 & 9.40 & 67.9 & 3.5 & 6.51 & 2.84 & 14 & -20 \\
NGC7619 & 9.40 & 0.09 & 11.64 & 317.0 & 4.9 & 58.0 & 5.20 & 51.5 & -23.7 \\
NGC7768 & 9.11 & 0.15 & 11.89 & 289.7 & 11.9 & 42.1 & 6.70 & 112.8 & -24.2 \\
        \hline
    \end{tabular}
\end{table*}

\begin{table*}
    \centering
    \caption{Data for the 26 Late Type Galaxies (LTGs). $\sigma$ and $\delta\sigma$ have units of km\,s$^{-1}$. All magnitudes are AB at 3.6 $\mu$m. Both effective radius and S\'{e}rsic index are the ``equivalent'' measures from \citet{Sahu2020}.}
    \begin{tabular}{|c|c|c|c|c|c|c|c|c|c|}
        \hline
        Name & $\log\left(\frac{\rm M_{\text{BH}}}{{\rm M_{\odot}}}\right)$ & $\delta\log\left(\frac{\rm M_{\text{BH}}}{{\rm M_{\odot}}}\right)$ & $\log\left(\frac{\rm M_{\text{sph}}}{{\rm M_{\odot}}}\right)$ & $\sigma$ & $\delta\sigma$ & $\text{R}_e(")$ & $n$ & Distance (Mpc) & Mag \\
        \hline
Circinus & 6.25 & 0.11 & 10.12 & 148.0 & 18.0 & 23.13 & 1.80 & 4.2 & -21.1 \\
IC2560 & 6.49 & 0.20 & 9.63 & 141.0 & 10.0 & 3.92 & 1.63 & 31.0 & -21.2 \\
NGC0224 & 8.15 & 0.16 & 10.11 & 154.0 & 4.0 & 173.6 & 1.30 & 0.8 & -21.8 \\
NGC0253 & 7.00 & 0.30 & 9.76 & 96.0 & 18.0 & 27.89 & 2.33 & 3.5 & -21.3 \\
NGC1097 & 8.38 & 0.04 & 10.83 & 195.0 & 5.0 & 11.39 & 1.52 & 24.9 & -23.1 \\
NGC1300 & 7.71 & 0.16 & 9.42 & 218.0 & 29.0 & 7.39 & 2.83 & 14.5 & -20.3 \\
NGC1320 & 6.78 & 0.29 & 10.25 & 110.0 & 10.0 & 2.23 & 2.87 & 37.7 & -21 \\
NGC1398 & 8.03 & 0.11 & 10.57 & 196.0 & 18.0 & 10.38 & 3.00 & 24.8 & -22.6 \\
NGC2960 & 7.06 & 0.17 & 10.44 & 166.0 & 16.0 & 2.19 & 2.86 & 71.1 & -21.7 \\
NGC2974 & 8.23 & 0.07 & 10.23 & 232.0 & 4.0 & 6.53 & 1.17 & 21.5 & -21.4 \\
NGC3031 & 7.83 & 0.09 & 10.16 & 152.0 & 2.0 & 42.98 & 3.46 & 3.5 & -21.1 \\
NGC3079 & 6.38 & 0.12 & 9.92 & 175.0 & 12.0 & 4.35 & 0.58 & 16.5 & -21.2 \\
NGC3227 & 7.88 & 0.14 & 10.04 & 127.0 & 6.0 & 8.34 & 1.90 & 21.1 & -21.5 \\
NGC3368 & 6.89 & 0.09 & 9.81 & 119.0 & 4.0 & 4.83 & 1.00 & 10.7 & -21.2 \\
NGC3627 & 6.95 & 0.05 & 9.74 & 127.0 & 6.0 & 3.92 & 2.10 & 10.6 & -21.5 \\
NGC4151 & 7.68 & 0.37 & 10.27 & 116.0 & 3.0 & 6.0 & 1.85 & 19.0 & -21.1 \\
NGC4258 & 7.60 & 0.01 & 10.05 & 133.0 & 7.0 & 26.4 & 2.60 & 7.6 & -21.3 \\
NGC4303 & 6.58 & 0.17 & 9.42 & 95.0 & 8.0 & 2.16 & 0.90 & 12.3 & -20.7 \\
NGC4388 & 6.90 & 0.11 & 10.07 & 100.0 & 10.0 & 14.3 & 1.15 & 17.8 & -20.6 \\
NGC4501 & 7.13 & 0.08 & 10.11 & 166.0 & 7.0 & 20.35 & 2.83 & 11.2 & -21.2 \\
NGC4594 & 8.81 & 0.03 & 10.81 & 226.0 & 3.0 & 41.36 & 4.24 & 9.6 & -22.1 \\
NGC4699 & 8.34 & 0.10 & 11.12 & 192.0 & 9.0 & 29.75 & 6.77 & 23.7 & -22.8 \\
NGC4736 & 6.78 & 0.10 & 9.89 & 107.0 & 4.0 & 9.65 & 1.03 & 4.4 & -20.4 \\
NGC4826 & 6.07 & 0.15 & 9.55 & 97.0 & 6.0 & 11.93 & 0.76 & 5.6 & -20.6 \\
NGC5055 & 8.94 & 0.10 & 10.49 & 101.0 & 3.0 & 43.52 & 1.76 & 8.9 & -21.6 \\
NGC7582 & 7.67 & 0.09 & 10.15 & 147.0 & 19.0 & 4.55 & 2.21 & 19.9 & -21.5 \\
        \hline
    \end{tabular}
\label{Table 2}
\end{table*}

\label{lastpage}
\end{document}